\documentclass{ws-procs9x6}

\def\eg{{\it e.g.}}
\def\ie{{\it i.e.}}
\def\micros{$\mu$s}
\def\micron{$\mu$m}

\newcommand{\perkkgd}{\ensuremath{\textrm{keV}^{-1}\textrm{kg}^{-1}\textrm{day}^{-1}}} 
\newcommand{\perkeekgd}{\ensuremath{\textrm{keVee}^{-1}\textrm{kg}^{-1}\textrm{day}^{-1}}} 
\newcommand{\perkgd}{/(kg day)} 
\def\gev{GeV/$c^2$}

\newcommand{\lapprox}{\ensuremath{\lesssim}}
\newcommand{\gapprox}{\ensuremath{\gtrsim}}

\begin{document}

\title{Introduction to Dark Matter Experiments}

\author{R. W. Schnee$^*$}

\address{Department of Physics, Syracuse University,\\
Syracuse, New York 12344, United States\\
$^*$E-mail: rwschnee@phy.syr.edu\\
}

\begin{abstract}
This is a set of four lectures presented at the Theoretical Advanced Study Institute (TASI-09) in June 2009.
I provide an introduction to experiments designed to detect WIMP dark matter 
directly, focussing on building intuitive understanding of the characteristics of potential WIMP signals and the experimental techniques.
After deriving the characteristics of potential signals in direct-detection experiments for standard WIMP models, I summarize the general experimental methods shared by most direct-detection experiments and review the advantages, challenges, and status of such searches (as of late 2009). 
Experiments are already probing SUSY models, with best limits on the  spin-independent coupling
below $10^{-7}$\,pb. 
\end{abstract}


\bodymatter

\section{Introduction}\label{section:introduction}
A variety of astrophysical observations~\cite{wmap2008,TytlerBBN2000}
indicate that 83\% of the matter in the Universe is nonbaryonic and dark,
presumably in the form of elementary particles produced in the early Universe. 
Because no such particles have yet been identified in particle accelerators,
these observations require new fundamental particle physics.
Weakly Interacting Massive Particles (WIMPs~\cite{steigmanturner}) form a particularly interesting 
generic class of new-particle candidates
because they naturally provide about the inferred amount of this nonbaryonic dark matter~\cite{leeweinberg}, a result dubbed the ``WIMP miracle.'' 

WIMPs would be produced thermally in the early Universe.
Because they interact only weakly,
 their annihilation rate would become insignificant as the
Universe expands, thus ``freezing out'' a relic abundance of the particles
(for a pedagogical discussion, see \eg\ Refs.~\refcite{jkg,HooperTASI2008,FengReview, weinerTASI2009}).
The expected WIMP density would be the same as that of the nonbaryonic dark matter if
the WIMP velocity-averaged annihilation cross section is 
$\sim$1\,pb, so that the WIMP mass is 
$\sim$100\,\gev.

Remarkably, extensions to the Standard Model
motivated  entirely by particle physics predict
particles with the same cross section and mass as these dark-matter WIMPs.
Detection of the $W$ and $Z$ bosons 
indicates that the electroweak symmetry
is spontaneously broken at a scale of $\sim 100$\,\gev.
Whatever physics 
solves the hierarchy problem associated with this symmetry breaking -- 
be it supersymmetry~\cite{jkg,ellis,gondolosusy,bertoneHS}, 
extra dimensions~\cite{ChengKKDM, ServantKKDM, AgasheKKDM, HooperUED, ArrenbergKK}, 
or something else --  
gives rise to additional particles.
If an appropriate (often independently motivated) discrete symmetry exists, the lightest such particle is stable.
This particle is then 
weakly interacting, massive, and stable -- it is a WIMP.
Thus, particle theorists are ``almost justified in saying that the problem of electroweak symmetry breaking predicts
the existence of WIMP dark matter''~\cite{Baltz2006LCC}.

Although the argument for WIMP dark matter is generic,
supersymmetry dominates the discussion 
as a particularly
well-motivated model. Supersymmetry interactions arise in theories of quantum gravity, 
stabilize the Higgs mass hierarchy
problem, predict the observed value of $\sin^2
\theta_{\mathrm W}$, and over a broad range
of parameter space predict cosmologically significant
relic WIMP densities. 

WIMPs can potentially be detected by three complementary methods.  
They may be produced and detected (indirectly) at accelerators such as the Large Hadron Collider (see \eg\ Ref.~\refcite{battaglia_lhcdm}).
Relic WIMPs may be detected indirectly when they clump in massive astrophysical objects, increasing their annihilation rate enough that their annihilation products may be detectable~\cite{PressIndirectSun,FreeseIndirectSun,KraussIndirect,bertoneHS}.
Many potential (or suggested~\cite{dixonhalo, heatinterp,heatinterp2001,finkbeiner_wmapwimp,pamela,pamela:Bergstrom,atic})
indirect signals are ambiguous, with alternate astrophysical explanations 
(see Refs.~\refcite{CarrRPPindirect2006,FengReview}
and references therein).
Some potential indirect signals, however, would be compelling.
Annihilation 
in the Sun or Earth would produce 
higher-energy neutrinos than 
any other known process.
These neutrinos could be 
observed in neutrino telescopes such as
IceCube~\cite{icecubeDM,icecubeDM22} or ANTARES~\cite{antaresDMucla2006,antaresDM}.  
Either FERMI
or ground-based air Cerenkov telescopes
may 
detect
distinctive gamma-ray features from the galactic center or from sub-halos~\cite{glast2007DM, glastDMbaltz2008,magicDMUCLA2004, veritas2008,indirectDMstep}.  

Relic WIMPs may 
also 
be detected directly when they scatter off nuclei in terrestrial detectors~\cite{goodmanwitten,primack}.  
This article offers an introduction to these direct-detection experiments. 
Section~\ref{section:signal} includes derivations and explanations of the characteristics of potential signals in direct-detection experiments for standard WIMP models.  
Section~\ref{section:directexp} 
summarizes the general experimental methods shared by most direct-detection experiments and 
discusses particular experiments briefly, emphasizing the relative advantages and different challenges and capabilities of the various approaches.

\section{WIMP-nucleus elastic scattering: from model to signal}
\label{section:signal}

Understanding experiments designed for direct detection of dark matter  begins with the observables of potential signals.  In this section we  consider the observables of any model that predicts standard WIMP-nucleus elastic scattering (see Neil Weiner's contribution to these proceedings~\cite{weinerTASI2009}
for discussion of more speculative models with non-standard scattering).
Following the reviews by Lewin and Smith~\cite{lewinsmith}, 
and Jungman, Kamionkowski and Griest~\cite{jkg}, this section derives how the observed WIMP interaction rate depends on energy, target, time, and direction.

\subsection{Spin-independent and spin-dependent cross sections}

Using Fermi's Golden Rule, we can divide the energy dependence of the differential WIMP-nucleon cross section into a term $\sigma_{0\mathrm WN}$ that is independent of the momentum transfer and a term $F^2(q)$ (known as the form factor) containing the entire dependence on the momentum transfer $q$:
\begin{equation}
\frac{d\sigma_{\mathrm{WN}}(q)}{dq^2} = \frac{1}{\pi v^2} {\left| \mathcal{M} \right|}^2 =
\frac{\sigma_{0\mathrm{WN}} F^2(q)}{4\mu_{A}^2 v^{2} } .
\label{eqnFermi}
\end{equation}
Here, $v$ is the velocity of the WIMP in the lab frame, and the WIMP-nucleus reduced mass $\mu_{A} \equiv M_{\chi}M_{A} / ( M_{\chi}+M_{A})$ in terms of the WIMP mass $M_{\chi}$ and the mass $M_{A}$ of a target nucleus of atomic mass $A$.
Since the WIMPs are nonrelativistic, the zero-momentum cross section for a WIMP of arbitrary spin and general Lorentz-invariant WIMP-nucleus cross section may be written in terms of a spin-independent  (mostly scalar) and a spin-dependent (mostly axial vector) term:
\begin{equation}
\sigma_{0\mathrm{WN}} = 
\frac{4\mu_{A}^2}{\pi} \left[ Z f_{\mathrm{p}} + (A-Z)f_{\mathrm{n}} \right]^2 +
\frac{32G_{\mathrm F}^2 \mu_{A}^2}{\pi} \frac{J+1}{J}
 \left( a_{\mathrm{p}}\langle S_{\mathrm{p}}\rangle +a_{\mathrm{n}}\langle S_{\mathrm{n}}\rangle \right)^{2}.
\label{eqnSISD}
\end{equation}
The proof of this claim makes a good exercise for the reader; solution may be found in 
Ref.~\refcite{kurylovkam}.
Here $f_{\mathrm{p}}$ and $f_{\mathrm{n}}$ ($a_{\mathrm{p}}$ and $a_{\mathrm{n}}$) are effective spin-independent (spin-dependent) couplings of the WIMP to the proton and neutron, respectively.  Together with the WIMP mass, $M_{\chi}$, these parameters contain all the particle physics information of the model under consideration.
The other parameters describe the target material: its atomic number $Z$,  total nuclear spin $J$, and the expectation values of the proton and
neutron spins within the nucleus $\langle S_{p,n}\rangle=\langle
N|S_{p,n}| N\rangle $.  For free nucleons, $\langle S_p\rangle= 
\langle S_n\rangle$= 0.5. Table~\ref{t1} from Ref.~\refcite{toveySD} lists values of  $\langle S_p\rangle$ and  
$\langle S_n\rangle$ for materials commonly used for dark matter searches,
although some are subject to significant nuclear-physics uncertainties.

\begin{table}[t]
\tbl{Values of the atomic number $Z$, the total nuclear spin $J$, and the expectation values of the proton and
neutron spins within the nucleus $\langle S_{p,n}\rangle$ for various nuclei with odd numbers of protons or neutrons, leading to the relative sensitivities to spin-dependent interactions shown,
from Refs.~\refcite{toveySD,jkg}
and the references contained therein.}
{\begin{tabular}{lcccrrcc}
\toprule
  & & Odd   &  &  &  & \underline{$4 \langle S_p\rangle^2 (J+1)$} & \underline{$4 \langle S_n\rangle^2 (J+1)$} \\ 
Nucleus &$Z$ & Nuc. &$J$ &$\langle S_p\rangle $ &$\langle S_n\rangle $
& $3J$ & $3J$ \\
\colrule
~~$^{19}$F	&9	&p	&1/2	&0.477	&-0.004	&9.1$\times 10^{-1}$	&6.4$\times 10^{-5}$ \\ 
~~$^{23}$Na	&11	&p	&3/2	&0.248	&0.020	&1.3$\times 10^{-1}$	&8.9$\times 10^{-4}$ \\ 
~~$^{27}$Al	&13	&p	&5/2	&-0.343	&0.030	&2.2$\times 10^{-1}$	&1.7$\times 10^{-3}$ \\ 
~~$^{29}$Si	&14	&n	&1/2	&-0.002	&0.130	&1.6$\times 10^{-5}$	&6.8$\times 10^{-2}$ \\ 
~~$^{35}$Cl	&17	&p	&3/2	&-0.083	&0.004	&1.5$\times 10^{-2}$	&3.6$\times 10^{-5}$ \\ 
~~$^{39}$K	&19	&p	&3/2	&-0.180	&0.050	&7.2$\times 10^{-2}$	&5.6$\times 10^{-3}$ \\ 
~~$^{73}$Ge	&32	&n	&9/2	&0.030	&0.378	&1.5$\times 10^{-3}$	&2.3$\times 10^{-1}$ \\
~~$^{93}$Nb	&41	&p	&9/2	&0.460	&0.080	&3.4$\times 10^{-1}$	&1.0$\times 10^{-2}$ \\ 
~~$^{125}$Te	&52	&n	&1/2	&0.001	&0.287	&4.0$\times 10^{-6}$	&3.3$\times 10^{-1}$ \\ 
~~$^{127}$I	&53	&p	&5/2	&0.309	&0.075	&1.8$\times 10^{-1}$	&1.0$\times 10^{-2}$ \\ 
~~$^{129}$Xe	&54	&n	&1/2	&0.028	&0.359	&3.1$\times 10^{-3}$	&5.2$\times 10^{-1}$ \\ 
~~$^{131}$Xe	&54	&n	&3/2	&-0.009	&-0.227	&1.8$\times 10^{-4}$	&1.2$\times 10^{-1}$ \\ 
\botrule


\end{tabular}
}
\label{t1}
\end{table}

For many models, $f_{\mathrm{p}} \approx f_{\mathrm{n}}$, so the spin-independent WIMP-nucleus cross section
\begin{equation}
\sigma_{0\mathrm{WN,SI}} \approx
\frac{4\mu_{A}^2}{\pi}f_{\mathrm{n}}^2 A^2.
\end{equation}
The dependence of this cross section on the target material may be factored out by rewriting this result  as
\begin{equation}
\sigma_{0\mathrm{WN,SI}} = \sigma_{\mathrm{SI}} \frac{ \mu_{A}^2}{\mu_{\mathrm{n}}^2}
 A^2,
\label{eqn:WIMP-nucleon}
\end{equation}
where $\mu_{\mathrm{n}}$ is the reduced mass of the WIMP-nucleon system, 
and the (target-independent) spin-independent cross section of a WIMP on a single nucleon 
\begin{equation}
\sigma_{\mathrm{SI}} \equiv  \frac{ 4\mu_{\mathrm{n}}^{2} f_{\mathrm n}^2}{\pi} .
\end{equation}
This WIMP-nucleon cross section $\sigma_{\mathrm{SI}}$ may be used to compare experimental results to theory and to each other.
A given model predicts particular combinations 
of $\sigma_{\mathrm{SI}} $ and $M_{\chi}$;
different experiments produce limits on $\sigma_{\mathrm{SI}} $ as functions of $M_{\chi}$ by translating limits on the WIMP-nucleus cross-section to limits on $\sigma_{\mathrm{SI}}$ using equation~\ref{eqn:WIMP-nucleon}.
The dependence on $\mu_{A}^{2}A^2$ in eqn.~\ref{eqn:WIMP-nucleon} indicates the advantage of experiments using relatively heavy target materials (but see the effects of the form factor in Sec.~\ref{sec:formfactor}). 
For a 50\,\gev\ WIMP incident on a target with $A=50$, $\mu_{A}^2/\mu_{\mathrm{n}}^2 = 625$, so 
the spin-independent WIMP-nucleus cross section is larger than the WIMP-nucleon cross section by a factor $>10^6$.

The situation for spin-dependent interactions is quite different~\cite{toveySD}.  First of all, contributions from the spin-dependent proton and neutron couplings often cancel, so it is important to quote limits on the spin-dependent interaction on neutrons separately from that on protons, each under the assumption that the other interaction is negligible.  Furthermore, while the coherent interaction on the nucleus results in a spin-independent cross section that scales with $A^2$ since the contribution of each nucleon adds inside the matrix element, the spin-dependent contributions of nucleons with opposite spins cancel, so that the total spin-dependent cross section depends on the net spin of the nucleus.  
As shown in Table~\ref{t1}, nuclei with even numbers of protons have nearly no net proton spin and essentially no sensitivity to spin-dependent interactions on protons, and nuclei with even numbers of neutrons similarly have almost no sensitivity to spin-dependent interactions on neutrons.  Argon, with even numbers of protons and neutrons for all significant isotopes,  is thus insensitive to spin-dependent interactions.  Many materials used as WIMP targets (Ge, Si, Xe) have even numbers of protons and hence are insensitive to spin-dependent interactions on protons; only some isotopes  of these targets (and hence only a fraction of the detector's active mass) have sensitivity to spin-dependent interactions on neutrons.  Typically, sensitivity to spin-dependent interactions on protons requires alternate target materials, often resulting in worse backgrounds or background rejection and lower sensitivity to spin-independent interactions.
The relative sensitivity of a material to spin-dependent interactions is summarized by its ``scaling factors''  
$4 \langle S_p\rangle^2 (J+1)/3J$ and 
$4 \langle S_n\rangle^2 (J+1)/3J$, 
which are listed in Table~\ref{t1}.
As with spin-independent limits, experimenters quote limits on target-independent quantities: the spin-dependent WIMP-proton cross section 
$\sigma_{\mathrm{SDp}}\equiv 24G_{\mathrm{F}}^2 \mu_{\mathrm{p}}^2 a_{\mathrm{p}}^2/\pi$ and the spin-dependent WIMP-neutron cross section 
$\sigma_{\mathrm{SDn}}\equiv 24G_{\mathrm{F}}^2\mu_{\mathrm{n}}^2 a_{\mathrm{n}}^2/\pi$. 

\begin{figure}[t]
\begin{center}
\psfig{file=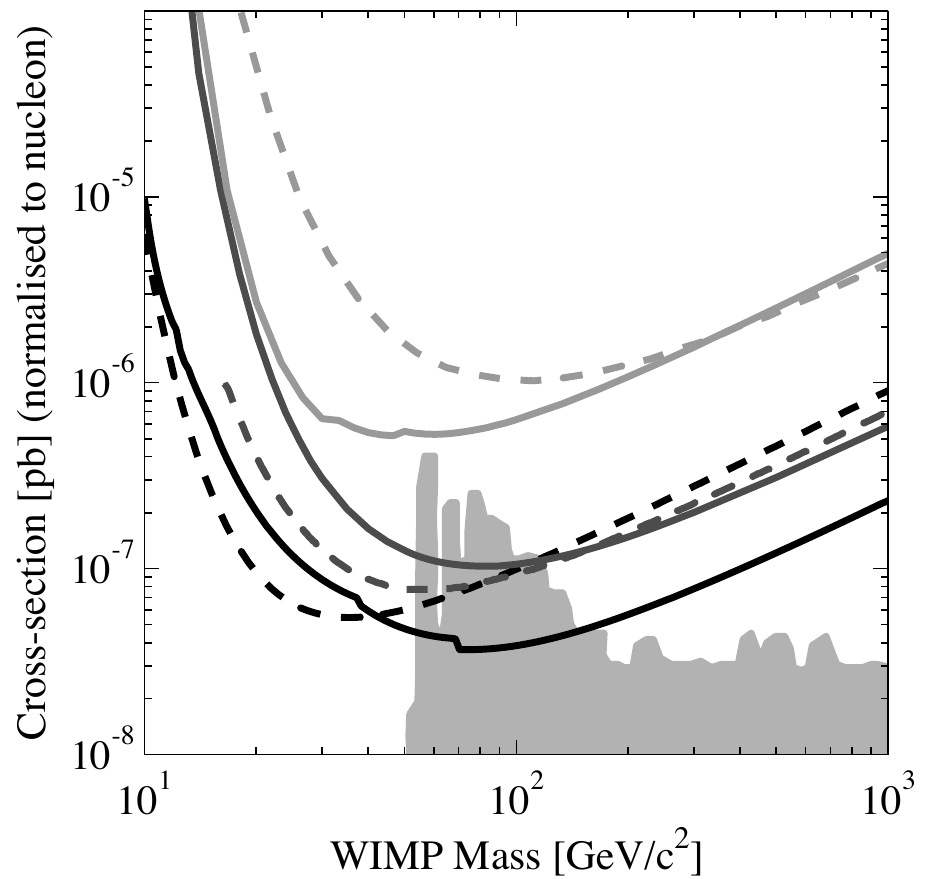,width=3in}
\end{center}
\caption{Upper limits on the spin-independent WIMP-nucleon coupling  $\sigma_{\mathrm{SI}}$ under the standard assumptions about the Galactic halo described in the text.  Most sensitive limits are from cryogenic experiments (solid) CDMS~\cite{CDMSscience} (black), EDELWEISS-II~\cite{edelweiss2009} (medium gray), and CRESST~\cite{CRESST2008} (light gray), and two-phase noble experiments (dashed) XENON10~\cite{xenon2007} (black), ZEPLIN-III~\cite{zeplin2009} (medium gray), and WArP~\cite{warp2007} (light gray).  Current experiments already exclude part of the parameter space of MSSM models (shaded)~\cite{baltzgondolo2004}.  Figure made using the Dark Matter Limit Plotter~\cite{Gaitskell:dmplotter}.}
\label{fig:SI}
\end{figure}

\begin{figure}
\begin{center}
\psfig{file=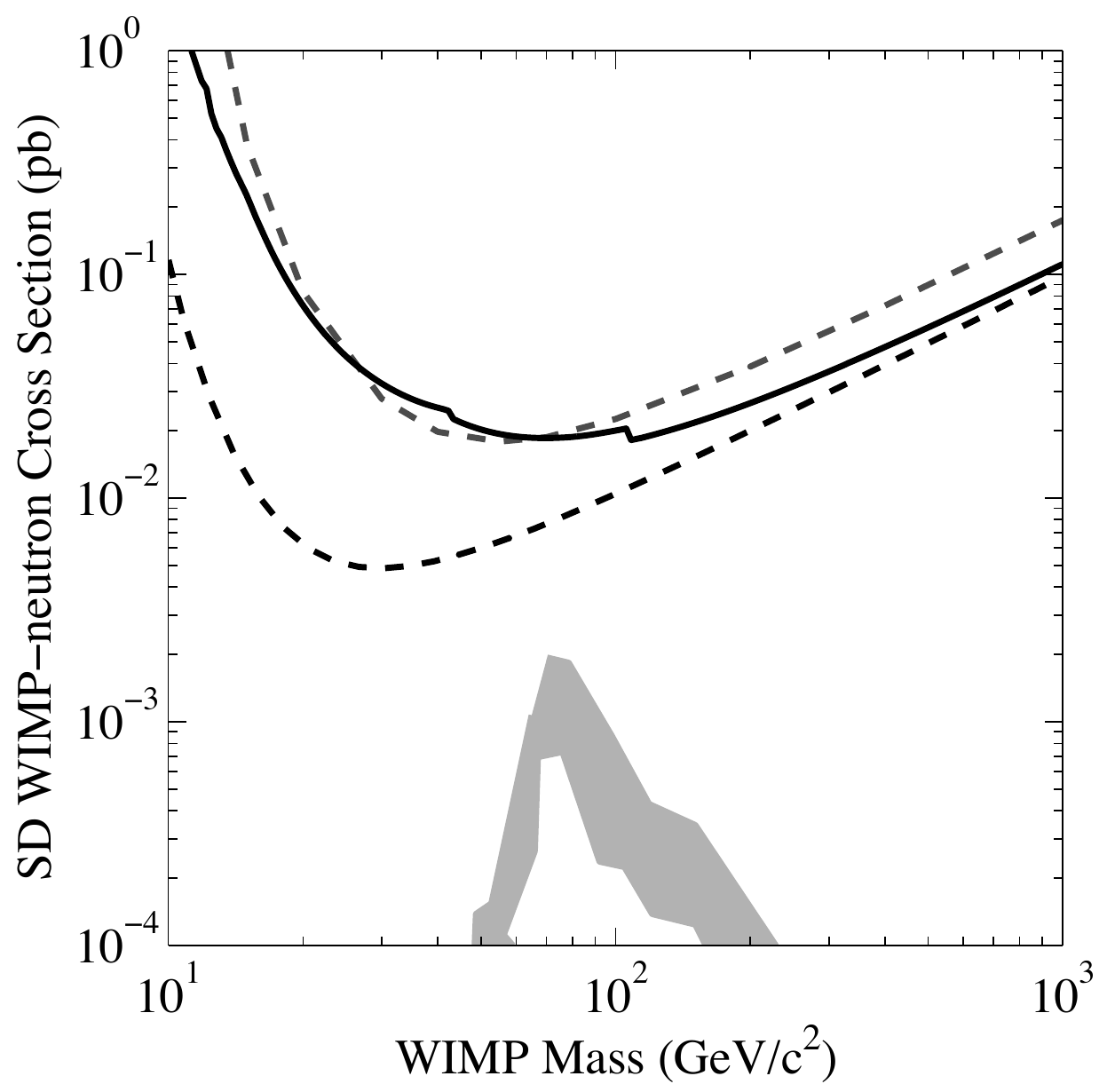,width=2.22in}
\psfig{file=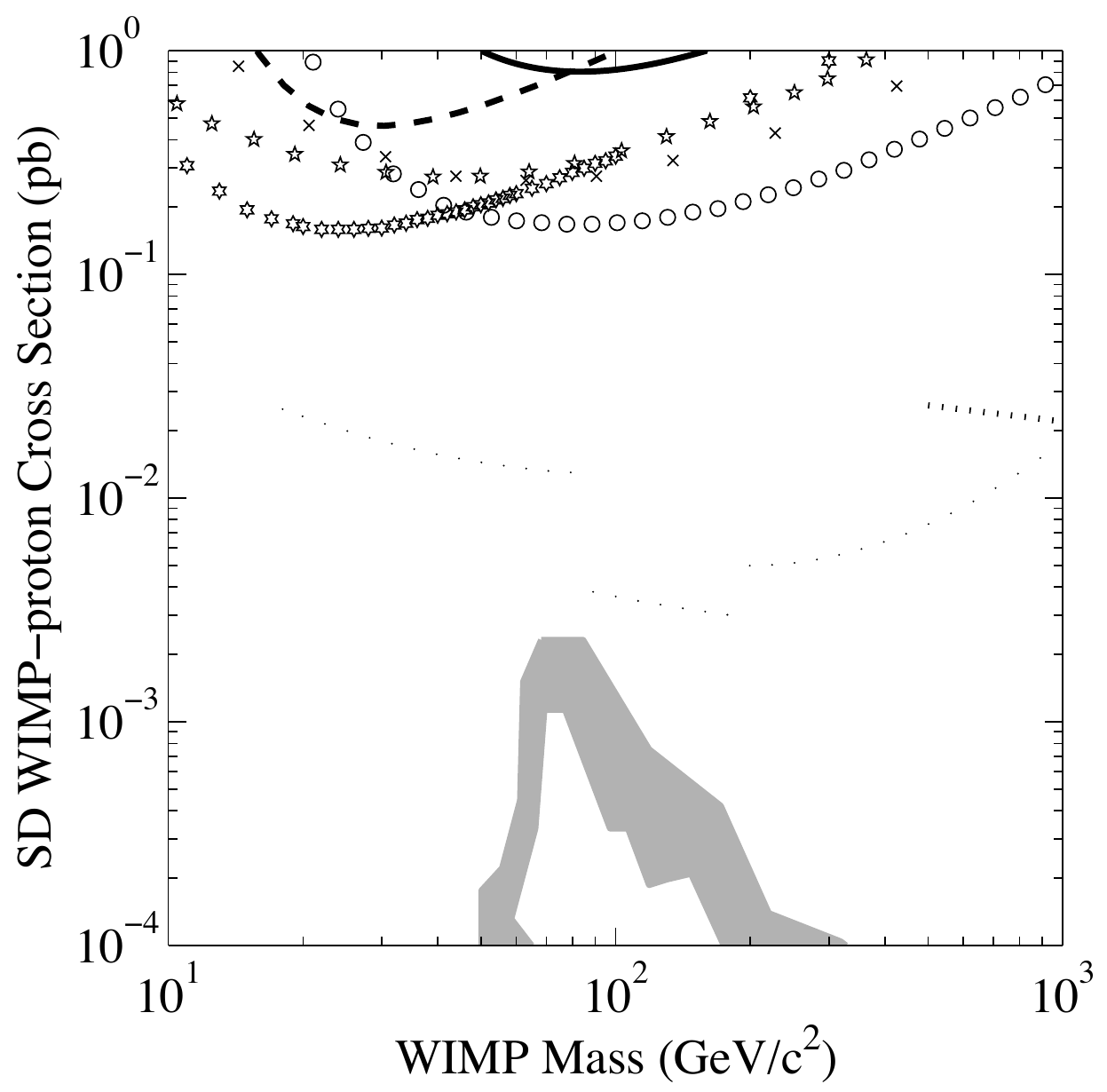,width=2.22in}
\end{center}
\caption{Upper limits on the spin-dependent WIMP-neutron coupling  $\sigma_{\mathrm{SDn}}$ (left) and the spin-dependent  WIMP-proton coupling  $\sigma_{\mathrm{SDp}}$ (right) under the standard assumptions about the Galactic halo described in the text.  The most sensitive limits on $\sigma_{\mathrm{SDn}}$ are from the same experiments shown in Fig.~\ref{fig:SI} (with the same linetypes):   XENON10~\cite{xenon2008SD} (black dashes), ZEPLIN-III~\cite{zeplin2009SD} (medium gray dashes), and CDMS~\cite{CDMSscience} (black solid).  Note ZEPLIN-III limits were calculated with a  scaling factor 2$\times$  smaller than that used for XENON10.
Due to the low intrinsic sensitivity of leading (Xe and Ge) experiments 
to spin-dependent interactions on protons, the most sensitive limits on $\sigma_{\mathrm{SDp}}$ are from experiments with only modest sensitivity to spin-independent interactions: PICASSO~\cite{picasso2009} (6-sided stars), COUPP~\cite{coupp2008SDlimits} (5-pointed stars), KIMS~\cite{kims2007} (circles), 
and NAIAD~\cite{naiad2005}  ($\times$).
Limits from indirect search experiments SuperKamiokande~\cite{superk2004} (points) and IceCube~\cite{icecubeDM22} (dotted) make additional assumptions about branching fractions to neutrinos. 
Current experiments do not exclude any part of the parameter space of the same MSSM models (shaded)~\cite{baltzgondolo2004} shown in Fig.~\ref{fig:SI}, despite the fact that the predicted spin-dependent cross sections are $\sim3000\times$ larger than the spin-independent ones.
Figure made using the Dark Matter Limit Plotter~\cite{Gaitskell:dmplotter}.}
\label{fig:SD}
\end{figure}

The lack of benefit from the coherent interaction for spin-dependent interactions
results in most models being more accessible experimentally via their spin-independent interactions than by their spin-dependent interactions.  As shown in Fig.~\ref{fig:SI} and Fig.~\ref{fig:SD}, current experiments are already constraining MSSM models based on their spin-independent couplings, but none is yet sensitive enough to constrain such models based on their spin-dependent couplings, despite the fact that spin-dependent couplings are typically $\sim3000\times$ larger than spin-independent couplings. 

\subsection{The WIMP recoil energy spectrum} 

It is illuminating to calculate the energy spectrum for the case of zero momentum-transfer (i.e.\ taking $F^2 \equiv 1$).  Furthermore, simply multiplying this spectrum by the energy dependence of $F^{2}(q)$, rather than including the form factor $F^2$ within the kinematic integral to follow,
is convenient and usually adequate.  

\begin{figure}[t]
\begin{center}
\psfig{file=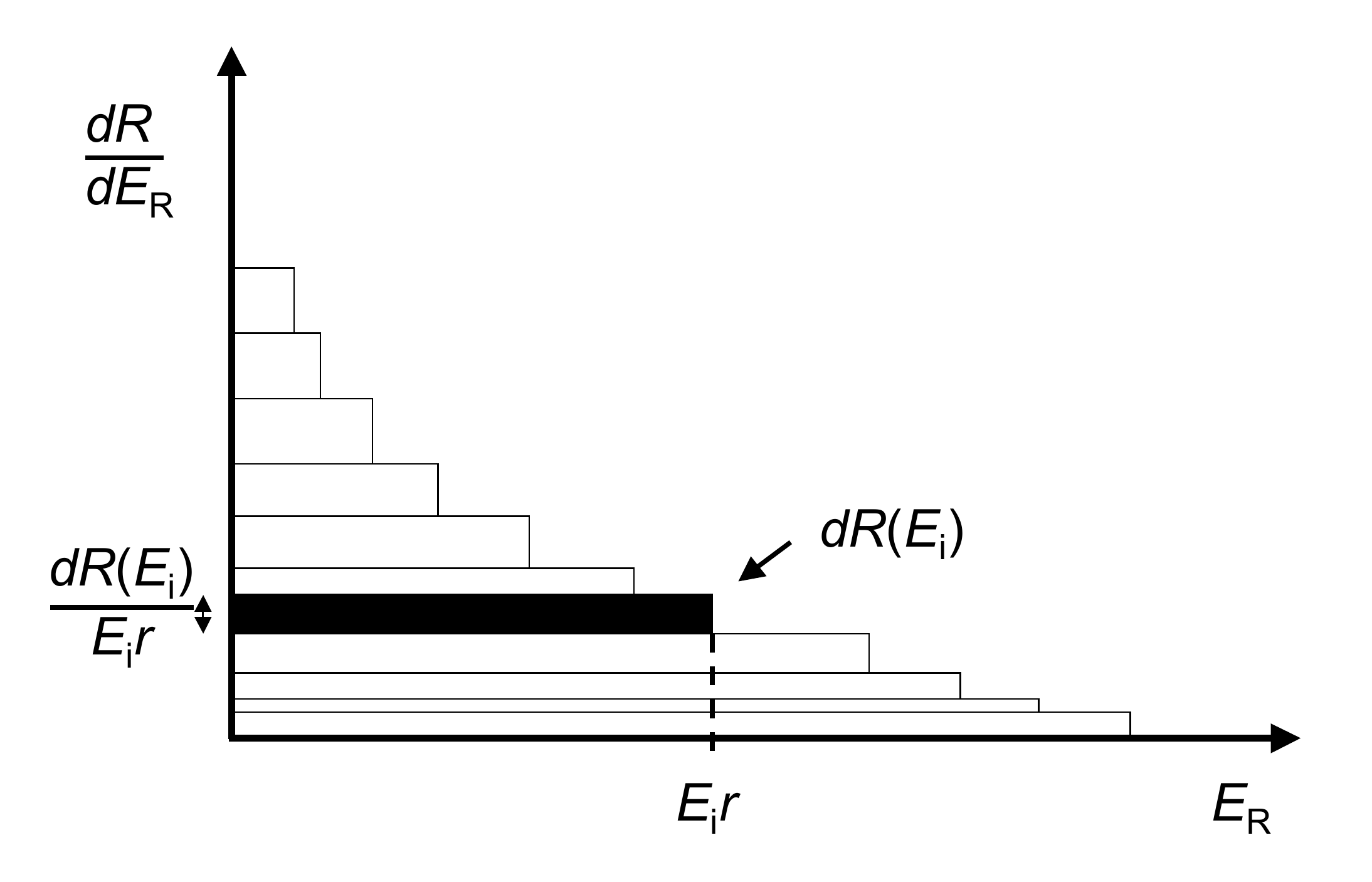,width=3.5in}
\end{center}
\caption{Plot showing schematically how contributions to the differential rate $dR/dE_{\mathrm{R}}$ for different values of the initial WIMP energy $E_{\mathrm{i}}$ add.  
We define total differential rate $dR/dE_{\mathrm{R}}$ of WIMPs with initial energy $E_{\mathrm{i}}$ to be $dR(E_{\mathrm{i}})$. 
For a WIMP initial energy $E_{\mathrm{i}}$, the recoil energy $E_{\mathrm{R}}$ is uniformly distributed between 0--$E_{\mathrm{i}}r$, so  
$dR(E_{\mathrm{i}})$ contributes equally to the rates of all recoils between 0--$E_{\mathrm{i}}r$, as 
depicted by the shaded area in the figure. The contribution to the differential rate at  a given recoil energy (the height of the shaded area in the figure) is simply the area divided by the length, or $dR(E_{\mathrm{i}})/(E_{\mathrm{i}}r)$.  
The total differential rate can then be found by summing all the boxes, i.e.\ integrating $dR(E_{\mathrm{i}})/(E_{\mathrm{i}}r)$ for all $E_{\mathrm{i}}$.
}
\label{fig:dRdE}
\end{figure}

The energy spectrum arises due to the familiar kinematics of elastic scattering.  In the center-of-momentum frame, the WIMP scatters off a nucleus through an angle $\theta$, with $\cos\theta$ uniformly distributed between $-1$ and $1$ for the isotropic scattering that occurs with zero-momentum transfer.  If the WIMP's initial energy in the lab frame $E_{\mathrm{i}} = M_{\chi}v^{2}/2$, the nucleus recoils with energy 
\begin{equation}
E_{\mathrm{R}} = E_{\mathrm{i}}r \frac{(1-\cos\theta)}{2}
\end{equation}
(in the lab frame), where 
\begin{equation}
r \equiv \frac{4\mu_{A}^2}{M_{\chi}M_{A} } =
 \frac{4M_{\chi}M_{A} } { \left( M_{\chi}+M_{A}\right)^{2} }
\end{equation}
is a dimensionless parameter related to the reduced mass $\mu_{A}$.  Note that $r\leq 1$, with $r=1$ only if $M_{\chi}=M_{A}$.
For this isotropic scattering, the recoil energy is therefore uniformly distributed between  0--$E_{\mathrm{i}}r $.
As shown in Fig.~\ref{fig:dRdE}, the differential contribution to the differential rate for a given initial WIMP energy
\begin{equation}
d \left( \frac{dR}{dE_{\mathrm{R}}}(E_{\mathrm{R}}) \right) =  \frac{dR(E_{\mathrm{i}})} {E_{\mathrm{i}} r}, 
\end{equation}
so
\begin{equation}
\frac{dR}{dE_{\mathrm{R}}}\left(E_{\mathrm{R}} \right) =
\int_{E_{\mathrm{min}}}^{E_{\mathrm{max}}}
 \frac{dR(E_{\mathrm{i}})} {E_{\mathrm{i}} r}.
\end{equation}
The maximum initial WIMP energy may be taken as infinity as an initial approximation, or more accurately may be based upon the Galactic escape velocity, $v_{\mathrm{esc}}$: 
$E_{\mathrm{max}} = M_{\chi}v_{\mathrm{esc}}^{2}/2$.
To cause a recoil of energy $E_{\mathrm{R}}$, the minimum initial WIMP energy 
$E_{\mathrm{min}} =E_{\mathrm{R}}/r$ (for head-on scattering, with $\theta = \pi$), and the minimum WIMP velocity $v_{\mathrm{min}} = \sqrt{2E_{\mathrm{min}}/M_{\chi}} = 
\sqrt{2E_{\mathrm{R}}/(rM_{\chi})}$.

To determine the rate of WIMP-nucleus scattering, it is helpful to imagine the motion of the target nucleus relative to WIMPs with velocity $v$ in the lab frame.  In time $dt$, each nucleus interacts with any WIMP inside  a volume $dV = \sigma v dt$, where $\sigma$ is the WIMP-nucleus cross section.  The number of WIMPs inside the volume moving with velocity $v$
\begin{equation}
dN = n_0 f(\vec{v} + \vec{v}_{\mathrm{E}})\sigma v dt ,
\end{equation}
where the local WIMP number density $n_0 = \rho_{\chi}/M_{\chi}$, where 
$\rho_{\chi}$ is the mass density of WIMPs in the galaxy, estimated from studies of Galactic dynamics to be about 0.3\,GeV/($c^2$ cm$^3$) (with wide systematic uncertainties~\cite{GatesRhoWIMP,bergstrombuckley1998,WidrowBlueprint,Garbari:2010sk}).
Note that the number density based on $\rho_{\chi}=0.3$\,GeV/($c^2$ cm$^3$) is really an upper limit, since Galactic dark matter may include species other than WIMPs.
We use the fact that the velocity $\vec{v}_{\mathrm{g}}$ of the WIMP in the galaxy is the vector sum of the WIMP velocity with respect to the Earth $\vec{v}$ and the velocity $\vec{v}_{\mathrm{E}}$ of the Earth with respect to the Galaxy.
We assume that the WIMPs' velocities in the frame of the Galaxy 
follow the Maxwellian distribution:  
\begin{equation}
f(\vec{v} + \vec{v}_{\mathrm{E}}) = 
\frac{e^{\left(-\vec{v}+\vec{v}_{\mathrm{E}} \right)^2/v_0^{2}} } {k}
\end{equation}
where $v_{0}= 220\pm20$\,km/s is the 
local circular velocity~\cite{KerrGalaxyRotation},
and $k$ is a normalization factor.  
This simple distribution is not expected to be especially accurate,
but it provides a useful standard.
See \eg\ Refs.~\refcite{Evans:triaxial, CerdenoGreen}
for discussion of  alternatives that are likely more accurate, and
see Section~\ref{detection} below for discussion of the impact of uncertainties on a WIMP discovery.

The differential interaction rate per kilogram of detector is then the product of the number of interactions per nucleon with the number of nuclei per kg of material:
\begin{equation}
dR = \frac{N_0}{A}n_0 f(\vec{v} + \vec{v}_{\mathrm{E}})\sigma v d^3\vec{v}
\end{equation}
where $N_0$ is Avogadro's number, so that $N_0/A$ is the number of nuclei per kilogram of material.

It is instructive (and reasonably accurate, as shown below) to consider the simplified case ignoring the Earth velocity and the Galaxy's escape velocity (\ie\ setting $v_{\mathrm{E}}=0$, $v_{\mathrm{esc}}=\infty$), for which the integral is trivial.  After setting 
\begin{equation}
R_0 \equiv \frac{2}{\sqrt{\pi}} \frac{N_0}{A} n_0 \sigma v , 
\label{eqnR0}
\end{equation}
we get
\begin{eqnarray}
\frac{dR}{dE_{\mathrm{R}}} (E_{\mathrm{R}}) &=& 
\int_{E_{\mathrm{R}}/r}^\infty
 \frac{1}{\left(\frac{1}{2}M_{\chi}v^2\right)r} 
\frac{R_0}{2\pi v_0^4} v e^{-v^2/v_0^2} \left(4\pi v^2 dv\right) \\
&=& \frac{R_0}{\left(\frac{1}{2}M_{\chi}v_0^2\right)r} 
\int_{v_{\mathrm{min}}}^\infty \frac{2}{v_0^2} e^{-v^2/v_0^2} v dv \\
&=& \frac{R_0}{E_{0}r} e^{-E_{\mathrm{R}}/E_{0}r} , 
\label{eqn:exponential}
\end{eqnarray}
where $E_0 \equiv M_{\chi}v_0^2/2$ is the most probable WIMP incident energy.
The mean recoil energy is easily seen: $\langle E_{\mathrm{R}} \rangle = E_0 r$.  Since $r\leq 1$, the mean recoil energy $\langle E_{\mathrm{R}} \rangle = E_0$ only if the WIMP mass is equal to the mass of the target nucleus; $\langle E_{\mathrm{R}} \rangle < E_0$ both for smaller and for larger WIMP masses.  
As an example, since $v_0 \approx 220$\,km/s $\approx (0.75\times 10^{-3})c$,
$M_{\chi} = M_{A} = 50$\,\gev\ would result in
\begin{equation}
\langle E_{\mathrm{R}} \rangle = E_{0}r = \frac{1}{2}M_{\chi}v_0^2 \approx 15 \mathrm{~keV}. 
\end{equation}
A different target mass would result in even lower  $\langle E_{\mathrm{R}} \rangle$.
This low energy sets the first challenge for direct detection experiments -- they must have low energy thresholds, much lower than past solar neutrino experiments for example.

From the exponential form of the approximate energy spectrum, 
we see that $R_0$ is the total WIMP rate.
If we plug known numerical values into equation~\ref{eqnR0}, we find
\begin{equation}
R_0 \approx \frac{500}{M_{\chi} \left(\mathrm{GeV}/c^2 \right)} 
\frac{\sigma_{0WN}}{1\mathrm{\,pb}}
\frac{\rho_{\chi}}{0.4\mathrm{\,GeV/cm}^3}\mathrm{events~kg}^{-1} \mathrm{day}^{-1} . 
\end{equation}
A 50\,\gev\ WIMP with a WIMP-nucleus cross section $\sigma_{0WN}= 1$\,pb 
(so that the spin-independent WIMP-nucleon cross section $\sigma_{\mathrm{SI}} \sim 10^{-6}$\,pb,
or the spin-dependent WIMP-nucleon cross section  $\sigma_{\mathrm{SDp,n}} \sim 10^{-3}$\,pb) results in about 10 events\perkgd.
Since the energy spectrum is a falling exponential, a low energy threshold is critical to detect most of these events; the fraction of events above an energy threshold $E_{\mathrm{th}}$ is $e^{-E_{\mathrm{th}}/E_{0}r}$.

\begin{figure}[t]
\begin{center}
\psfig{file= 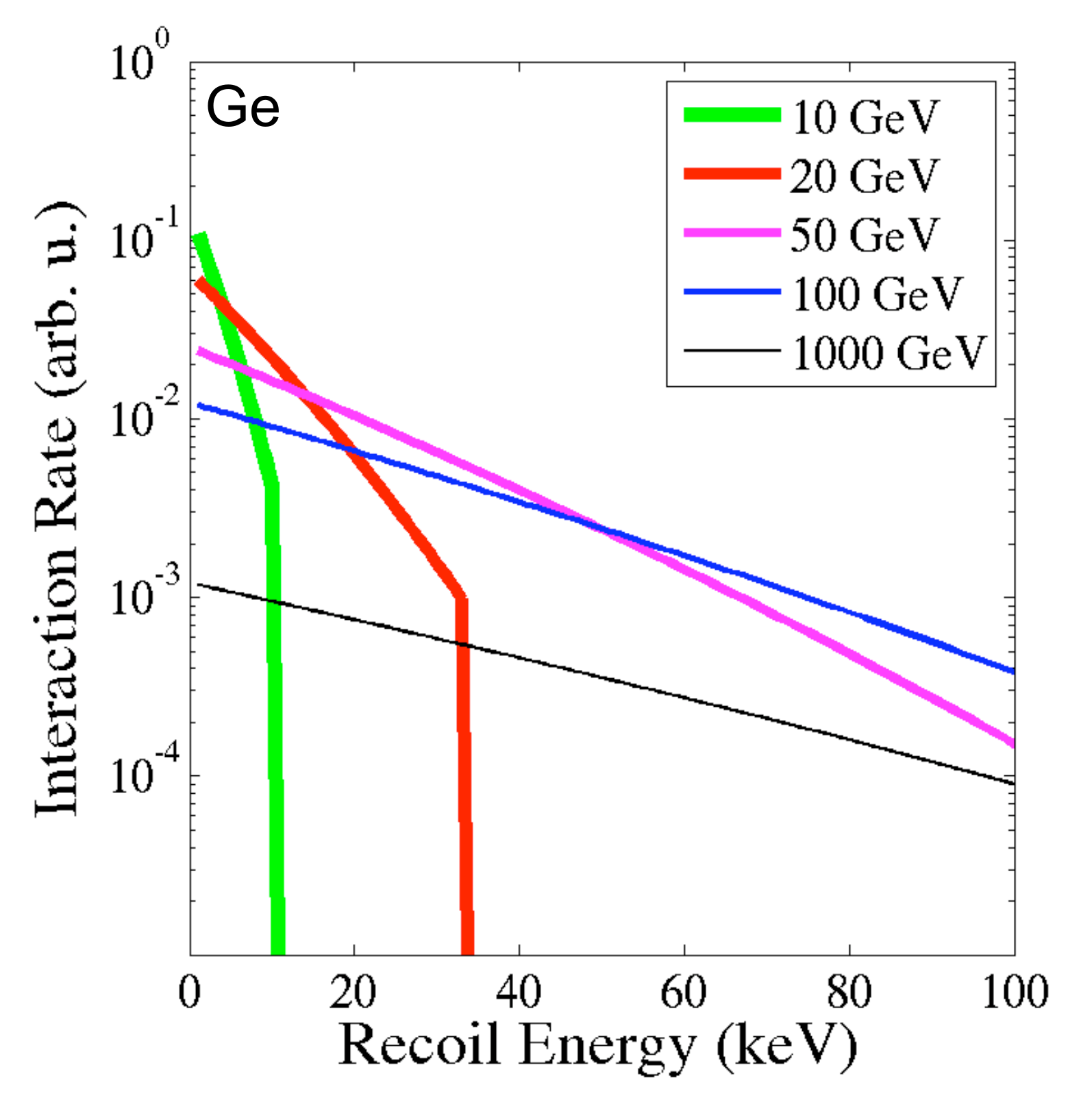,width=2.5in}
\end{center}
\caption{Expected interaction rate on Ge as a function of recoil energy for different 
WIMP masses $M_{\chi}$ (as shown in the legend in units of \gev), for  most probable WIMP velocity $v_0= 270$\,km/s, on the high end of possible values.
The spectra of lower-mass WIMPs are softer and cut off at lower energy due to the Galactic escape velocity.
The spectra of WIMPs heavier than the target nucleus are nearly identical.  
The rate is 10$\times$ larger for a 100\,\gev\ WIMP than for a 1000\,\gev\ WIMP since there would be 10$\times$ more  100\,\gev\ WIMPs than 1000\,\gev\ WIMPs.
Higher-mass spectra deviate from straight lines due to non-unity form factor $F^2$ (see Section~\ref{sec:formfactor}).
}
\label{fig:EnergySpectrumVsM}
\end{figure}

The 
dependence of the energy spectrum on the WIMP mass may be seen easily from equation~\ref{eqn:exponential}.  The mean recoil energy
\begin{equation}
\langle E_{\mathrm{R}} \rangle = rE_{0} \propto \frac{v_0^2}{\left(1 + M_A/M_{\chi}\right)^2} \propto \left\{
\begin{array}{rl}
M_{\chi}^2 & \text{if } M_{\chi} \ll M_A\\
\text{constant} & \text{if } M_{\chi} \gg M_A
\end{array} \right. .
\label{eqn:SpectrumOnMass}
\end{equation}
Heavy WIMPs all yield about the same energy spectrum.
This result holds for calculations made including the Earth velocity, Galaxy escape velocity, and nuclear form factor, as shown in Fig.~\ref{fig:EnergySpectrumVsM}.

WIMPs with velocities above the Galaxy's escape velocity are likely to have already escaped.
The finite escape velocity ($\sim540$\,km/s $\approx 2 \times 10^{-3} c$ according to the RAVE survey~\cite{RAVEescapespeed}) alters the recoil spectrum slightly and produces a cut-off at
\begin{equation}
E_{\mathrm{max}} =  \frac{1}{2}rM_{\chi}v_{\mathrm{esc}}^2 \approx 100\mathrm{~keV}. 
\label{eqn:Emax}
\end{equation}
The cutoff energy has the same dependence on the WIMP mass as the mean recoil energy 
(see equation~\ref{eqn:SpectrumOnMass}) since 
\begin{equation}
E_{\mathrm{max}} =  \frac{ v_{\mathrm{esc}}^2} {v_0^2} \langle E_{\mathrm{R}} \rangle 
 \approx 6 \langle E_{\mathrm{R}} \rangle .
\end{equation}
Hence, higher-mass WIMPs produce recoils that are easier to detect and have cutoff energies so high as to usually be negligible.  The cutoff energy, however, is significant for low-mass WIMPs (see Fig.~\ref{fig:EnergySpectrumVsM}); experiments will have no sensitivity at all to WIMPs of low enough masses due to the cutoff.  Since the Galactic escape velocity is not known especially well, 
caution should be taken when drawing conclusions that may be sensitive to the the number of WIMPs with velocities at or near the assumed cutoff energy.  It should also be noted that for historical reasons (in order to quote limits using the same assumptions as previous experiments) most experimenters routinely assume the ``standard halo'' described in Ref.~\refcite{DonatoHaloRotation}, 
which uses a value $v_{\mathrm{esc}} = 650$\,km/s, somewhat above the 90\% upper limit quoted in the more recent RAVE survey~\cite{RAVEescapespeed}.

The dependence of the energy spectrum on target mass $M_A$ (ignoring the form factor $F^2$) is entirely through the $r$ parameter in Eqn.~\ref{eqn:SpectrumOnMass} or Eqn.~\ref{eqn:Emax}.  
For a given WIMP mass, the cut-off energy $E_{\mathrm{max}}$ and the mean recoil energy $E_0$ are largest for targets whose masses most closely match the WIMP mass.  As shown below, including the form factor makes the energy spectra of more massive targets softer.

\begin{figure}[t]
\begin{center}
\psfig{file=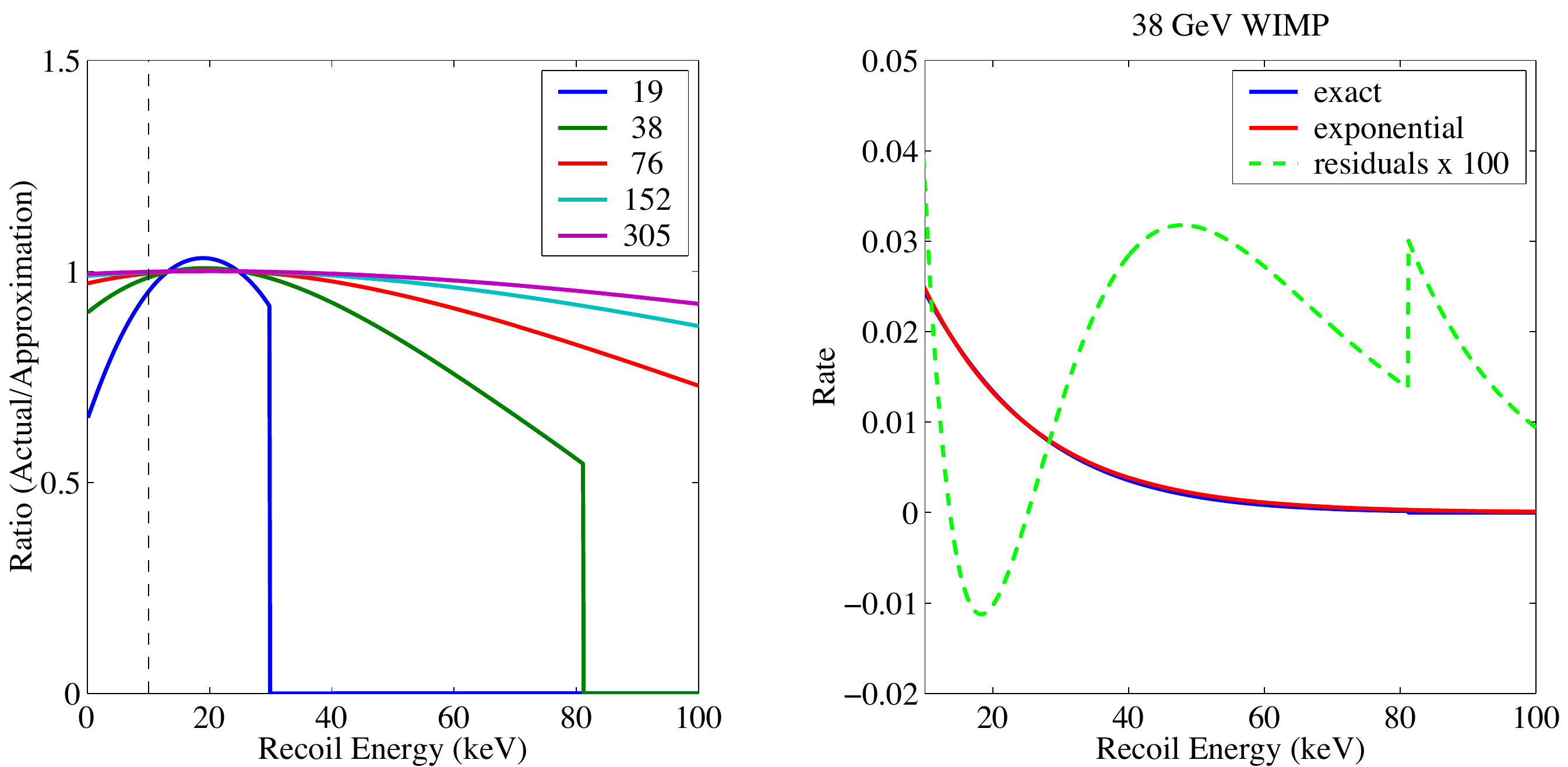, width=2.07in}
\psfig{file=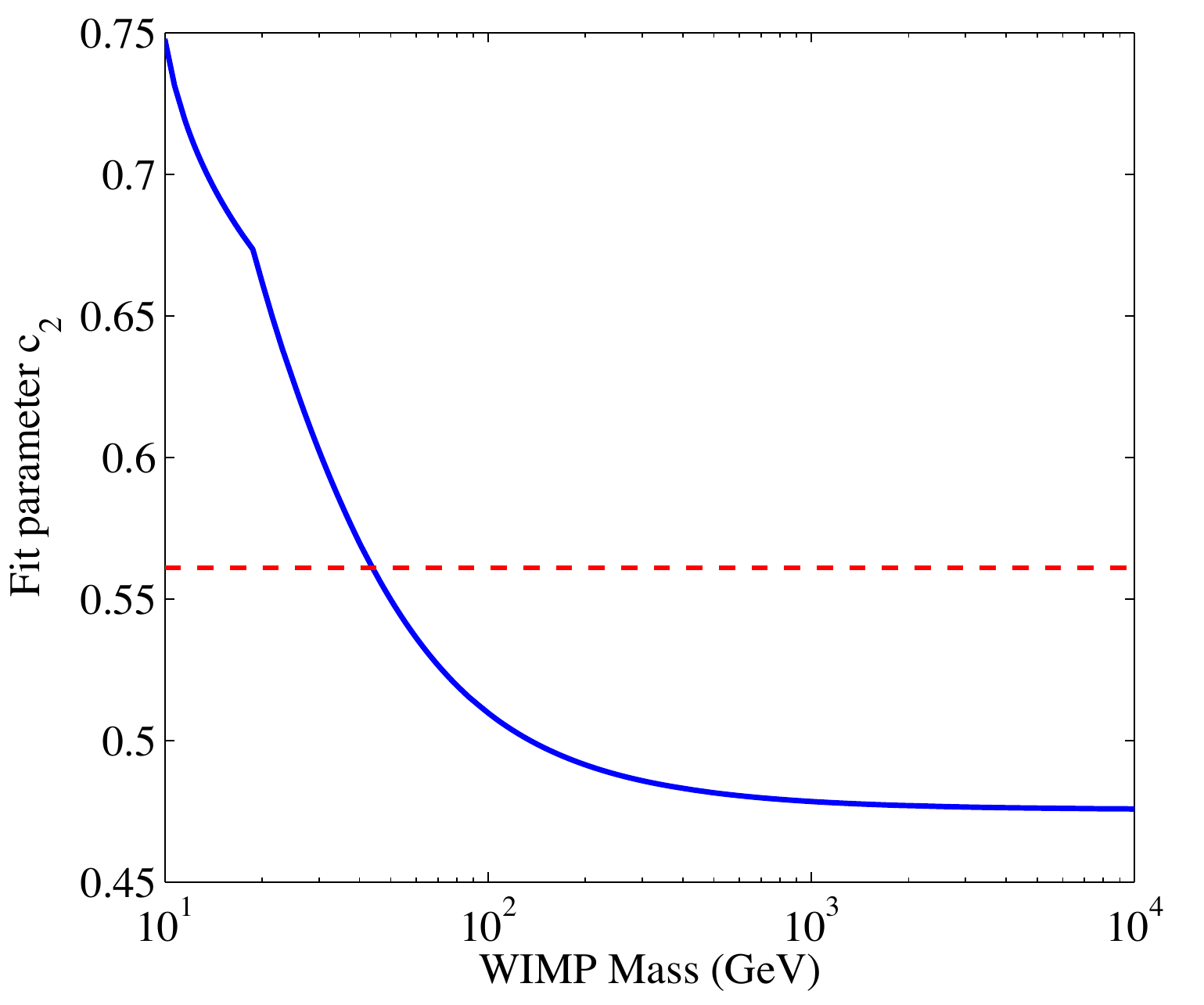, width=2.37in}
\caption{
{\it Left:}
Comparison of the spectrum using the exponential approximation (eqn.~\ref{eqndRdEexpo} with no cut off energy $E_{\mathrm{max}}$) to the 
full calculation (eqn.~\ref{eqndRdEfull}) for a WIMP with mass $M = 38$\,GeV (solid overlapping curves).
The residuals are shown 
(dashed) after scaling by 100.  
At higher energies, the fractional deviations become larger but the 
absolute deviations remain small.
The shape of the residuals $> 80$\,keV is from applying the cutoff energy $E_{\mathrm{max}}$ in the full calculation but not to the exponential approximation.
The approximation is slightly less accurate for lower-mass WIMPs.
{\it Right:}
Dependence of spectral fit parameter $c_{2}$ on WIMP mass 
$M_{\chi}$ for a Ge detector with a 10-keV threshold.  Use of the advised approximate value~\cite{lewinsmith}, $c_{2}=0.561$ (dashed), produces not more than a 50\% error for all WIMP masses.
}
\label{figexponential}
\label{figc2}
\end{center}
\end{figure}

The full calculation of the energy spectrum for WIMP-nucleus elastic scattering including the effects of both escape velocity and the earth's velocity is left as an exercise for the reader (for an almost complete solution see Lewin and Smith~\cite{lewinsmith}).  The result for recoil energies such that 
$v_{\mathrm{min}}(E_{\mathrm{R}})  < v_{\mathrm{esc}}  - v_{\mathrm{E}}$
is
\begin{equation}
\frac{dR}{dE_{\mathrm{R}}}  \approx
 \frac{R_0}{E_{0}r} \left\{ \frac{v_0\sqrt{\pi}} {4v _{\mathrm{E}}}
\left[ 
\mathrm{erf} \left( \frac{v_{\mathrm{min}} + v_{\mathrm{E}}} {v_0} \right)
- \mathrm{erf} \left( \frac{v_{\mathrm{min}} -v_{\mathrm{E}}} {v_0} \right)
\right]
-e^{-v_{\mathrm{esc}}^{2}/v_0^2}
\right\},
\label{eqndRdEfull}
\end{equation}
while for 
$ v_{\mathrm{esc}}  - v_{\mathrm{E}} < v_{\mathrm{min}}(E_{\mathrm{R}})  < v_{\mathrm{esc}}  + v_{\mathrm{E}}$
the terms inside curly braces become~\cite{ArrenbergiDM,savageDAMA}
\begin{equation*}
\left\{ \frac{v_0\sqrt{\pi}} {4v _{\mathrm{E}}}
\left[ 
\mathrm{erf} \left( \frac{v_{\mathrm{esc}} } {v_0} \right)
- \mathrm{erf} \left( \frac{v_{\mathrm{min}} -v_{\mathrm{E}}} {v_0} \right)
\right]
-\frac{v_{\mathrm{esc}} +v_{\mathrm{E}} -v_{\mathrm{min}}} {2v_{\mathrm{E}}}e^{-v_{\mathrm{esc}}^{2}/v_0^2}
\right\} .
\end{equation*}
This energy spectrum 
is reasonably approximated by another falling exponential:
\begin{equation}
\frac{dR}{dE_{\mathrm{R}}} (E_{\mathrm{R}}) \approx
c_1 \frac{R_0}{E_{0}r}
e^{-c_{2} E_{\mathrm{R}}/E_{0}r} ,
\label{eqndRdEexpo}
\end{equation}
as shown in Figure~\ref{figexponential}. 
Here 
 $c_1 \approx 0.75$ and $c_2 \approx 0.56$, although both depend on WIMP and target masses, day of year, and the energy range of interest~\cite{lewinsmith}.
For example, the right panel of Fig.~\ref{figc2} shows the dependence of $c_{2}$
on the WIMP mass 
$M_{\chi}$ for a Ge target.

Since $c_1/c_2=1.3$, the Earth's motion increases the interaction rate by $\sim$30\% in addition to making the energy spectrum harder, as should be expected from analogy to a car's driving through the rain resulting both in more raindrops hitting the front windshield and in the drops hitting 
with more force on average.  Despite the wide use of this analogy, it must be noted that the effect of Earth moving through the WIMP rain is not nearly as pronounced as when one drives a car through rain, since the WIMPs are moving with velocities comparable to the Earth's.


The Earth's velocity in the Galaxy of course is not constant over a year, but varies due to the small velocity of the Earth around the Sun.  As a function of the day of the year $t$,
 \begin{equation}
v_{\mathrm{E}}(t)  \approx 232 + 15 \cos \left( 2\pi \frac {t-152.5}{365.25} \right) \mathrm{~km/s},
\label{eqnvE}
\end{equation}
with maximum occurring at $t = 152.5$ days, or June~2.
From equation~\ref{eqndRdEfull}, one can show that $dR/dv_{\mathrm{E}} \approx R/2v_{\mathrm{E}}$,
so the 6\% annual modulation in the Earth velocity from equation~\ref{eqnvE} causes about a 3\% annual modulation in the total WIMP interaction rate.  Note that this result is true only when considering all interactions, even those down to zero recoil energy.  The interaction rate above a (non-zero) experimental threshold energy can be as big as 7\%.

The motion of the Earth in the Galaxy towards the constellation Cygnus makes the WIMP flux in the lab frame sharply peaked, resulting in a higher rate of recoils from the direction of Cygnus.
For the standard halo model considered here (and neglecting the escape velocity)~\cite{SpergelDirection},
 \begin{equation}
\frac{dR}{dRd\cos\psi} \approx \frac{1}{2} \frac{R_0}{E_{0}r} \exp\left[ - \left( 
\frac{ v_{\mathrm{E}} \cos \psi - v_{\mathrm{min}} } {v_0} \right)^2 \right], 
\end{equation}
where $\psi$ is the recoil angle in the laboratory relative to the direction of Cygnus.
Since the Earth speed is comparable to the mean WIMP speed, 
the rate in the forward direction is roughly an order of magnitude larger than the rate in the backward direction~\cite{SpergelDirection}.

\subsection{Nuclear Form Factors}
\label{sec:formfactor}

\begin{figure}[t]
\begin{center}
\psfig{file=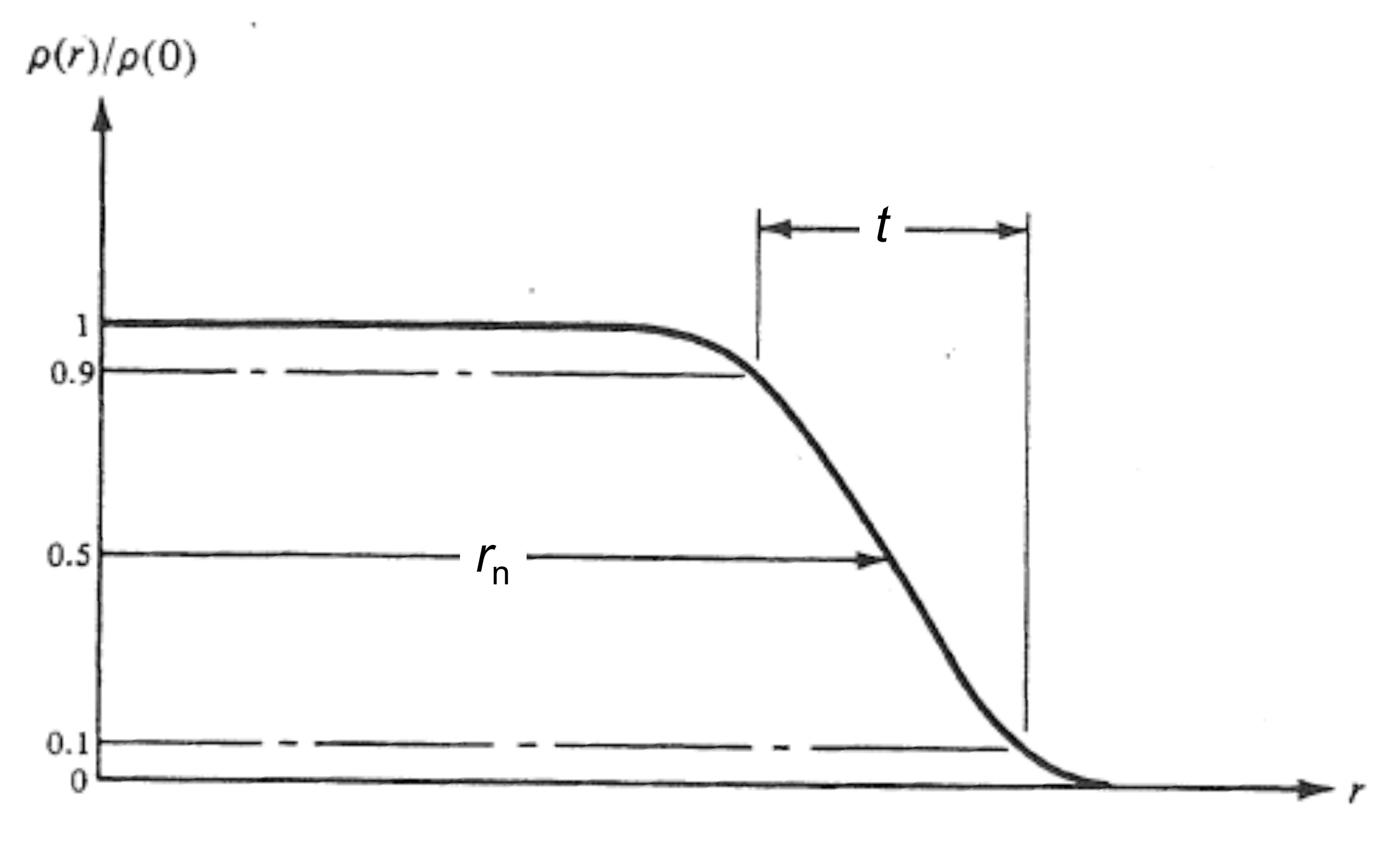, width=10cm}
\caption{Assumed density of scattering centers for spin-independent interactions, as proposed by Helm~\cite{Helm}.  Density is constant within the nuclear radius $r_{\mathrm{n}}$ then decreases to zero over a skin thickness $s$ (the related 10\%--90\% thickness $t$ is shown in this diagram). The Fourier transform of this distribution yields the Woods-Saxon form factor used for spin-independent scattering.
}
\label{fig:HelmDiagram}
\end{center}
\end{figure}

Under the approximation of plane-wave (Born) scattering,
 \begin{equation}
\mathcal{M}(\vec{q}) = f_{\mathrm{n}} A \int d^{3}x \rho(\vec{x}) e^{i\vec{q} \cdot \vec{x}}.
\end{equation}
We may identify the momentum-dependent part of this interaction, the form factor
\begin{equation}
F(\vec{q}) = \int d^3x \rho(\vec{x}) e^{i\vec{q} \cdot \vec{x}},
\end{equation}
as the Fourier transform of the scattering site positions.
For spin-independent interactions, a good approximation~\cite{lewinsmith} is the Woods-Saxon form factor
\begin{equation}
F(q) = \frac{3\left[ \sin(qr_{\mathrm{n}}) - qr_{\mathrm{n}}\cos(qr_{\mathrm{n}}) \right]}
{\left(qr_{\mathrm{n}}\right)^{3} } e^{-(qs)^{2}/2},
\end{equation}
which is the Fourier transform of a solid sphere of radius $r_{\mathrm{n}}$ with a skin thickness $s$, as shown in Figure~\ref{fig:HelmDiagram}.  In practice, Lewin and Smith~\cite{lewinsmith} recommend values of $s = 0.9$\,fm and
\begin{equation}
r_{\mathrm{n}}^2 = \left(1.23A^{1/3} - 0.60\mathrm{~fm}\right)^2 + \frac{7}{3}(0.52\pi\mathrm{~fm})^2 - 5s^2 .
\end{equation}

For spin-dependent interactions, the situation is more complicated.  A first approximation starts with a thin shell of valence nucleons,
\begin{equation}
F(q) = \frac{\sin(qr_{\mathrm{n}}) } {qr_{\mathrm{n}} },
\end{equation}
but must be extended with detailed nuclear-physics calculations~\cite{lewinsmith}.

\begin{figure}[t]
\begin{center}
\psfig{file=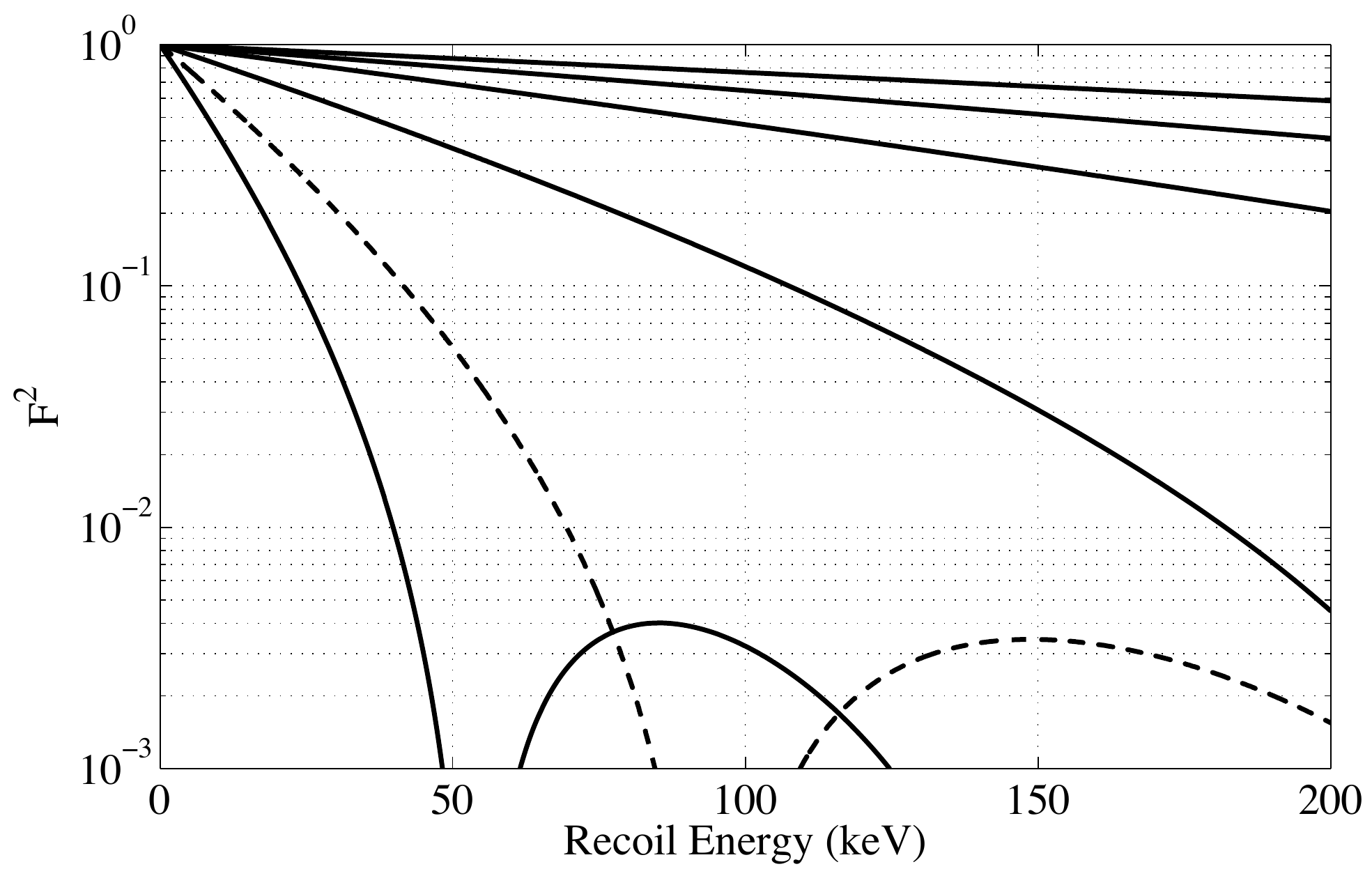, width=10cm}
\caption
{Spin-independent form factors $F^2$ as a function of recoil energy for targets of 6 atomic masses $A$.  From top to bottom on plot, materials are Ne ($A = 20$; F or Na are similar), Si ($A=  28$), Ar ($A= 40$), Ge ($A= 73$), Xe (dashed, $A= 131$, I is similar), and  W ($A=183$).
}
\label{fig:SIform}
\end{center}
\end{figure}

\begin{figure}[t]
\begin{center}
\psfig{file=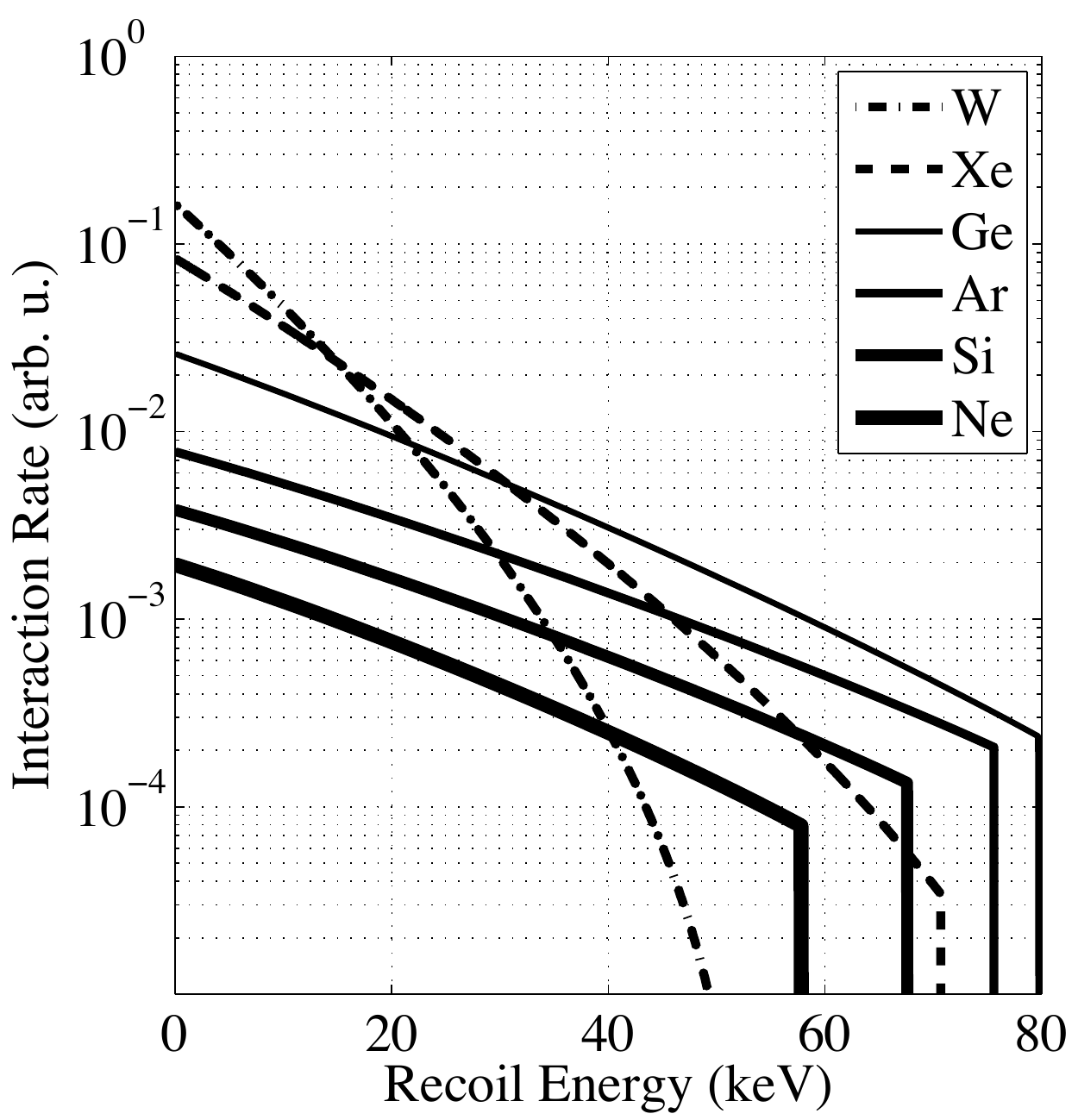, width=2.22in}
\psfig{file=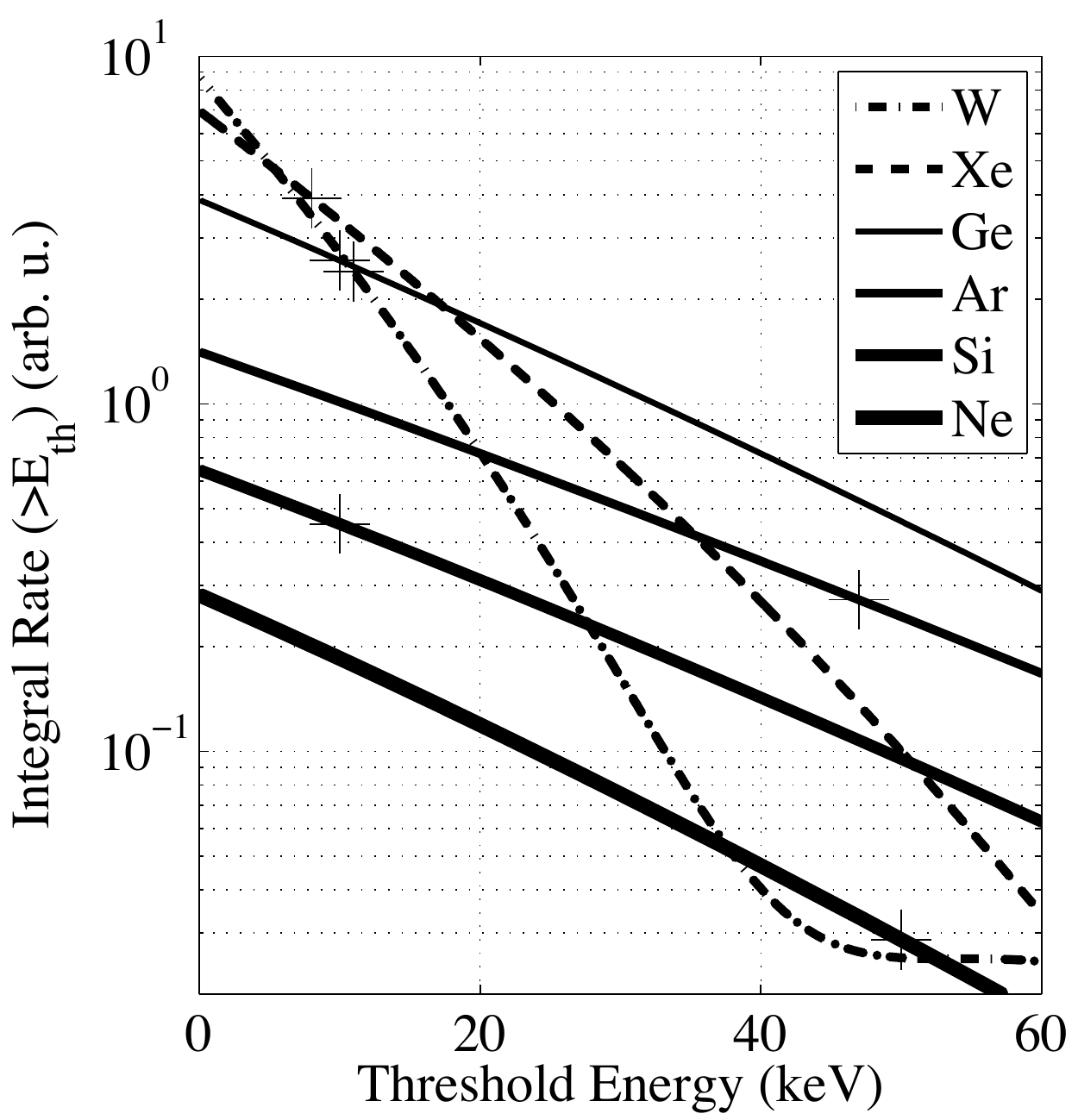, width=2.22in}
\caption
{Spin-independent interaction rates (per detector exposure) as a function of recoil energy for a 
WIMP on targets of 6 atomic masses $A$.  From top to bottom on each plot, materials are  W ($A=183$), Xe (dashed, $A= 131$, I is similar), Ge ($A= 73$), Ar ($A= 40$), Si ($A=  28$), and Ne ($A = 20$, F or Na are similar). {\it Left}: Differential rate for a 60--\gev\ WIMP.  High-$A$ materials have a higher rate at low energies, since the rate $\propto \mu_{A}^{2}A^2$, but loss of coherence greatly decreases the rate in these materials at high energies.  
As $A$ increases towards $M_{\chi}$, the mean energy and cutoff energy both increase due to kinematics, while loss of coherence offsets the increase in the mean energy.  As $A$ increases past $M_{\chi}$, the energy spectrum becomes softer and the cutoff energy decreases.
{\it Right}:
Integral rate above the energy threshold indicated for a 100--\gev\ WIMP.  Although energy thresholds vary from experiment to experiment, typical energy thresholds for each material are indicated by $+$ signs on each curve.  With these thresholds, the 100-\gev\ WIMP would produce the highest signal rate in Xe, with rates in W and Ge about 40\% lower.  I follows about the same curve as Xe, typically with a $3\times$ higher threshold and half the rate.  Rates in Si are  $\sim9\times$ lower than in Xe, rates in Ar are $\sim14\times$ lower, and rates in Ne (or Na or F with this threshold) are $\sim100\times$ lower.
}
\label{fig:rates}
\end{center}
\end{figure}

In either case, $F(q)<1$ when the de Broglie wavelength $\lambda < r_{\mathrm{n}}$ and the WIMP ceases to interact coherently with the entire nucleus.  Since the nuclear radius $r_{\mathrm{n}} \approx A^{1/3}$\,fm, 
this criterion may be rewritten
\begin{equation}
\lambda = \frac{\hbar}{q} = \frac{\hbar c}  { \sqrt{ 2M_{A}c^{2}E_{\mathrm{R}} }  } = 
\frac{197 \mathrm{~MeV~fm} } {\sqrt{2AE_{\mathrm{R}}(\mathrm{keV}) } } < A^{1/3} . 
\end{equation}
Hence, coherence is lost when
\begin{equation}
E_{\mathrm{R}}> \frac{2 \times 10^4}{A^{5/3}} \mathrm{~keV} \sim 100~\mathrm{keV}.
\end{equation}
The strong dependence on $A$ indicates that coherence is lost much earlier for high-$A$ targets, as shown in Fig.~\ref{fig:SIform}. This loss of coherence significantly reduces the advantage of using particularly heavy target materials; practically speaking use of materials heavier than Ge yield only modest increases in overall rate, far short of the $A^2$ increase that would occur without the loss of coherence.  Since the loss of coherence makes these high-$A$ targets intrinsically insensitive to high-energy depositions, it is particularly critical that experiments with high-$A$ materials achieve low energy thresholds.  Figure~\ref{fig:rates} shows the relative rates for the same WIMP in several different targets.

\begin{figure}
\begin{center}
\psfig{file=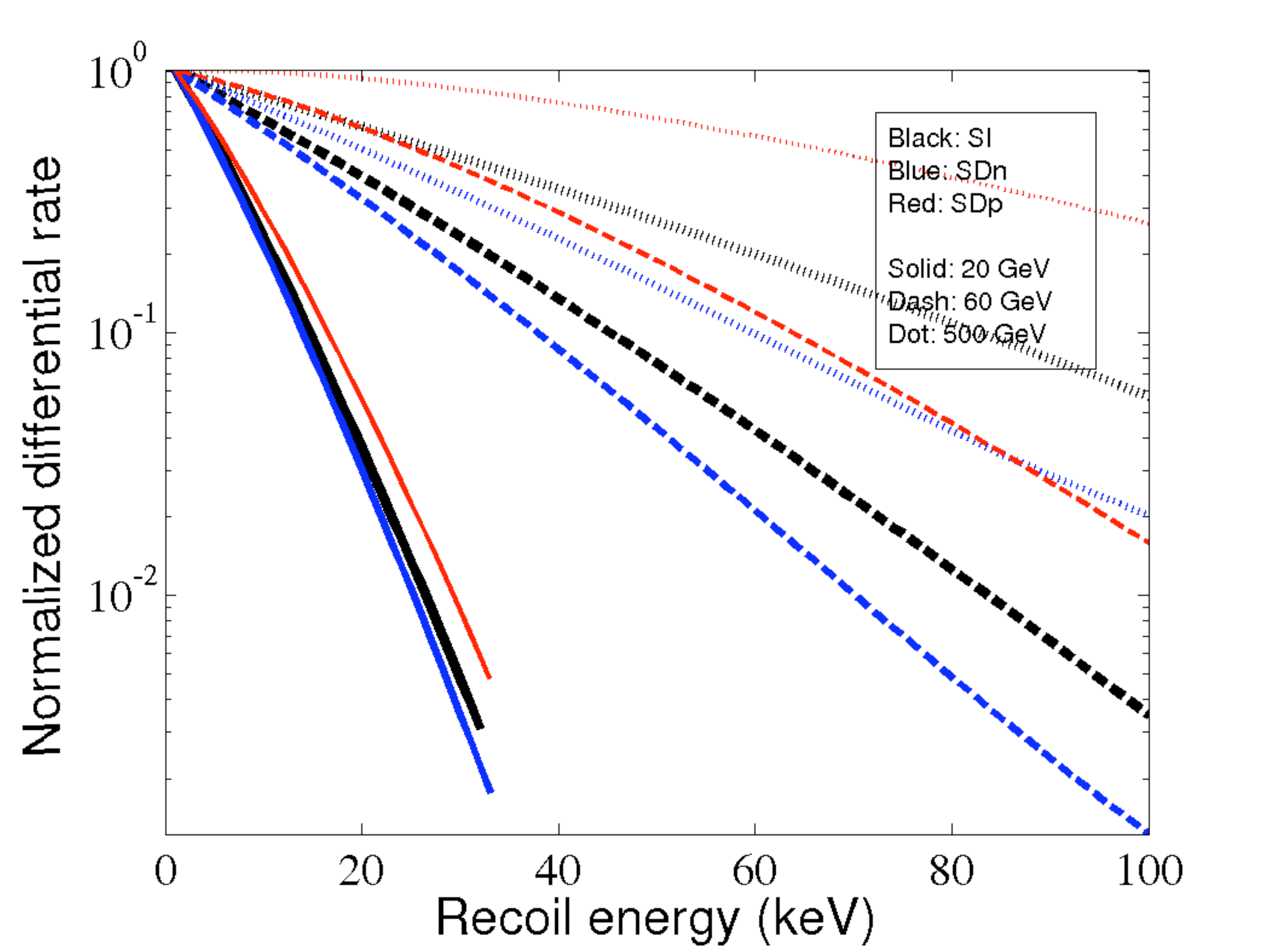, width=8cm}
\caption
{
Expected energy spectra for a 20-\gev (solid), 60-\gev (dashed), or 500-\gev (dotted) WIMP interacting on Ge via the neutron-spin-dependent (lowest), spin-independent (middle), or proton-spin-dependent (top) interaction.  For relatively massive WIMPs for which the loss of coherence is significant, a high-statistics detection could identify the type of interaction via the spectral shape.  Figure provided by J.~ Filippini. }
\label{fig:SISDspectra}
\end{center}
\end{figure}

\subsection{Implications of a detection}
\label{detection}
 
Because the spin-independent, proton-spin-dependent, and neutron-spin-dependent form factors are different for a given target, it is possible in principle to 
distinguish the type of interaction by the energy spectrum on a single target isotope.
Differences are insignificant for low-mass WIMPs since all form factors are essentially unity.  However, as shown in Fig.~\ref{fig:SISDspectra}, differences for high-mass WIMPs may be significant with a sufficiently high-statistics detection (particularly if the WIMP mass is known independently).  

Detections with several target materials would reveal the type of interaction more clearly.
Comparing the interaction rate in different materials would indicate the interaction type since the rates scale differently in each (Figure~\ref{fig:rates} shows the scaling for spin-independent interactions, while Table~\ref{t1} shows the material-dependent scaling factors for spin-dependent interactions).   
Furthermore, detection with different target materials provides a useful confirmation of the detection, especially if a consistent WIMP mass is determined from each.

\begin{table}
\tbl{Projected limits (at the 99\% confidence level) on WIMP mass for a 60-\gev\ WIMP on Ge based on statistical uncertainties only~\cite{JacksonGaitskellSchnee}.  Also listed is the minimum mass $M_{\mathrm{min}}$ for which there is no upper limit on the WIMP mass.}
{\begin{tabular}{cccc}\toprule
Events     & Lower  &  Upper & \\
Detected & Limit     &  Limit   &  $M_{\mathrm{min}}$ \\
\colrule
\hphantom{00}10	 & 30 \gev & none       & \hphantom{0}50 \gev \\
\hphantom{0}100  	 & 45 \gev & 101 \gev & 100 \gev \\
                      1000	 & 55 \gev & \hphantom{0}69 \gev   & 250 \gev \\
\botrule
\end{tabular}
}
\label{tab:MassLimits}
\end{table}

\begin{figure}[t]
\begin{center}
\psfig{file= 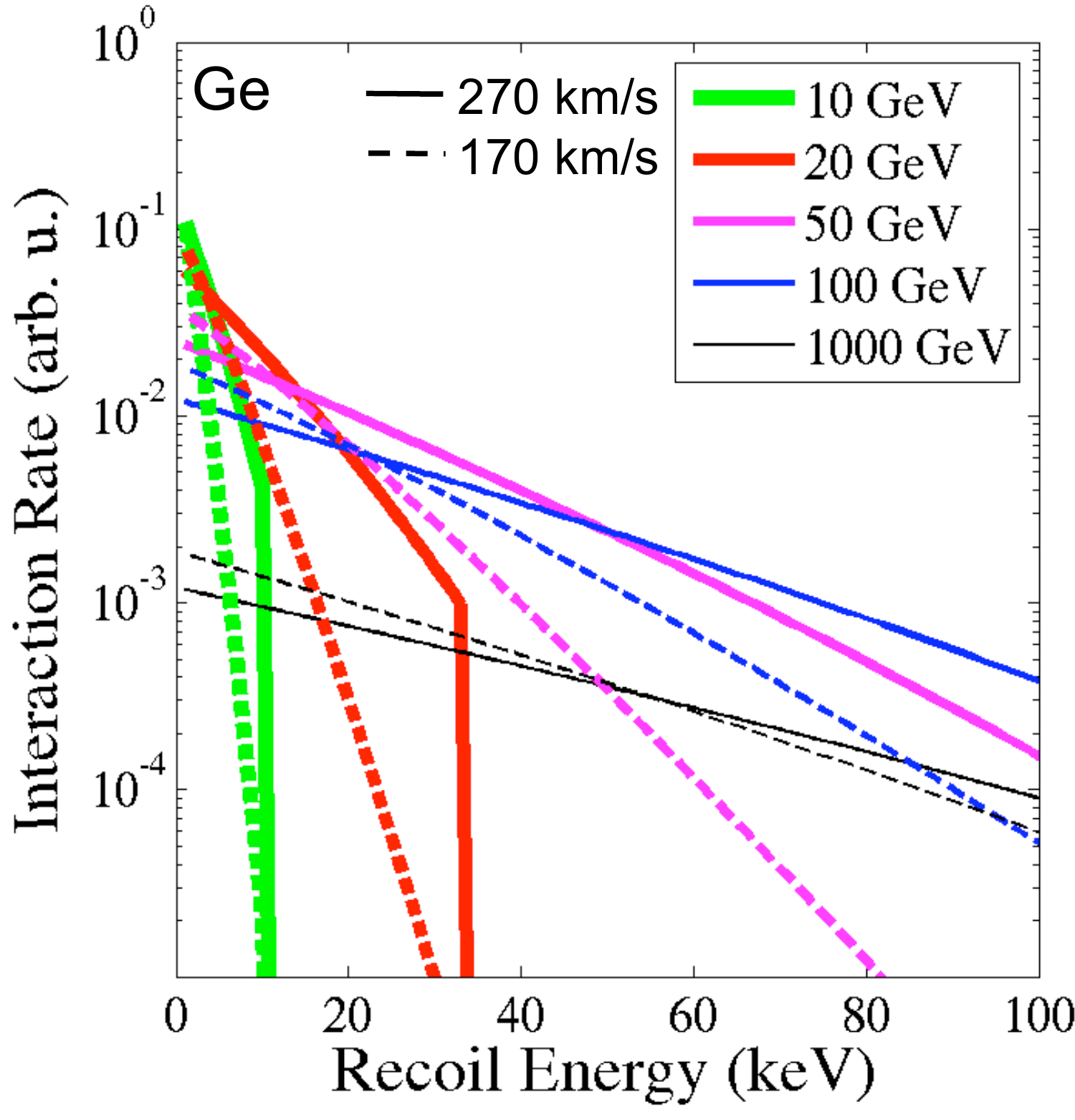,width=2.18in}
\psfig{file= 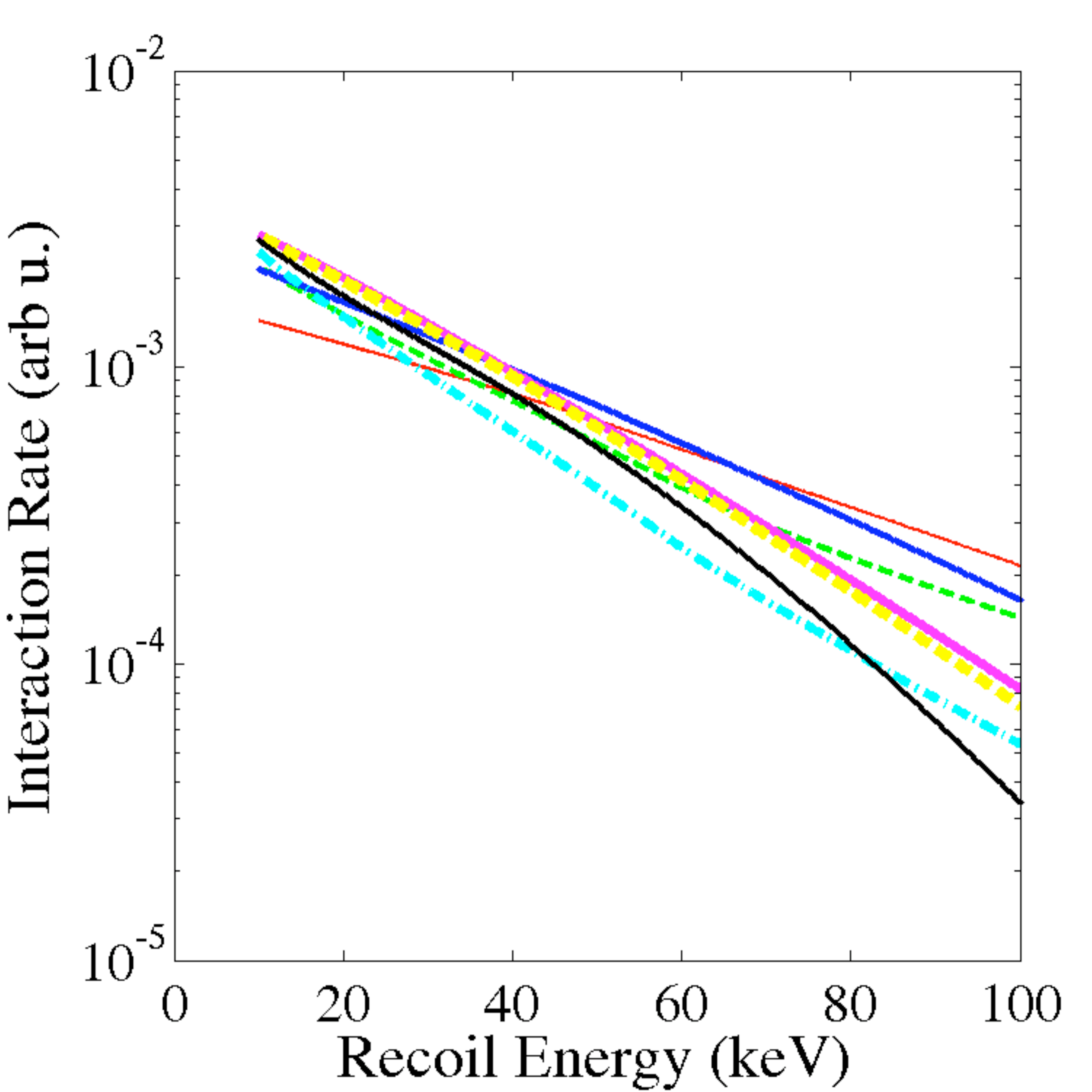,width=2.26in}
\end{center}
\caption{{\it Left}: Expected interaction rate on Ge as a function of recoil energy for different 
WIMP masses $M_{\chi}$ (as shown in the legend in units of \gev), for  most probable WIMP velocity $v_0= 270$\,km/s (solid) and
$v_0= 170$\,km/s (dash), which span the extremes of possible values.
For a large-statistics detection of a low-mass WIMP, the 30\% uncertainty on $v_0$ would result in  a similar uncertainty on the WIMP mass, as seen by the similarity in spectral shapes \eg\ between the 10\,\gev\ WIMP with $v_0=270$\,km/s and the 100\,\gev\ WIMP with  $v_0=170$\,km/s. 
{\it Right}:
Expected interaction rate on Ge for a 60-\gev WIMP as a function of recoil energy for 
isothermal, triaxial, and Evans halo models with various parameters.  For an initial WIMP discovery, uncertainty in the WIMP velocity distribution will increase uncertainty on the WIMP mass inferred from the detection.
Ultimately, these differences in energy spectra may allow inference of the correct halo model.
}
\label{fig:EnergySpectrumVsMv0}
\label{fig:HaloModels}
\end{figure}

Measurement of the WIMP recoil spectrum would provide constraints on the WIMP mass, as can be seen from Fig.~\ref{fig:EnergySpectrumVsM}.  However, since 
heavy WIMPs all yield about the same energy spectrum, as shown by Eqn.~\ref{eqn:SpectrumOnMass}, detection of a heavy WIMP would provide only weak constraints on its mass, other than that it must be relatively heavy.  Table~\ref{tab:MassLimits} lists how well a WIMP mass may be determined by a detection for a spin-independent interaction on Ge if the WIMP velocity distribution is known. 
The uncertainties on the type of interaction and WIMP velocity distribution would contribute additional uncertainty on an inferred WIMP mass.  As shown in Fig.~\ref{fig:SISDspectra}, the different form factors may help a lower-mass WIMP with one predominant interaction produce a spectrum more similar to that of a higher-mass WIMP. 
For a Maxwellian distribution, the uncertainty on $v_0$ translates into uncertainty on $M_{\chi}$ even for low-mass WIMPs
(see Fig.~\ref{fig:EnergySpectrumVsMv0}).
For small $M_{\chi}$, since
$\langle E_{\mathrm{R}} \rangle  \propto v_0^2 M_{\chi}^2$ (from Eqn.~\ref{eqn:SpectrumOnMass}), 
\begin{equation}
\frac{\Delta M_{\chi}}{ M_{\chi}} = \frac{\Delta v_{0} } { v_0 } ,
\end{equation}
so systematic uncertainties on the WIMP mass due to halo uncertainties are of order 20\%.
Ultimately, sufficient measurement of the energy spectrum may allow better determination of $v_0$ (especially if the WIMP mass is determined independently from collider data) and even identification of the full WIMP velocity distribution, and hence the shape of the dark matter halo.  The right panel of Fig.~\ref{fig:HaloModels} shows the differences in the expected spectrum due to different halo models~\cite{copiprivate}.
Although detectors that are sensitive to the direction of the WIMP are in the prototype stage (see Section~\ref{section:directional}), a high-statistics detection with a detector capable of the determining the recoil direction would allow the detailed determination of the WIMP velocity distribution~\cite{MorganDirectional1,MorganDirectional2}, essentially ushering in an age of WIMP astronomy.

Finally, better measurements of WIMP mass from colliders may be combined with information from direct detection to better constrain the WIMP-nucleon cross section (and hence particle-physics parameters). 
For many models, the LHC will constrain the WIMP mass to 10\%.
However, it is difficult to measure WIMP properties well.
If the LHC determines the WIMP mass, direct detection can determine WIMP-nucleon cross section much better than LHC alone~\cite{Baltz2006LCC}.

\section{WIMP Direct Detection Experiments}
\label{section:directexp}
\label{section:methods}

Direct-detection experiments have already limited the expected WIMP-nucleon interaction rate to fewer than 1 event per 10 kilograms of target material per day (10 kg-day). With such a small event rate, it is a daunting task to search for a WIMP interaction amongst the background interactions from cosmic rays and natural radioactivity, which typically number in the millions per kg-day (see 
Ref.~\refcite{formaggioBackgroundsAnnRev} for a detailed review of the principle sources of background for underground experiments).

Because it is not possible to distinguish a single neutron scatter from a WIMP scatter if the neutron does not scatter in additional active material, neutrons provide a particularly dangerous background for WIMP-search experiments.  Material with lots of hydrogen, such as polyethylene or clean water, acts as shielding for neutrons by reducing the neutrons' energies enough that they cannot cause a recoil above threshold (due to simple kinematics, more massive elements do not significantly reduce the energy of a a scattering neutron).  Neutrons produced by $(\alpha,n)$ reactions (from  uranium and thorium in rock walls, for example) may be effectively shielded in this way since these neutrons start with relatively low energies and have high interaction cross sections. For every 13\,cm of polyethylene, this low-energy neutron background is reduced by an order of magnitude~\cite{meihime}, with a thickness of $\sim$40\,cm or more needed for current experiments.  

\begin{figure}[t]
\begin{center}
\psfig{file= 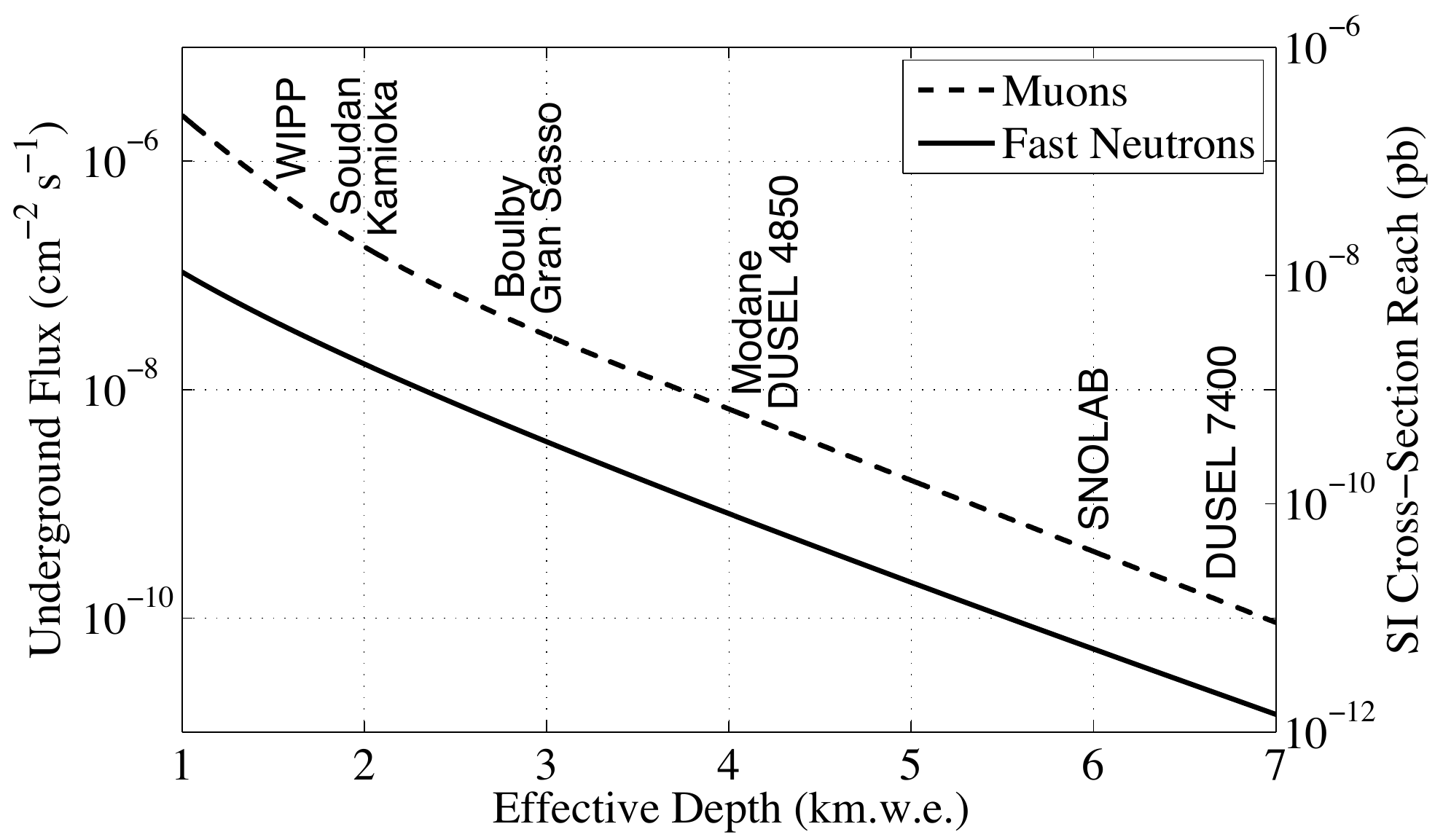,width=3.5in}
\end{center}
\caption{Flux of muons (dashes) and muon-induced neutrons (solid) as functions of depth underground, measured in terms of the equivalent thickness of water in km below a flat surface that is needed to provide equal shielding.  Effective depths of primary underground facilities for dark matter experiments are listed (Canfranc is similar to Soudan or Kamioka).  Although the neutron background resulting from a given fast neutron flux is highly dependent on the experimental setup and materials, the curve of neutron flux (still solid) referred to the right-hand axis shows the limit on sensitivity reach due to neutron backgrounds for one possible experimental setup~\cite{meihime}.  At depths below about 10 km w. e., the muon flux is $3\times 10^{-12}$\,cm$^{-2}$\,s$^{-1}$, dominated by neutrino-induced muons~\cite{formaggioBackgroundsAnnRev}.}
\label{fig:muondepth}
\end{figure}

Such shielding is not effective for more energetic neutrons, such as those produced by cosmic-ray muons.  To reduce this critical background (and others from cosmic rays), all experiments are located underground, with all but prototype experiments located deep underground.  Table~\ref{TableUnderground} lists the depths and locations of the principle underground laboratories for dark matter experiments.  
Since denser rock provides a greater effective depth than less dense rock, depths are standardly
listed in terms of the thickness of water (\eg\ meters of water equivalent, or mwe) that would provide the same integrated density as the actual overhead rock.  For facilities under mountains, usually the mean effective depth is quoted, which inaccurately suggests a lower muon flux than in actuality, since shorter pathlengths dominate the muon flux. 
As shown in Fig.~\ref{fig:muondepth}, depth is effective for reducing backgrounds due to the energetic neutrons produced by cosmic-ray muons. In addition, most experiments are surrounded, or at least covered, by an active muon veto to allow rejection of energetic neutrons if the muon progenitor passes close to the experiment.  Designs for most future experiments use large instrumented water tanks to provide both shielding for low-energy neutrons and identification of fast neutrons or muons that traverse the shield.

\begin{table}
\tbl{Locations, depths, and effective depths~\cite{meihime} of primary underground facilities for dark matter experiments~\cite{formaggioBackgroundsAnnRev}.  Both the 4850-foot (currently the Sanford Lab) and planned 7400-foot DUSEL spaces are listed.}
{\begin{tabular}{lrrcc}\toprule
		  & Depth  & Depth  &  \\
Laboratory &  (m)~     &  (mwe) &  website \\
\colrule
WIPP, AZ, U.S. & 600 & 1600 & www.wipp.energy.gov/science/index.htm \\
Soudan, MN, U.S. & 710 & 2000  &  www.hep.umn.edu/soudan \\
Canfranc, Spain & $\leq$860 & $\leq$2450  & www.unizar.es/lfnae \\
Kamioka, Japan & 1000 & 2050 & www-sk.icrr.u-tokyo.ac.jp \\
Boulby, U.K. & 1100 & 2800 &   hepwww.rl.ac.uk/ukdmc/pix/boulby.html \\
Gran Sasso, Italy & 1400 & 3100 & www.lngs.infn.it \\
Modane, France & 1760 & 4200  &  www-lsm.in2p3.fr \\
Sudbury, Canada & 2160 & 6000  & www.sno.phy.queensu.ca \\
DUSEL, SD, U.S. & 1500 & 4300 & www.dusel.org \\
 (planned) & 2260 & 6200 & \\
\botrule
\end{tabular}
}
\label{TableUnderground}
\end{table}

\begin{figure}[t]
\begin{center}
\psfig{file= 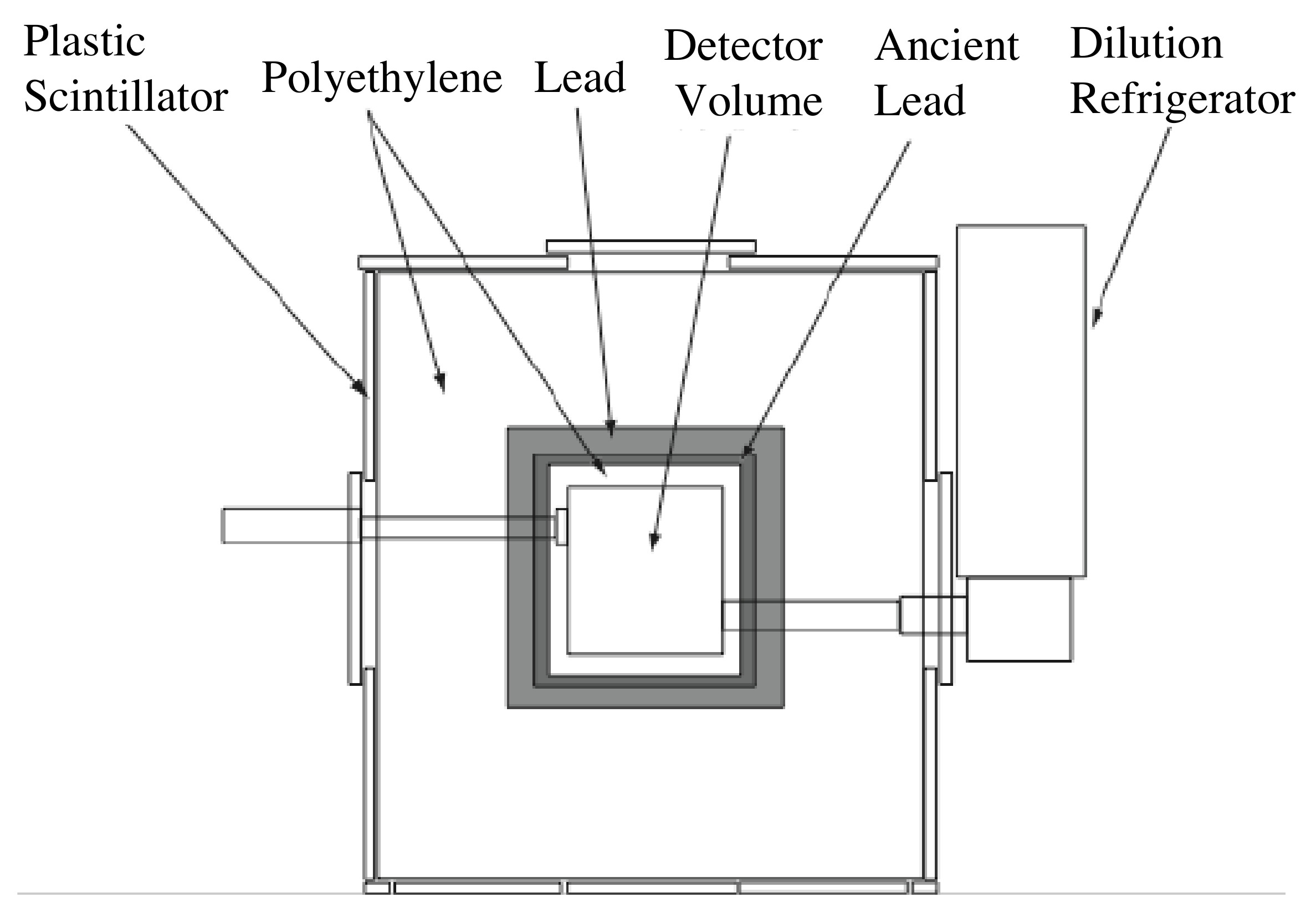,width=3.5in}
\end{center}
\caption{Sliced side view of a typical shielding setup (here for the CDMS experiment~\cite{r118prd}).
Outermost scintillator paddles act as a veto against events due to muons.  A thick polyethylene shield moderates the flux of neutrons from  $(\alpha,n)$ reactions by 5 orders of magnitude.  Lead (grey) reduces the photon backgrounds by 4 orders of magnitude, with the inner ancient lead liner (dark grey) reducing the background from electron bremsstrahlung from $^{210}$Bi, a daughter of $^{210}$Pb present in modern lead.  Inner polyethylene reduces the neutron background from fast neutrons that penetrate the outer polyethylene and interact in the lead.  Additional polyethylene within the lead shield would improve the neutron moderation, but at a significant increase in the amount and expense of the lead shielding.
Materials that cannot easily be made radiopure (here the dilution refrigerator used to cool the cryogenic detectors) must be shielded from the detectors. 
}
\label{fig:shielding}
\end{figure}

Experiments take additional precautions against other sources of backgrounds,
which otherwise would cause $\sim10^4$ events\,\perkkgd.
Low-radioactivity copper, which is straightforward to produce, or lead with an inner liner of ancient lead (for which radioactive isotopes present at its smelting have decayed away) is used to reduce the background from photons, typically by 4--5 orders of magnitude. 
Any air near the detectors is purged of radon.
Figure~\ref{fig:shielding} shows a typical shielding setup around an experiment. 
Materials that surround or constitute the detector must be ultra-low-radioactivity, requiring they be screened for possible contamination.  Residual radioactivity in the detectors or their shielding typically is the dominant source of background in experiments, with radioactivity on detector surfaces (typically from plateout of radon daughters) a particular problem.  

In addition to shielding backgrounds, experments reject events that are more likely to be due to backgrounds such as photons, electrons, or alpha-particles.  For example, WIMPs 
interact so weakly that they 
never interact more than once in a detector, 
allowing experiments to reject multiple-scatter events.
Most detectors allow rejection of some multiple-scatter events with negligible loss in efficiency to WIMPs, often through the use of arrays of detector modules, so that if two separate modules have energy depositions, the event must be a multiple-scatter.  For liquid nobles, multiple scatter may result in pulses separated in time or energy deposited in places sufficiently separated in space so as to allow identification by event reconstruction.
Similarly, WIMPs interact uniformly throughout a detector, so it pays to cut interactions near detector surfaces, where more background interactions occur.
Most experiments use some form of event reconstruction to form a ``fiducial'' volume by rejecting events inferred to occur near the detector surface.  When comparing detector masses, it is most appropriate to consider this fiducial mass.

Most significantly, WIMPs tend to interact with an atom's nucleus, while the dominant radioactive backgrounds (everything except neutrons) interact with electrons, so  experiments that 
discriminate between interactions causing an electron to recoil and those that cause a nuclear recoil can reject virtually the entire radioactive background.
There are three ways to discriminate between electron recoils and nuclear recoils.  
Each is based on the fact that, for $E_{\mathrm{R}} \approx 10$\,keV, an electron recoils with $v\approx 0.3c$, whereas a nucleus recoils with $v \approx 7\times 10^{-4}c$, depositing its energy much more densely over a very short track.
Threshold detectors such as COUPP~\cite{coupp2005} (described in more detail in Section~\ref{section:threshold}) require a dense energy deposition to trigger
and therefore
are nearly immune to electron recoils, whose deposited energy is almost never dense enough to trigger. 
For other experiments, the pulse timing 
is different for electron recoils than for nuclear recoils. 
Finally, depending on the material, recoil energy may be converted into light, ionization, and/or phonons.  
Experiments that measure two of these forms may discriminate against electron-recoil backgrounds because the
relative amount of energy in 
the two forms is different for nuclear recoils than for electron recoils.

These different measures of recoil energy and the differing response of electron and nuclear recoils may introduce an ambiguity in quoted energy.
To avoid this ambiguity, most (but not all) experimenters are explicit about which signal (or combination of signals) is used to determine an event's energy. 
The unit ``keVee'' quantifies a measured
signal in terms of the energy (in keV) of an electron recoil that would
generate it, while ``keVr'' 
 indicates the energy of a nuclear recoil that would generate the signal observed.

The energy scale for keVee is generally easy to establish since photon backgrounds (or calibration sources) typically produce mono-energetic features at known energies, although sometimes extrapolations of the scale to low energy are required.
The nuclear-recoil energy scale is more difficult, due to a lack of such features.
In practice, two methods are used.  Neutron
scattering experiments  allow the nucleus's recoil energy 
to be inferred from the incoming neutron energy and the neutron's measured angle of scattering,
while simultaneously measuring the signal size.   Unfortunately, multiple-scatter backgrounds are usually bad enough that significant simulations are needed to obtain accurate results, especially for the low energies that can be measured only with forward scattering.
Alternatively, comparing simulation results to the observed shape of an energy spectrum from a neutron source with a broad energy may yield the energy scale.  
Often, the ratio of a signal in keVr to keVee is called the signal's quenching factor,
 \begin{equation}
QF \equiv E \mathrm{(keVr)} / E \mathrm{(keVee)} .
\label{eqnQF}
\end{equation}
(although sometimes the inverse of this quantity is called the quenching factor).
If an electron recoil would produce a larger signal than a nuclear recoil of the same energy (as is usually the case), the quenching factor of the signal $QF >1$.

There are large fundamental differences in light, ionization, and phonon signals.
Light signals are the fastest, with ns timing possible, 
but  only $\sim$10 photons are produced per keV. 
In order to take advantage of the excellent discrimination potential of timing using light signals, efficient light collection is critical.
Ionization is somewhat better,  with $\sim$100 quanta  per keV, while a whopping $\sim$10,000 phonons are produced per keV.
Experiments that detect phonons therefore have fundamentally better energy resolution and 
energy-based discrimination capabilities compared to other experiments.

Reduction of backgrounds is critical in order to maximize sensitivity reach.
If backgrounds are kept negligible, the search sensitivity of an experiment is directly
proportional to the target exposure (target mass $M \times$ exposure time $t$). 
If the expected background is non-negligble but can be estimated with negligible uncertainty by some means, it may be statistically subtracted (explicitly or implicitly), with the resulting Poisson errors causing 
the sensitivity improvement to be proportional to $\sqrt {MT}$~\cite{gaitskellucla1996}. 
In practice, most dark matter experiments have been background-dominated, without means to estimate the backgrounds accurately.  In these cases, the experiments are unable to take full advantage of their target exposure, as increasing exposure would result in little or no sensitivity improvement due to the systematic uncertainties in any background subtraction.
The importance of systematics also makes pursuing different techniques critical, since these technologies 
tend to have
different systematics, thus providing critical cross checks for a detection claim.

\begin{table}
\tbl{Characteristics of selected dark matter experiments~\cite{gaitskellreview2004}, including fiducial mass $M$ and whether scintillation light ($\gamma$), phonons ($\phi$), ionization ($q$), or another form of energy is detected, and whether the experiment's primary mission is neutrinoless double-beta decay ($\beta\beta$).}
{\begin{tabular}{lllcrcl}\toprule
		       &                 &      Readout  &             T      &     M~ &             & Search \\
Experiment     & Location & ($\gamma, \phi, q$) & (K) & (kg) & Target & Dates \\
\colrule
NAIAD & Boulby         & $\gamma$ & 300  &   50 & NaI & 2001--2005 \\  
DAMA/NaI & Gran Sasso & $\gamma$ & 300  &  87 & NaI & 1995--2002 \\ 
DAMA/LIBRA & Gran Sasso & $\gamma$ & 300  &  233 & NaI & 2003-- \\ 
ANAIS & Canfranc     & $\gamma$ & 300  & 11 & NaI & 2000--2005 \\ 
ANAIS & Canfranc     & $\gamma$ & 300  & 100 & NaI & 2011-- \\ 
KIMS   & Yangyang   & $\gamma$ & 300  & 35 & CsI & 2006--2007 \\ 
KIMS   & Yangyang   & $\gamma$ & 300  & 104 & CsI & 2008-- \\ 
CDMS II & Soudan &  $\phi, q$ & $<1$ & 1 & Si   & 2001--2008 \\ 
                &                 &                                         &               & 3 & Ge & 2001--2008 \\ 
SuperCDMS & Soudan   & $\phi, q$& $<1$ & 12 & Ge & 2010--2012 \\ 
SuperCDMS & SNOLAB & $\phi, q$ & $<1$ & 120 & Ge & 2013--2016 \\ 
GEODM         & DUSEL   & $\phi, q$ & $<1$ & 1200 & Ge & 2017-- \\ 
EDELWEISS I & Modane &  $\phi, q$ & $<1$ & 1 & Ge & 2000--2004 \\ 
EDELWEISS II & Modane & $\phi, q$& $<1$ & 4 & Ge & 2005-- \\ 
CRESST II & Gran Sasso &  $\phi, \gamma$ & $<1$ & 1 & CaWO$_4$ & 2000-- \\
EURECA & Modane & $\phi,  q$ & $<1$ & 50 & Ge & 2012--2017 \\ 
                  &                 & $\phi, \gamma$  & $<1$ & 50 & CaWO$_4$ & 2012--2017 \\
SIMPLE & Rustrel & Threshold & 300 & 0.2 & Freon &  1999-- \\ 
PICASSO & Sudbury & Threshold & 300  & 2 & Freon & 2001-- \\ 
COUPP   & Fermilab & Threshold & 300  & 2 &  Freon & 2004--2009 \\
COUPP   & Fermilab & Threshold & 300  & 60 &  Freon & 2010-- \\
TEXONO      & Kuo-Sheng & $q$, $\beta\beta$ & \hphantom{0}77   & 0.02 & Ge & 2006-- \\ 
CoGeNT       & Chicago & $q$, $\beta\beta$ & \hphantom{0}77   & 0.3 & Ge & 2005-- \\
                       & Soudan & $q$, $\beta\beta$ & \hphantom{0}77   & 0.3 & Ge & 2008-- \\
MAJORANA & Sanford & $q$, $\beta\beta$ & \hphantom{0}77   & 60 & Ge & 2011-- \\
ZEPLIN III    & Boulby & $\gamma, q$ & 150  & 7 & LXe & 2004-- \\ 
LUX              & Sanford & $\gamma, q$ & 150  & 100 &  LXe & 2010-- \\ 
XMASS        & Kamioke & $\gamma, q$ & 150  & 3 & LXe & 2002--2004 \\
XMASS        & Kamioke & $\gamma, q$ & 150  & 100 &  LXe & 2010-- \\ 
XENON10   & Gran Sasso & $\gamma, q$ & 150  & 5 &  LXe & 2005--2007 \\ 
XENON100 & Gran Sasso & $\gamma, q$ & 150  & 50 &  LXe & 2009-- \\ 
WArP            & Gran Sasso & $\gamma, q$ & \hphantom{0}86  & 3 &  LAr & 2005--2007 \\
WArP            & Gran Sasso & $\gamma, q$ & \hphantom{0}86  &140 &  LAr & 2010-- \\
ArDM            & CERN          & $\gamma, q$ & \hphantom{0}86  & 850 &  LAr & 2009-- \\ 
DEAP-1        & SNOLAB     & $\gamma$     & \hphantom{0}86  & 7     &  LAr & 2008-- \\
MiniCLEAN & SNOLAB     & $\gamma$     & \hphantom{0}86  & 150   &  LAr & 2012-- \\
DEAP-3600 & SNOLAB     & $\gamma$     & \hphantom{0}86  & 1000 &  LAr & 2013-- \\
DRIFT-I  & Boulby & Direction & 300  & 0.17 & CS$_2$ & 2002--2005 \\
DRIFT-2 & Boulby & Direction & 300  & 0.34 &  CS$_2$ & 2005-- \\
NEWAGE & Kamioka & Direction & 300  & 0.01 &  CF$_4$ & 2008-- \\
MIMAC     & Saclay     & Direction & 300  & 0.01 &  many & 2006-- \\
DMTPC & MIT & Direction & 300  & 0.01 &  CF$_4$ & 2007-- \\

\botrule
\end{tabular}
}
\label{TableExperiments}
\end{table}

The basic techniques include threshold detectors that nucleate if a sufficient energy deposition occurs (see Section~\ref{section:threshold}), ultrapure scintillators (see Section~\ref{section:dama}), masses of liquid nobles view by light detectors, with or without an electric field to collect ionization (see Section~\ref{section:nobles}), solid-state detectors cooled to mK temperatures in order to detect phonons, as well as light or ionization (see Section~\ref{section:cryo}), gaseous detectors for measuring the direction and energy density of each recoil (see Section~\ref{section:directional}), and others that do not fit into these categories.
There are dozens of WIMP-search experiments in progress or development worldwide.
Table~\ref{TableExperiments}, based on the more extensive but slightly dated table in 
Ref.~\refcite{gaitskellreview2004}, lists a selection of them.  
In the sections that follow, I describe the various types of WIMP-search experiments, with the aim of introducing how the technique works and describing its advantages and challenges.
I have not included discussion of many interesting results that have been made public during the preparation of this manuscript.
For more detailed reviews of dark matter experiments, see 
Refs.~\refcite{chardin2004review,gaitskellreview2004,spooner2007review, directional2009,bertonePDM}.

\subsection{Threshold Detectors}
\label{section:threshold}

As mentioned above,
one promising technology uses a superheated liquid (in bulk~\cite{coupp2005}  or as droplets within a matrix~\cite{simple2008,picasso2005long}) as a threshold detector.  
By tuning thermodynamic parameters (\eg\ temperature and pressure), the detector may be made insensitive to the low energy density deposited by a minimum-ionizing electron recoil.  Only a dense energy deposition, such as from a nuclear recoil, will provide enough energy to 
cause nucleation (smaller depositions result in sub-critical bubbles that are squashed to nothing by their surface pressure).  
A drawback of these detectors is that they provide limited energy information on events.
The 
attraction of these detectors is that they could allow inexpensive scaling to very large masses with  a broad range of materials and without need of cryogens or photon shielding.  Already, the ability of these experiments to use materials sensitive to spin-dependent interactions on protons has allowed them to set world-leading limits~\cite{coupp2008SDlimits,simple2005,picasso2009}, as shown in Fig.~\ref{fig:SD}.

The primary background for these experiments other than neutrons (which plague all experiments) is alpha decays from radioactive contamination (primarily from radon daughters) in the detectors.  The PICASSO collaboration has recently shown that these alpha decays are rejectable~\cite{picasso2008}.  Since a recoiling alpha particle is lighter and hence has a much longer stopping distance than a recoiling nucleus, an alpha particle may produce more than a single nucleation along its track.  Although these individual nucleations grow together quickly, the amplitude of the high-frequency signal, detectable through microphones on the detector,  allows discrimination from the nuclear recoils of WIMPs.  If the rejection proves to be strong enough, it will dramatically increase the achievable sensitivity of experiments based on the superheated liquid technique.

One of these experiments, COUPP~\cite{coupp2005}, uses a large volume of superheated liquid and hence is similar to the classic bubble chambers used in high-energy experiments in the 1960s.
A principle challenge is keeping the detector stable, since any nucleation in the chamber requires recondensing the liquid, with a significant deadtime. 
The COUPP collaboration has met the principle challenge of preventing nucleation due to micro-cracks in vessel walls~\cite{coupp2005}.  They are currently running a 60-kg chamber with higher-purity materials and capability of rejecting alpha events, and have shown preliminary results indicating an impressive reduction in backgrounds, including a significant reduction in the rate of alphas from impurities in the quartz of the vessel that otherwise would limit the livetime of a larger device~\cite{couppDM2010}.

\subsection{Noble Liquid Detectors}
\label{section:nobles}

\begin{figure}[t]
\begin{center}
\psfig{file= 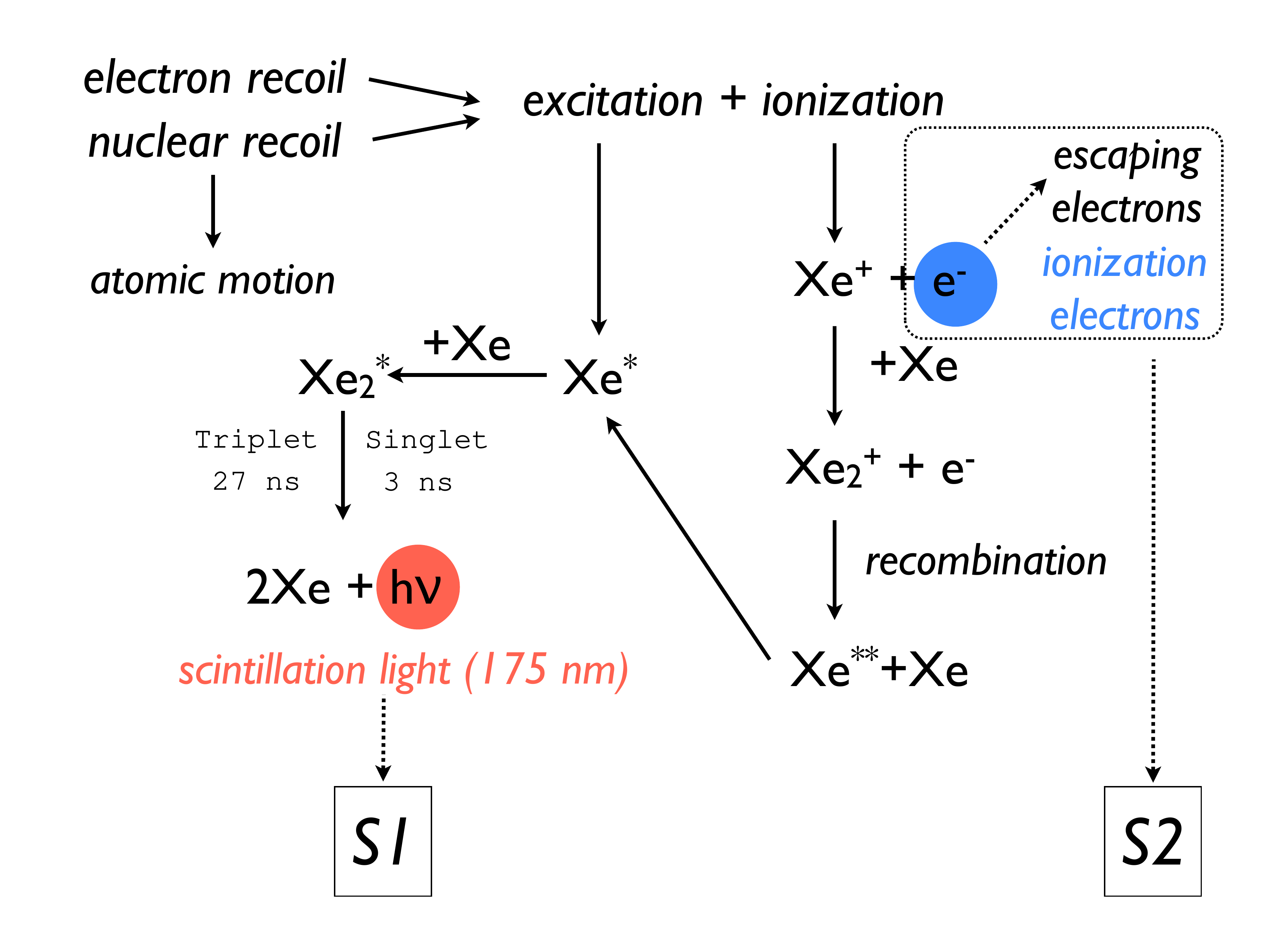,width=3.5in}
\end{center}
\caption{Diagram of the processes leading to primary scintillation (``S1'') light in a liquid noble detector (here Xe), and (if the detector is dual-phase) to secondary (``S2'') light proportional to the amount of ionization.  Recoils dissipate energy as atomic motion, excitation, and ionization.  Both excitation and ionization result in excited dimers, Xe$_2^*$, in either a longer-lived triplet state or a shorter-lived singlet.  Numerical values are for Xe; see Table~\ref{tab:nobles} for properties of Ar and Ne.
Figure based on Ref.~\refcite{manzurXeLeff}. 
}
\label{fig:nobles}
\end{figure}

\begin{table}
\tbl{Properties of noble liquid detectors~\cite{AprileBaudisPDM,NikkelNeon2006}.}
{\begin{tabular}{lrrr}\toprule
Property (unit) &  Xe~     &  Ar~ &  Ne~ \\
\colrule
Atomic Number & 54 & 18 & 10 \\
Mean relative atomic mass & 131.3 & 40.0 & 20.2 \\
Boiling Point $T_{\mathrm{b}}$ (K)& 165.0 & 87.3 & 27.1 \\
Melting Point $T_{\mathrm{m}}$ (K) & 161.4 & 83.8 & 24.6 \\
Liquid density at $T_{\mathrm{b}}$ (g cm$^{-3}$) & 2.94 & 1.40 & 1.21 \\
Volume fraction in Earth's atmosphere (ppm) & 0.09 & 9340 & 18.2 \\
Cost/kg$^{\text a}$ & \$1000 & \$2 & \$90 \\
Scintillation light wavelength (nm) & 175 & 128 & 78 \\
Triplet lifetime (ns) & 27 & 1600 & 15000 \\
Singlet lifetime (ns) & 3 & 7 & $<$18 \\
Electron mobility (cm$^2$ V$^{-1}$ s$^{-1}$) & 2200 & 400 & low \\
Scintillation yield (photons/keV) & 42 & 40 & 30 \\
\botrule
\end{tabular}
}
{\begin{tabnote}
$^{\text a}$ price subject to change; the author does not guarantee any price listed.
\end{tabnote}} 
\label{tab:nobles}
\end{table}

Detectors using noble liquids (and/or gases) also show great promise for WIMP detection
 and have the advantage of relatively easy scaling to large masses. 
Figure~\ref{fig:nobles} shows the basic physics behind these detectors.
A recoil in liquid Xe, for example,  induces both ionization and excitation of Xe atoms (in addition to wasting some energy increasing atomic motion). Both excitation and ionization lead to production of either a singlet or triplet state of an excited dimer (Xe$_2^*$). De-excitation of either state produces emission of a 175-nm photon that is not absorbed by the noble liquid.  The triplet state has a longer lifetime than the singlet state (27\, ns vs. 3\,ns for Xe).  The dense energy depositions from nuclear recoils result in fewer triplet decays and faster recombination, so nuclear recoils have a faster pulse shape than electron recoils. As shown in Table~\ref{tab:nobles}, the effect is particularly pronounced in Ar and Ne, leading to extremely good discrimination in Ar and Ne based on timing alone.  
Additional discrimination is possible in Xe or Ar based on the relative amount of primary scintillation versus ionization (low electron mobility makes measuring ionization in Ne impractical).   For a given energy deposition, nuclear recoils produce less ionization than electron recoils, and much of the ionization of nuclear recoils is ``quenched'' through recombination.

Experiments with noble liquids must overcome significant challenges.  
Since there are relatively few light quanta (typically 5--10\,pe/keV collected),
maintaining discrimination to low energies is difficult.
Stringent purity levels ($\sim10^{-9}$ impurities) must be achieved in order to prevent absorption of the scintillation light or attachment of drifting electrons. 
The experiments must overcome radioactive backgrounds 
such as $^{85}$Kr in Xe or especially $^{39}$Ar in liquid Ar (which produces 1 decay per second per kg of natural Ar). 

\begin{figure}[t]
\begin{center}
\psfig{file= 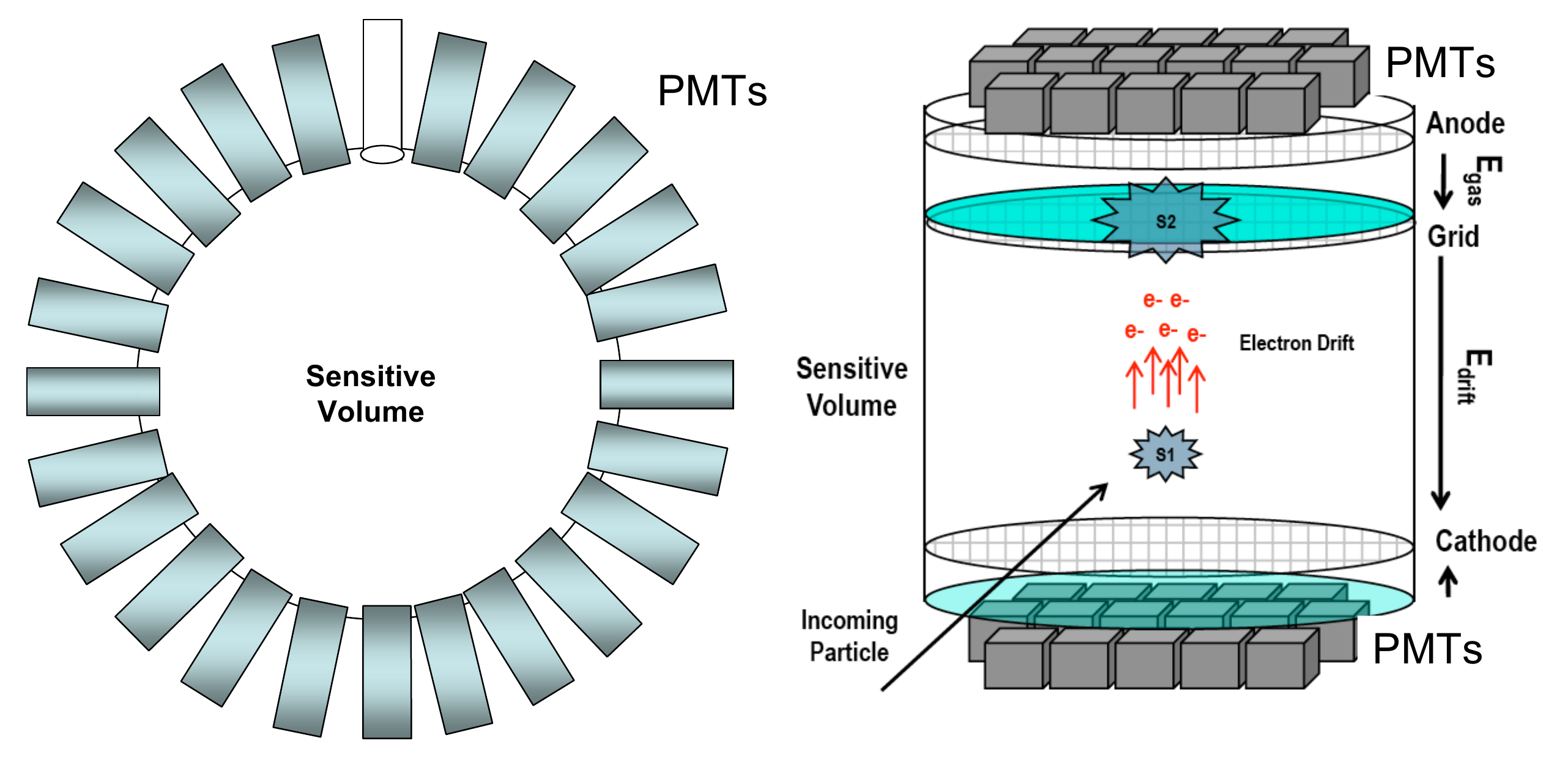,width=4.5in}
\end{center}
\caption{{\it Left}: Sketch of a typical modern single-phase detector geometry.  A spherical volume of noble liquid is surrounded by an array of close-packed photomultiplier tubes (PMTs) to maximize light collection and simplify position reconstruction.  A ``neck'' leads up from the top of the detector to allow insertion of calibration sources. 
{\it Right}: Sketch of a typical dual-phase detector geometry.  Noble gas lies above the liquid with transition from liquid to gas occurring at the grid.  PMTs on top or bottom of the cylinder allow mm reconstruction accuracy in the x-y plane of the detector.  Interaction in the sensitive volume between the cathode and the grid produces primary scintillation (``$S_1$'') light and ionization electrons.  These electrons drift upwards in the strong electric field produced by the anode and cathode, producing secondary (``$S_2$'') light proportional to the amount of ionization, due to electroluminescence in the gas. As depicted, the $S_2$ signal is much larger than the $S_1$ signal.  The drift time of the electrons causes a delay between the $S_1$ and $S_2$ signals that allows mm reconstruction of the position in the vertical axis of the detector.  Interactions occurring in liquid below the cathode produce $S_1$ signals but no $S_2$ signal, since the direction of the electric field drifts electrons from these interactions to the bottom of the detector.
}
\label{fig:singledualnoble}
\end{figure}

There are two basic types of noble detectors: single-phase (usually liquid), which detect only the primary (``$S_1$'') light signal, and dual-phase time-projection chambers, which employ a large electric field to drift ionization electrons upwards out of the liquid and into a region where the noble is in its gas phase (see Fig.~\ref{fig:singledualnoble}).  There, the electrons produce a large secondary (``$S_2$'') light signal by electroluminescence that is proportional to the amount of ionization.  The ratio of this secondary light to the primary scintillation ($S_2/S_1$) provides additional discrimination with typical background leakage $10^{-2}$--$10^{-3}$ at 50\% acceptance.
With Xe, dual-phase detectors are the only way to get appreciable discrimination against electron-recoil backgrounds since timing discrimination is weak.
However, the primary advantage of the dual-phase set-up is that drifting the electrons yields mm-accurate position information on the interaction (compared to cm-accurate for single-phase detectors), providing much better rejection of events due to contaminants external to the detector or on its walls.  
There are costs to these advantages.
Drifting electrons is very slow ($\sim 50$\,$\mu$s) compared to collecting scintillation light. 
Single-phase detectors may stand a higher event rate, important for calibration and increasingly important as detectors become larger.  
Since they don't need an electric field, single-phase experiments can achieve better light collection, allowing lower energy thresholds and/or better pulse-shape discrimination.  Their simple design allows easiest scaling to very large masses.
In contrast, noble liquid experiments that measure ionization are inherently not as scalable since large detectors would require extremely high voltages and high purity for efficient collection of electrons;
Ar detectors that measure ionization would be limited 
by the high event pile-up due to the relatively slow electron drift speed.  

\begin{figure}[t]
\begin{center}
\psfig{file= 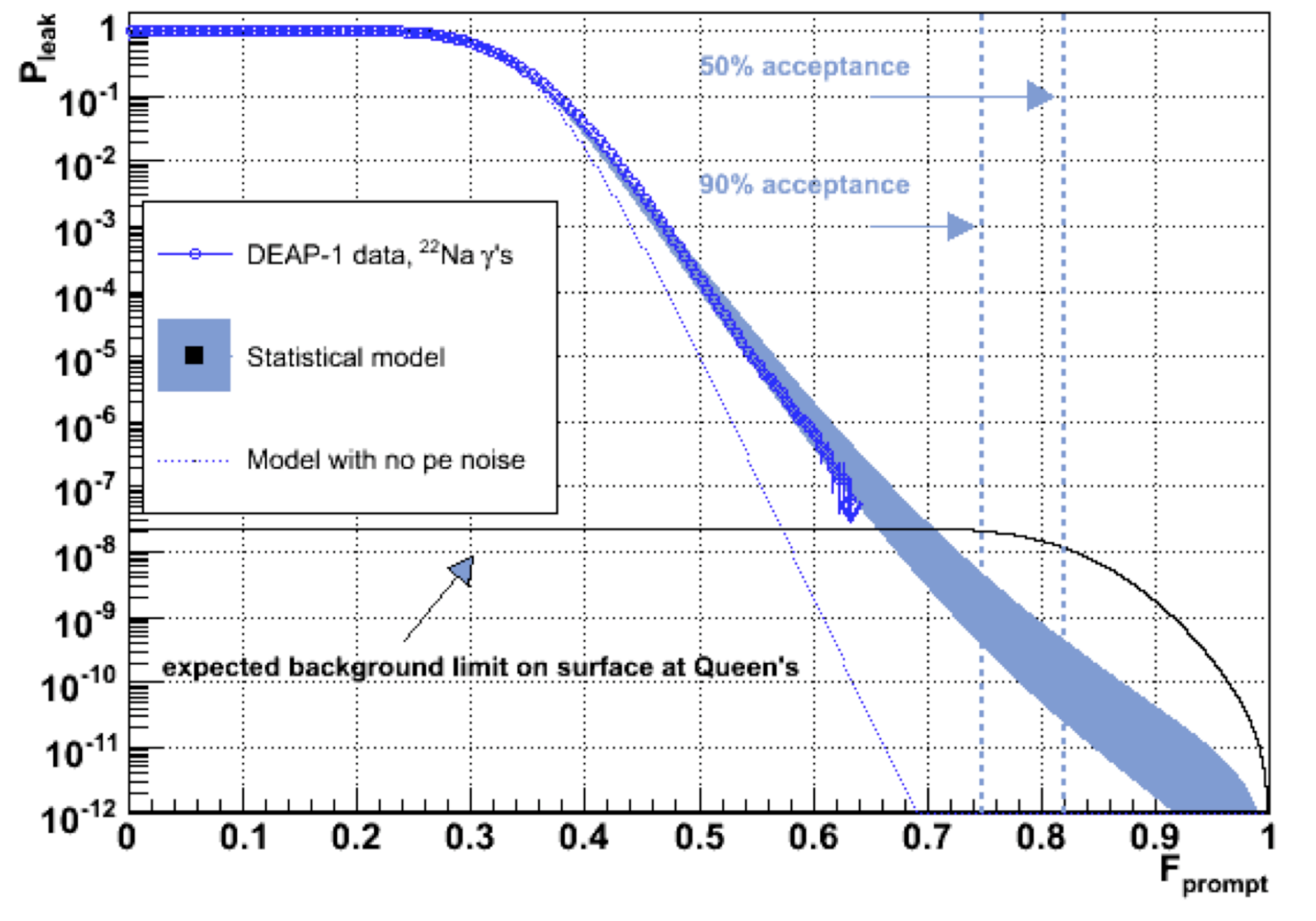,width=3.5in}
	\caption{Results of calibration of the DEAP-1 detector with a $^{22}$Na gamma-ray source at Queen's University, demonstrating $>10^7$:1 rejection of electron recoils based on their low fraction of prompt light $F_{\mathrm{prompt}}$.  Nuclear recoils have median $F_{\mathrm{prompt}} = 0.82$ (dashed vertical line labeled ``50\% acceptance'').  
Plotted is the probability $P_{\mathrm{leak}}$ that an electron recoil would have a fraction of prompt light  greater than $F_{\mathrm{prompt}}$. 
None of the $1.53\times10^{7}$ electron-recoil events (circles with error bars indicating uncertainties due to counting statistics) have $F_{\mathrm{prompt}}>0.64$, safely below the value needed to accept 90\% of WIMPs (dashed vertical line labeled ``90\% acceptance'').  Data show agreement with predictions from simulations when a statistical model of photoelectron noise is included (shaded swath); the model without photoelectron noise is also shown (dotted curve).  The full model predicts $\sim10^9$:1 rejection of electron recoils at 90\% acceptance of WIMPs.
Calibration  is underway deep underground at SNOLAB to measure rejection beyond that demonstrable at Queen's due to expected neutron backgrounds (solid curve).  Figure courtesy of C.~Jillings.
}
\label{fig:DEAP1PSD}
\end{center}	
\end{figure}

Three single-phase experiments are under construction and should take data in 2010 or 2011.  
XMASS~\cite{xmass2009}, which uses Xe,
will take advantage of self-shielding to create a low-background 100-kg fiducial volume within
800\,kg of instrumented Xe. 
The collaboration appears to be able to achieve sufficient reduction of backgrounds in the bulk Xe, most notably reduction of $^{85}$Kr to 3\,ppt by distillation~\cite{xmassKr}.
Experiments of the DEAP/CLEAN collaboration take advantage of the large timing difference between nuclear recoils and electron recoils in argon or neon. 
As shown in Figure~\ref{fig:DEAP1PSD}, the
 DEAP-1 experiment has already demonstrated better than $10^{7}$:1 
rejection of backgrounds~\cite{DEAP1PSD},  
and an analytic model predicts better than billion-to-one rejection with 90\% efficiency for WIMPs.
Construction of the 100-kg fiducial-mass MiniClean detector~\cite{McKinsey2007152} and the 1000-kg fiducial-mass DEAP-3600 detector~\cite{deap2006}
are both in progress and should start taking data in 2011--2013.

\begin{figure}[t]
\begin{center}
\psfig{file= 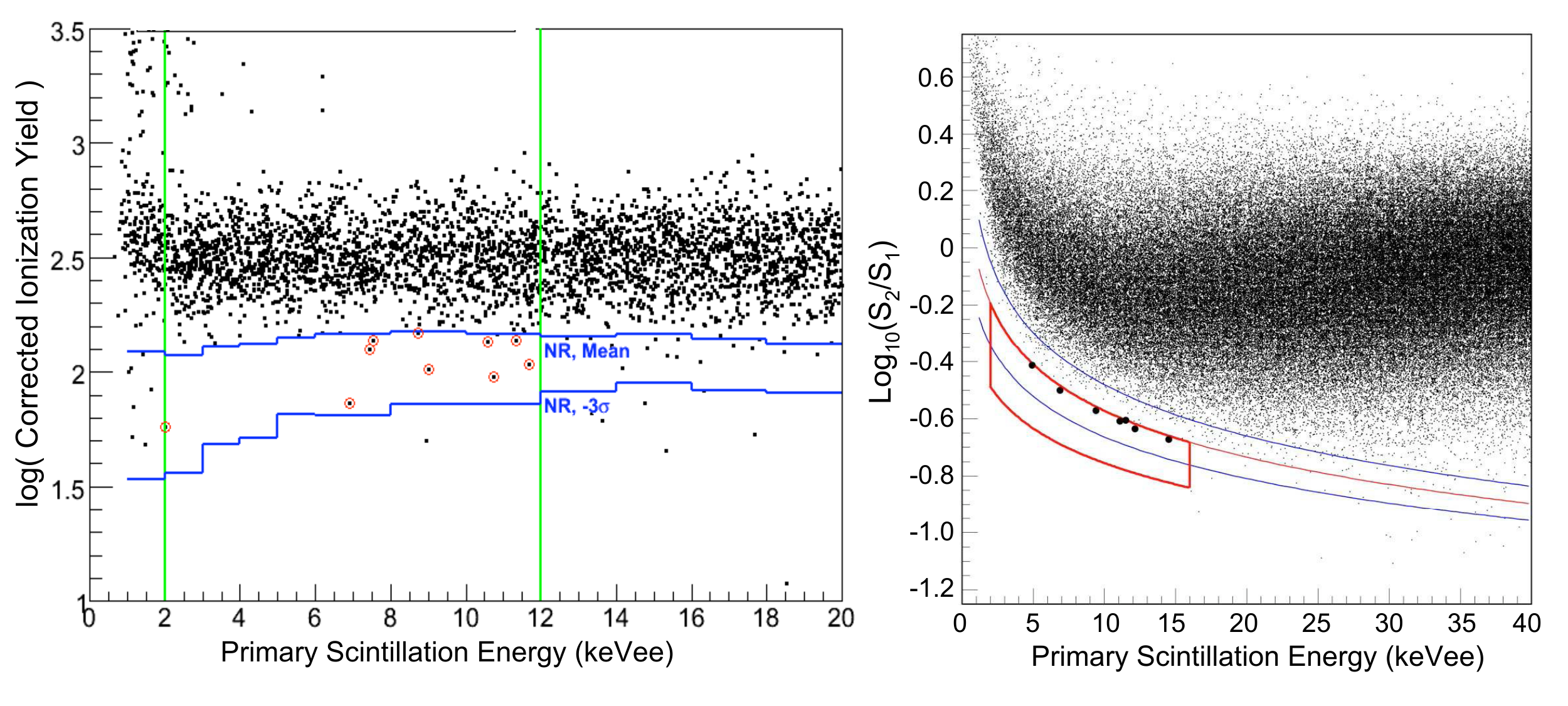,width=4.5in}
	\caption{Plots of events observed with dual-phase Xe detectors. {\it Left}: Corrected ionization yield (which is $\log(S_2/S_1)$ rescaled so that photons lie at $\log(S_2/S_1)_{\mathrm{corr}} = 2.5$) and primary scintillation $S_1$ energy for XENON-10 events passing blind data-selection cuts.  10 WIMP candidate events (circled) lie between 2--12\,keVee with ionization yield between the median and $-3\sigma$ values for nuclear recoils (horizontal lines as function of $S_1$ bin), consistent with backgrounds from electron-recoils including multiple-scatter events.
	 Plot from 
	 Ref.~\refcite{xenon2007}.
{\it Right}: Ionization yield $\log(S_2/S_1)$  and primary scintillation $S_1$ energy for ZEPLIN-III events passing data-selection cuts.  Curves depicting the median (central, red curve) and $\pm1\sigma$ (outer, blue curves) ionization yield for nuclear recoils are shown.  7 WIMP candidate events (large filled circles) lie between 2--16\,keVee with ionization yield between the median and $-2\sigma$ values expected for nuclear recoils.  All are near the top edge of the nuclear-recoil candidate region, consistent with expectations from background.
Plot from Ref.~\refcite{zeplin2009}.
}
\label{fig:XeTPC}
\end{center}	
\end{figure}

As shown in 
Figure~\ref{fig:SI}, the {\small XENON-10} experiment~\cite{xenon10long} 
has the most sensitive current results among noble-liquid experiments~\cite{xenon2007}, including the most stringent spin-independent limits of all experiments for WIMP masses below 40\,\gev, and, as shown in Fig.~\ref{fig:SD}, the most stringent neutron-spin-dependent limits of all experiments~\cite{xenon2008SD}. 
These results are from a ``blind'' analysis, in which data-selection cuts including energy range were all set before looking at the parameters of candidate WIMP events in the data.
Such blind analyses are critical for dark matter searches when data-selection cuts are made in a large dimensional space, otherwise potentially allowing bias to lead to fine-tuned cuts that preferentially select or omit the candidate events with little effect on the reported expected background leakage or WIMP efficiency.
This analysis yields about 50\% efficiency for nuclear recoils for 58.6\,days exposure of a detector of 5.4\,kg fiducial mass. The expected background based on the expected Gaussian tails of the electron-recoil $\log(S_2/S_1)$ distribution was $7.0^{+2.1}_{-1.0}$ events.  An additional, unestimated background resulted from multiple-scatter events where one scatter occurs in the sensitive volume and the other in the liquid below the cathode (see Fig.~\ref{fig:singledualnoble}).  Since the latter interaction produces $S_1$ light but no $S_2$ light, these events have reduced  $\log(S_2/S_1)$ values. In the blindly chosen energy region, 2--12\,keVee (corresponding to 4.5--27\,keV$_{\mathrm{R}}$ for the assumed nuclear quenching factor, but to a higher energy range for more recent measurements~\cite{AprileLeff2009,manzurXeLeff}), 10 presumed background events passed cuts (see Fig.~\ref{fig:XeTPC}), with studies after unblinding suggesting that most were multiple-scatter events with one interaction below the cathode.

A similar experiment, ZEPLIN-III~\cite{zeplin3long}, obtained very similar results~\cite{zeplin2009,zeplin2009SD} with a different geometry.  While XENON-10 was relatively long and thin, with a 20\,cm diameter
 and a 15\,cm maximum drift length resulting in a drift field of 0.7\,kV\,cm$^{-1}$, ZEPLIN-III is short and squat, with a diameter of 38.6\,cm and a maximum drift length of 3.5\,cm, allowing a larger drift field of 3.9\,kV\,cm$^{-1}$ and better resulting $S_2/S_1$ discrimination~\cite{zeplin3long,zeplin2009}.
 Their analysis was not blind, instead including strong efforts to remove the multiple-scattering events that limited XENON-10.  A difference in phototube performance between low-rate WIMP-search running and high-rate calibrations made estimation of expected backgrounds difficult, but the 7 observed events all appeared near the edge of the $S_2/S_1$ cut position and so appeared consistent with background, as shown in Fig.~\ref{fig:XeTPC}.
Limits for XENON-10 are significantly better at low WIMP masses than those for ZEPLIN-III because XENON-10 observed no events between 2--7 keVee, and ZEPLIN-III assumed a much more conservative scintillation efficiency for nuclear recoils in Xe. 
  
Two follow-up experiments (50-kg fiducial-mass XENON-100~\cite{xenon100IDM2008} and 100-kg LUX~\cite{lux2010}) 
should have science results in 2010 or 2011, although construction of LUX has been greatly delayed due to the need to drain the flooded Sanford (future DUSEL) site.
If 
backgrounds are well below the rate seen in XENON-10, 
sensitivity $\sim$$10^{-9}$\,pb may be achieved.
Prototypes for dual-phase Ar detectors WArP~\cite{warp2007} and ArDM~\cite{ArDM2006} have also been constructed.
Use of argon with low $^{39}$Ar content~\cite{GalbiatiUGargon} would allow operation of massive detectors without event pile-up.
Running the 3.2\,kg fiducial WArP prototype 
resulted in no candidate events above 42\,keV in a 96.5 kg day exposure~\cite{warp2007}.  
The WArP collaboration is building a 140-kg experiment that could have results in 2010 or 2011.  
The ArDM collaboration is constructing a 1-ton detector at CERN and will consider a deep underground operation after its successful commissioning~\cite{ArDM2010}.
Relatively poor light collection appears likely to limit the detector to a $30$\,keV energy threshold without reach beyond that of current experiments.
Plans exist for rapid increases in mass to ton or multi-ton liquid noble detectors~\cite{xenonDM2010,maxDM2010,LZDM2010,darwinDM2010}.

\subsection{Cryogenic Detectors}
\label{section:cryo}

Ultrapure Ge semiconductor detectors operated at the temperature of liquid nitrogen, 77\,K, were used for the first searches for dark matter particles in the 1980s~\cite{Ahlen1987GeDM,Caldwell1988GeDM}.  Because these  detectors measured only the ionization of energy depositions, they had no discrimination between electron-recoil backgrounds and nuclear recoils. Although cryogenic detectors with strong background discrimination are now used more widely, ionization-only detectors with improved designs are operated today because of the very low  energy thresholds they may achieve.  
The TEXONO collaboration has achieved an energy threshold of 220\,eVee in four 5\,g detectors~\cite{texono2007}, while the CoGeNT collaboration achieved nearly as low a threshold for a run of 8.4\,kg\,days exposure~\cite{cogent2008}.

The key advantage of most cryogenic detectors is the ability to measure the energy deposited as phonons, vibrations of the crystal lattice.  The initial phonons produced are not at equilibrium. Some detectors collect these ``athermal'' phonons, which contain information on the location and type of recoil that occurred.  On timescales of ms,  essentially all other forms of energy depositions are converted to heat and the phonons thermalize, resulting in a temperature increase of the detector $\Delta T = E/C \sim 1$\,$\mu$K, where $E$ is the energy deposited and $C$ is the detector's heat capacity.  Detectors insensitive to athermal phonons measure this temperature increase with sensitive thermistors, such as neutron-transmutation-doped Ge or transition-edge sensors (see \eg\ 
Ref.~\refcite{GerbierGasconPDM}). 

The EDELWEISS experiment uses thermistors attached to Ge crystals  at cryogenic temperatures 
(20\,mK) to measure phonons, in addition to measuring ionization using a small applied electric field.  
Backgrounds have been dominated by low-energy electrons interacting near the surfaces of the detectors~\cite{edelweiss2005,edelweiss2008} because such interactions result in incomplete collection of the ionization charges, mimicking nuclear recoils. 
The problem occurs because the electrons and holes generated by an interaction
are sufficiently energetic to diffuse
against the applied electric field into the nearby electrode, 
causing a fraction of the event ionization to be ``lost'' if the interaction is near the surface~\cite{tsltd9}.

\begin{figure}[t]
\begin{center}
\psfig{file=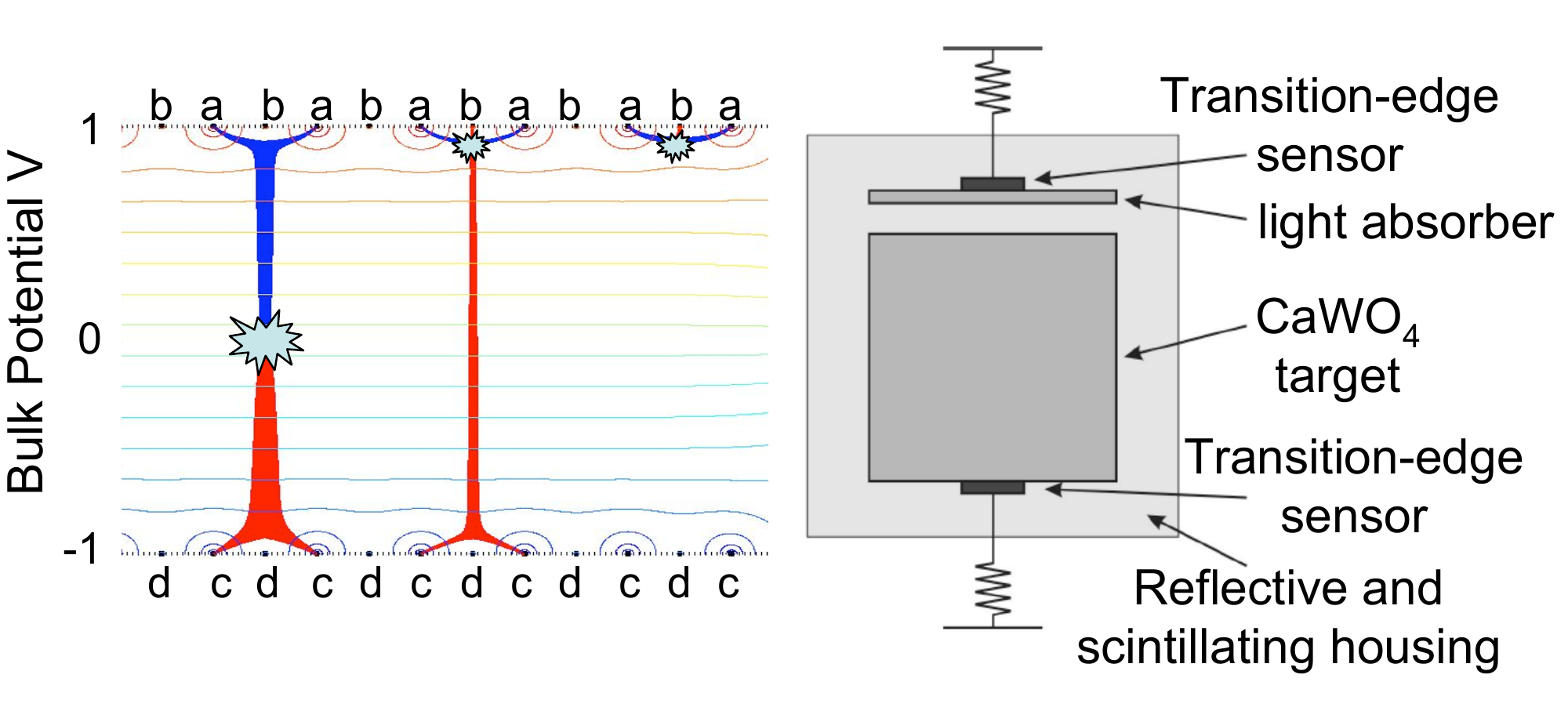,width=4.5in}
\end{center}
 \caption{
 Sketches of the EDELWEISS interdigitated and CRESST cryogenic detectors.
{\it Left}: Side view of a cylindrical interdigitated detector.  Both top and bottom sides have two sets of electrodes, with each commonly labeled electrode (a,b, c, or d) connected in parallel.
The voltage biases chosen  (\eg\ $V_{\mathrm{a}}=2$\,V, $V_{\mathrm{b}}=-1$\,V, $V_{\mathrm{c}}=-2$\,V, $V_{\mathrm{d}}=1$\,V) produce an axial field in the bulk of the detector, perpendicular to the depicted equipotential lines (with voltages labeled by the vertical axis),
while the field close to the surface links two adjacent electrodes and is
therefore approximately parallel to the surface.
A bulk event (leftmost in figure) is thus identified by the collection of electrons and holes on
the ``fiducial'' electrodes $a$ and $c$.  An event near enough to the surface for charge collection to be incomplete (middle or rightmost in figure) would result in charge collection in one of the two ``veto'' electrodes $b$ or $d$, allowing its rejection.  
Figure based on 
Refs.~\refcite{edelweissDM2010,edelweissID}. 
 {\it Right}: Sketch of CRESST phonon and scintillation detector.  Phonons produced in the CaWO$_4$ target are detected by the attached transition-edge sensor, which is connected by a thermal impedance to a constant-temperature cryostat, as in other phonon detectors.  
 Light emitted by the CaWO$_4$  target is detected by an adjacent silicon-on-sapphire detector, also with an attached TES  connected by a thermal impedance to the cryostat.  Figure from 
 Ref.~\refcite{cresst2009}.   
  }
 \label{fig:ID}
\end{figure}

The new EDELWEISS ``interdigitated'' detectors~\cite{edelweissID} promise to provide excellent discrimination against this dominant background.  As shown in Fig.~\ref{fig:ID}, each side of the detector is interleaved with oppositely charged electrodes.  Events in the bulk of the detector result in the collection of electrons and holes only on
the ``fiducial'' electrodes, whereas events near the surface result in some charge collection in the veto electrodes.  Overall, calibration indicates rejection of surface events has only $10^{-5}$ inefficiency, sufficient for 80\,kg\,years exposure at achieved backgrounds.
Recent running of an array of ten 400\,g detectors
for an effective exposure of 144\,kg\,d resulted in one nuclear-recoil candidate above 20\,keV,
leading to the competitive limits shown in Fig.~\ref{fig:SI}~\cite{edelweiss2009}.  
Continued running through Spring 2010 is planned, at which point additional detectors with larger fiducial masses will be commissioned and run.

The CRESST experiment also uses cryogenic detectors, but measures 
phonons and scintillation light, 
which provides as good rejection of surface events as it does rejection of photons in the bulk
since light production is not reduced for events near the crystal surface.
The main CRESST detectors are scintillating (300\,g) CaWO$_4$ crystals, with the W providing the heaviest nucleus of any dark matter search.
As shown in Fig.~\ref{fig:ID}, a transition-edge sensor deposited on the detector surface measures the phonon energy.  Adjacent to each crystal is a small (2\,g) silicon-on-sapphire light detector equipped with another transition-edge sensor.  The energy resolution of the phonon measurement is excellent, better than 1\,keV over the full energy spectrum~\cite{cresstbackgrounds}.  The energy resolution of the light measurement is worse, $\sim$10\,keVee, due to the small fraction of energy converted to light.  Nuclear recoils produce far less light than electron recoils, especially for interactions with the heavy W nuclei ($QF\approx40$).  This difference provides excellent discrimination between electron recoils and nuclear recoils, but also means that 
 only high-energy WIMP events would produce any detectable light at all, resulting either in a high energy threshold, or in potential susceptibility that something causing phonons but no light 
 (\eg\ crystal relaxation) may mimic a WIMP signal. 

Following major upgrades to the experiment's shielding, detector support, and electronics, the CRESST collaboration ran  
detectors in 2007 in a commissioning phase for CRESST-II, yielding a total exposure of 30.6 kg-days of W exposure from two detector modules~\cite{cresst2009}.
Three events consistent with W recoils were found in the energy region 10--40\,keV, with the high end of the energy range determined by the cut-off in the WIMP spectrum due to the form factor for W (see Fig.~\ref{fig:SIform}).  As shown in Fig.~\ref{fig:SI}, resulting limits 
for this small exposure are only about 20$\times$ less constraining than the world's best.  The cause of these events, and of additional events with energies $>40$\,keV, is unknown and under investigation.

For the longer timescale, the EDELWEISS and CRESST collaborations and others have formed a new collaboration, EURECA~\cite{eureca2006}, dedicated to a cryogenic experiment at or near a ton of detector mass. 
Plans are for a 100-kg experiment in 2012 and a ton scale experiment in 2018.

\begin{figure}[t]
\begin{center}
\psfig{file=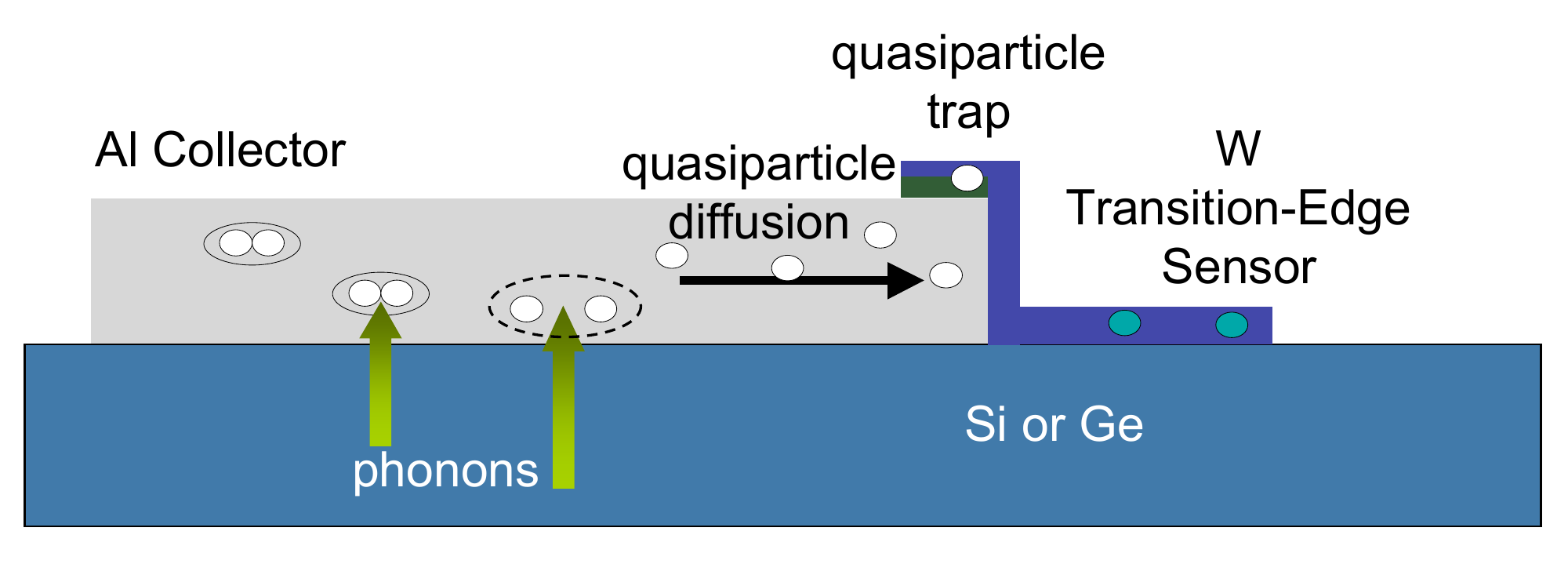,width=4.5in}
\end{center}
 \caption{
 Sketch of a 10-$\mu$m-long CDMS athermal phonon sensor on the surface of a much larger Ge or Si target.
 Phonons produced in the Ge or Si target break Cooper pairs in the superconducting Al, producing quasiparticles, which diffuse into the overlap region, becoming trapped in the W transition-edge sensor.  }
 \label{fig:zip}
\end{figure}

The CDMS~II experiment uses cryogenic Ge or Si ionization-and-phonon detectors  that discriminate 
against the otherwise dominant surface electron recoils by collecting 
phonons before they thermalize~\cite{r118prd}.
Each quadrant of the detector's top surface includes thousands of phonon sensors connected in parallel.
As shown in Fig.~\ref{fig:zip},
athermal phonons produced by the interaction 
propagate to the detector
surface, where most of them 
are absorbed in 
superconducting aluminum pads. 
Quasiparticles generated in the aluminum
by the phonons breaking Cooper pairs
diffuse in $\sim$10~\micros\ through the aluminum 
to a tungsten transition-edge sensor.
The aluminum fins allow a large phonon collection area while keeping the transition-edge sensors small, so that even a small amount of energy produces a large temperature change. Each tungsten sensor is kept in the middle of its sharp superconducting-to-normal transition, so a small increase in temperature greatly increases its resistance.  As a result these detectors have excellent energy resolution. 
Comparison of phonon-pulse arrival times in the four independent channels 
allows localization of the interaction position in the xy-plane of the detector. 

The ionization yield allows near-perfect separation of nuclear recoils from bulk electron recoils, and 
the shape, timing, and energy partition of the phonon pulses allow rejection of events occurring near the detector
surfaces.
This rejection works because the athermal phonons from electron recoils are faster than those from nuclear recoils, particularly if the electron recoils occur near a detector surface.  
Accepting only events with 
both 
slow phonon pulses and low ionization yield rejects over 99.5\% of the surface events 
while keeping over half of the nuclear-recoil events in an analysis tuned to  
maximize the discovery potential of the search.

\begin{figure}[t]
\begin{center}
\psfig{file=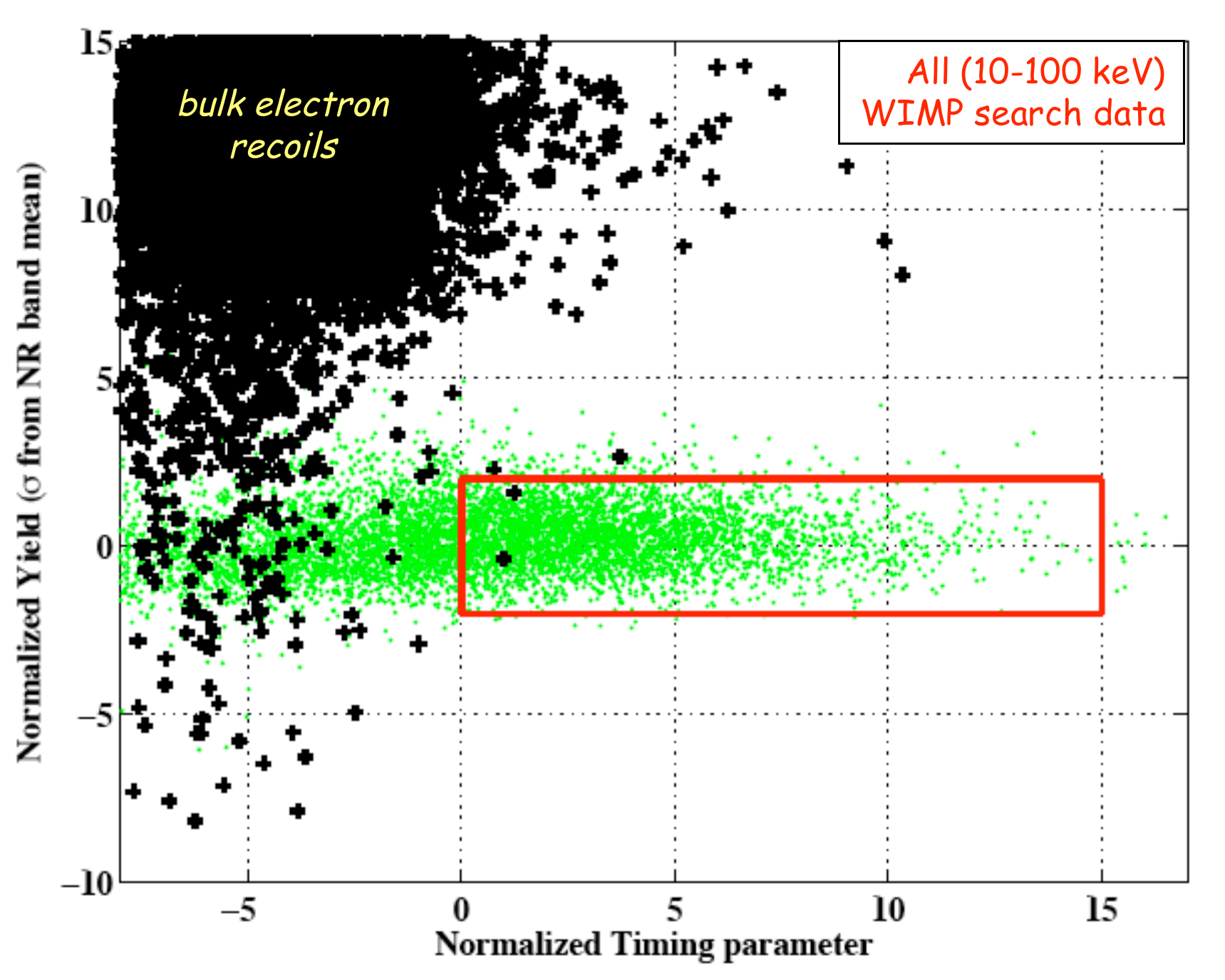,width=4.5in}
\end{center}
 \caption{
Normalized ionization yield (number of standard deviations from the mean nuclear-recoil yield)
versus normalized timing parameter (timing relative to acceptance region) for WIMP-search events ($+$) and neutron calibration events (grey dots) from the final exposure of the CDMS~II experiment.
Only events between 10--100\,keV, consistent with all signal criteria (excluding yield and timing) are shown.Ê 
Two WIMP candidate events lie within the solid box indicating the signal region.
Figure courtesy Zeesh Ahmed.
 }
 \label{fig:cdmsYvsT}
\end{figure}

Data taken with $\sim$4\,kg of Ge detectors in 2007--2008 resulted in a total exposure to WIMPs of 612\,kg-days.
An analysis in which data-selection cuts were set blind, based on events from calibration sources or other events that could not be from WIMPs, resulted in an expected background of $0.8 \pm 0.1 $(stat)$ \pm 0.2$(syst) misclassified surface electron recoils  and $\sim0.1$ events from neutrons.
Two WIMP candidates,
at recoil energies of 12.3\,keV and 15.5\,keV, were observed (see Fig.~\ref{fig:cdmsYvsT}).
Because the probability to have observed two or more 
background events in this exposure is 23\%, these results 
are not significant evidence for WIMP interactions~\cite{CDMSscience}.
Combined with the results of previous exposures, resulting spin-independent limits are the world's most constraining for WIMP masses $\gapprox40$\,\gev, as shown in Fig.~\ref{fig:SI}.  

A follow-up experiment, SuperCDMS~\cite{supercdmsLTD12,supercdmsBruch2010}, is being commissioned using the same infrastructure but 
improved, thicker (inch-thick instead of cm-thick) detectors with reduced exposure to radon daughters, and hence lower surface backgrounds.
The larger detectors result in $\sim4\times$ the total mass in Ge, with sensitivity reach to $4\times10^{-9}$\,pb, at which point the neutron backgrounds of the current setup at Soudan likely would begin to dominate.  
An experiment with 100\,kg at a deeper site, SNOLAB, is planned.
Research and development on the manufacturability of the 1.5\,ton
GEODM project~\cite{geodmDM2010}, is in progress.  It may be possible to make detectors 20$\times$ larger than the CDMS detectors by exploiting inexpensive, dislocation-free Ge crystals, which are unusable when run at liquid nitrogen temperatures but appear to work at 50\,mK. 
Multiplexing would simplify the readout electronics and reduce the heatload on the cryogenic systems.

\subsection{DAMA}
\label{section:dama}

The only claimed detection of WIMPs is from the DAMA experiments~\cite{dama2000, damanuovocimento,dama2004,damalibra,damalibra2008}.  
DAMA does not distinguish between WIMP signal and background on an event-by-event basis.  
Instead, the presence of WIMPs may be inferred  from the annual modulation
in the rate of the lowest-energy single-scatter interactions, assuming that the backgrounds do not modulate significantly.
The DAMA/NaI apparatus~\cite{damanuovocimento}, consisting of 87.3\,kg of NaI scintillator crystals, was run from 1996--2002.
The current, DAMA/LIBRA apparatus~\cite{damalibra} consists of an array of 
25 NaI scintillator crystals, each 9.7\,kg.
One is not operational, resulting in a total target mass of 232.8\,kg.
Each crystal is viewed by 2 phototubes through suprasil-B lightguides, with excellent achieved light collection (5.5--7.5 photoelectrons/keVee).
Extreme efforts have been taken to avoid contamination, including
etching of parts followed by 
installation within a high-purity nitrogen atmosphere using breathing apparatus.
Photon and neutron shielding is fairly standard, although it may be notable that the detectors are separated by a fair mass of copper, and there is no  
surrounding scintillator veto.

\begin{figure}[t]
\begin{center}
\psfig{file=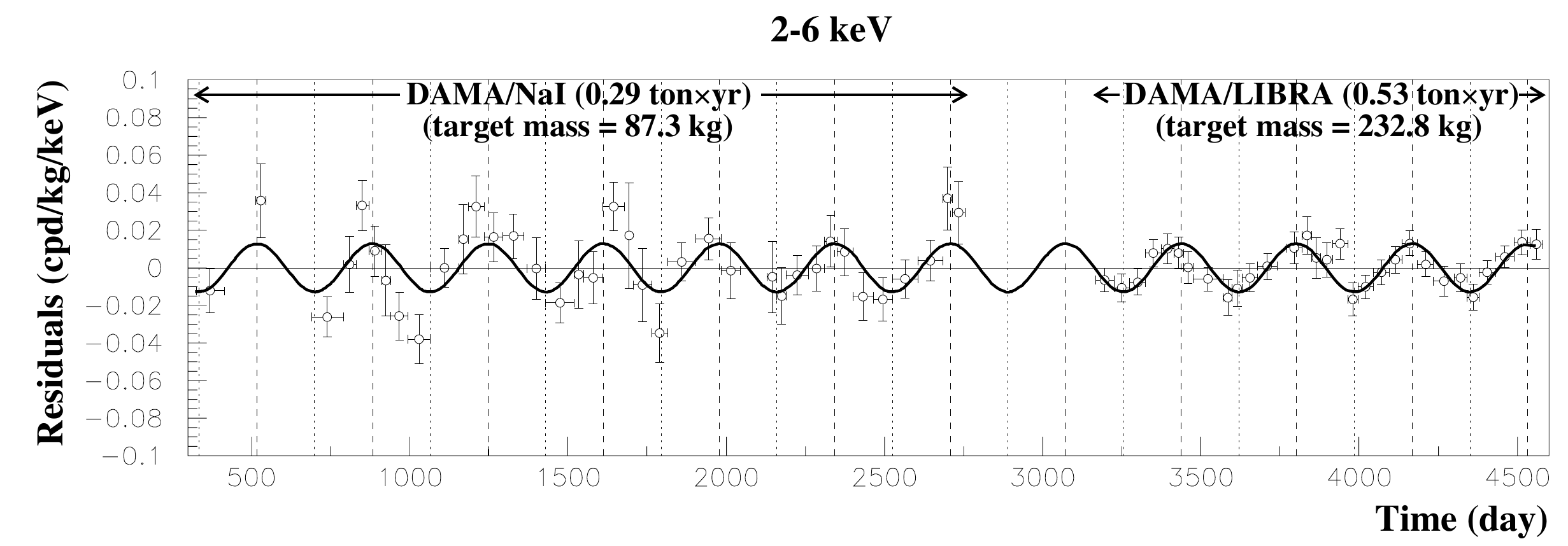,width=4.5in}
\end{center}
 \caption{
Sum of residuals of the single-hit scintillation events in the 2--6\,keVee energy interval, after subtracting time-averaged rates in each energy bin in each detector, as a function of days since January 1, 1996, for the DAMA/NaI and DAMA/LIBRA experiments.
The experimental errors are vertical bars, and the associated time bin widths are horizontal bars. 
The superimposed curve represents the cosinusoidal function $A \cos \omega(t-t_0)$  
with modulation amplitude  $A=(0.0129 \pm 0.0016)$ \perkeekgd
 obtained by best fit over the whole data while constraining the period $T 
 =2\pi/{\omega} 
 =  1$\,yr and phase $t_0 = 152.5$\,day (June 2$^{nd}$). 
The dashed vertical lines correspond to the maximum of the signal (June 2$^{nd}$), while the dotted vertical lines correspond to the minimum.
Figure from 
Ref.~\refcite{damalibra2008}.
 }
 \label{fig:DAMAmodulation}
\end{figure}

\begin{figure}[t]
\begin{center}
\psfig{file=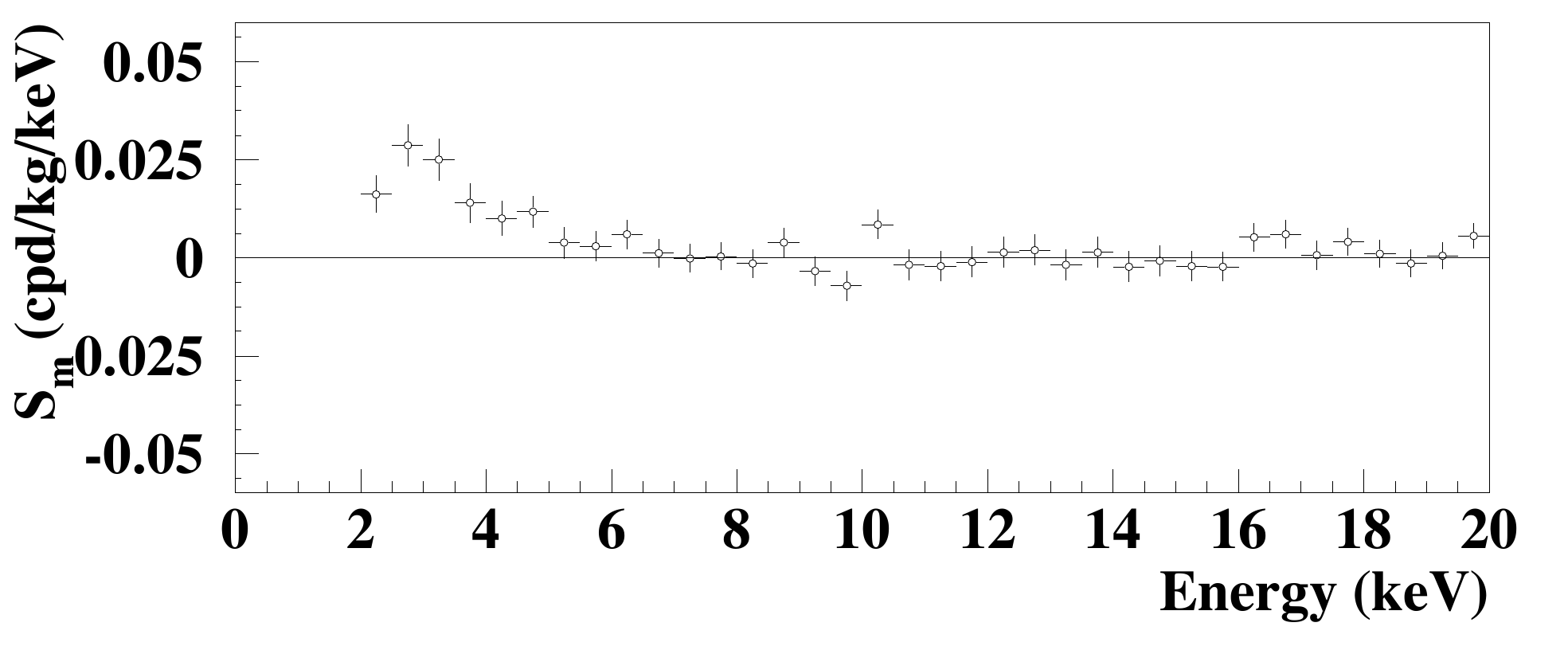,width=3.5in}
\end{center}
 \caption{
Energy distribution of the binned modulation amplitude $S_m$
for the
total exposure 
of DAMA/NaI and DAMA/LIBRA.
A clear modulation is present in the lowest energy region, 
while $S_{m}$ values compatible with zero are present just above. 
Figure from 
Ref.~\refcite{damalibra2008}.
 }
 \label{fig:DAMAenergy}
\end{figure}

As shown in Fig.~\ref{fig:DAMAmodulation}, the annual modulation in DAMA's event rate is compelling.
The combined 0.82\,ton-years of exposure result in a best-fit amplitude $A = 0.0131 \pm 0.0016$\,\perkeekgd, period $T =0.998 \pm 0.003$\,years, and phase $t_0 = 144 \pm 8$\,days (consistent with $t_0=152.5$\,days as expected for a standard halo), with a significance of $8.2\sigma$.  As expected for standard WIMP elastic scattering discussed in Section~\ref{section:signal}, the modulation amplitude (see Fig.~\ref{fig:DAMAenergy}) is significant only at low energies ($\lapprox5$\,keVee).  The shape of the energy spectrum of the modulation is not well constrained at low energies; since the lowest-energy bin has a somewhat lower rate than the next higher bins, a spectrum that is monotonically falling with increasing energy, or one that rises and then dips (such as a monoenergetic line near 3\,keVee) each provides an adequate fit to the data~\cite{damalibra2008}.  Finally, no modulation is seen in the rate of low-energy multiple-scatter events, providing 
evidence that the signal is not a simple modulation of the background.

\begin{figure}
\begin{center}
\psfig{file=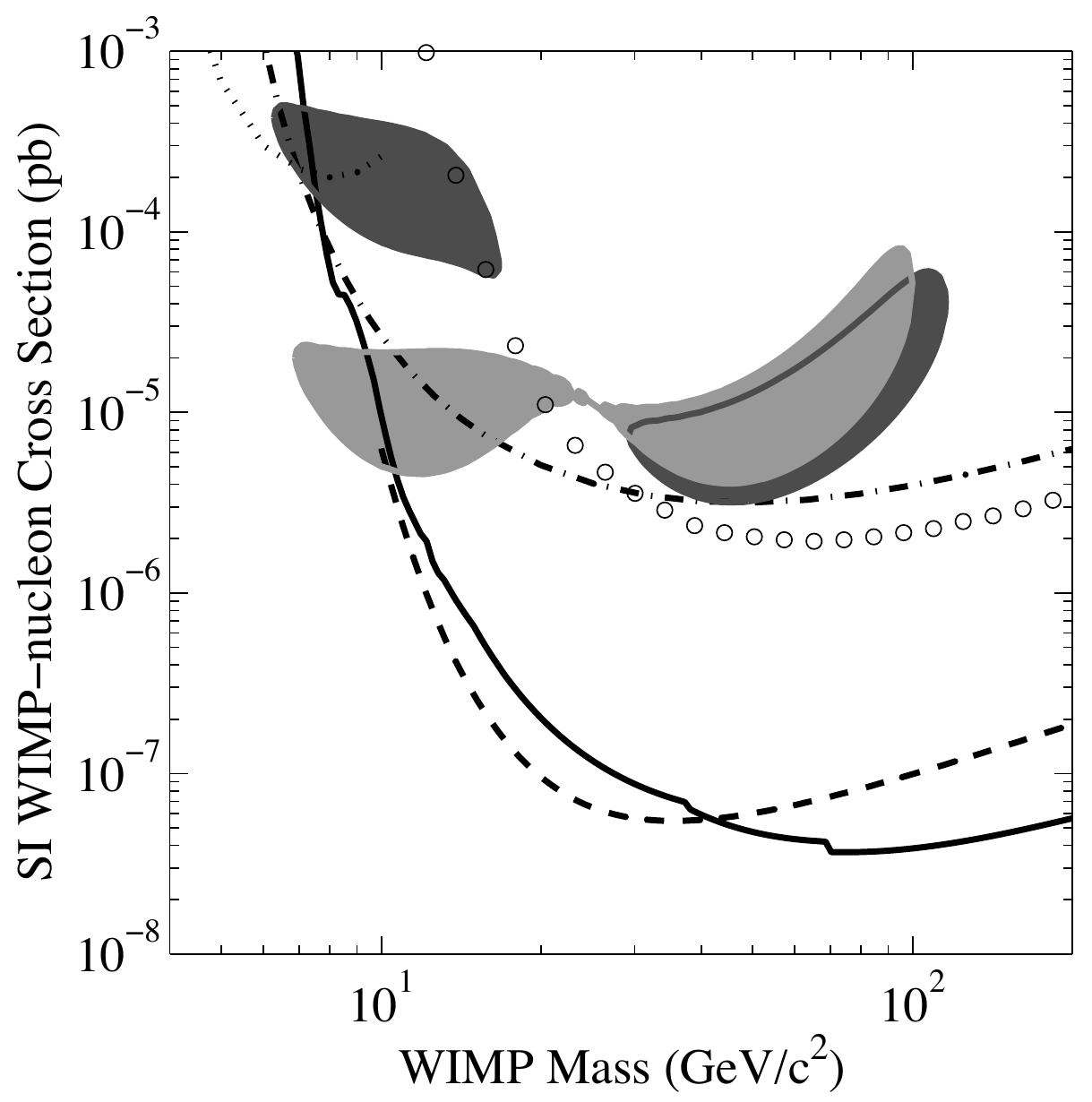,width=2.22in}
\psfig{file=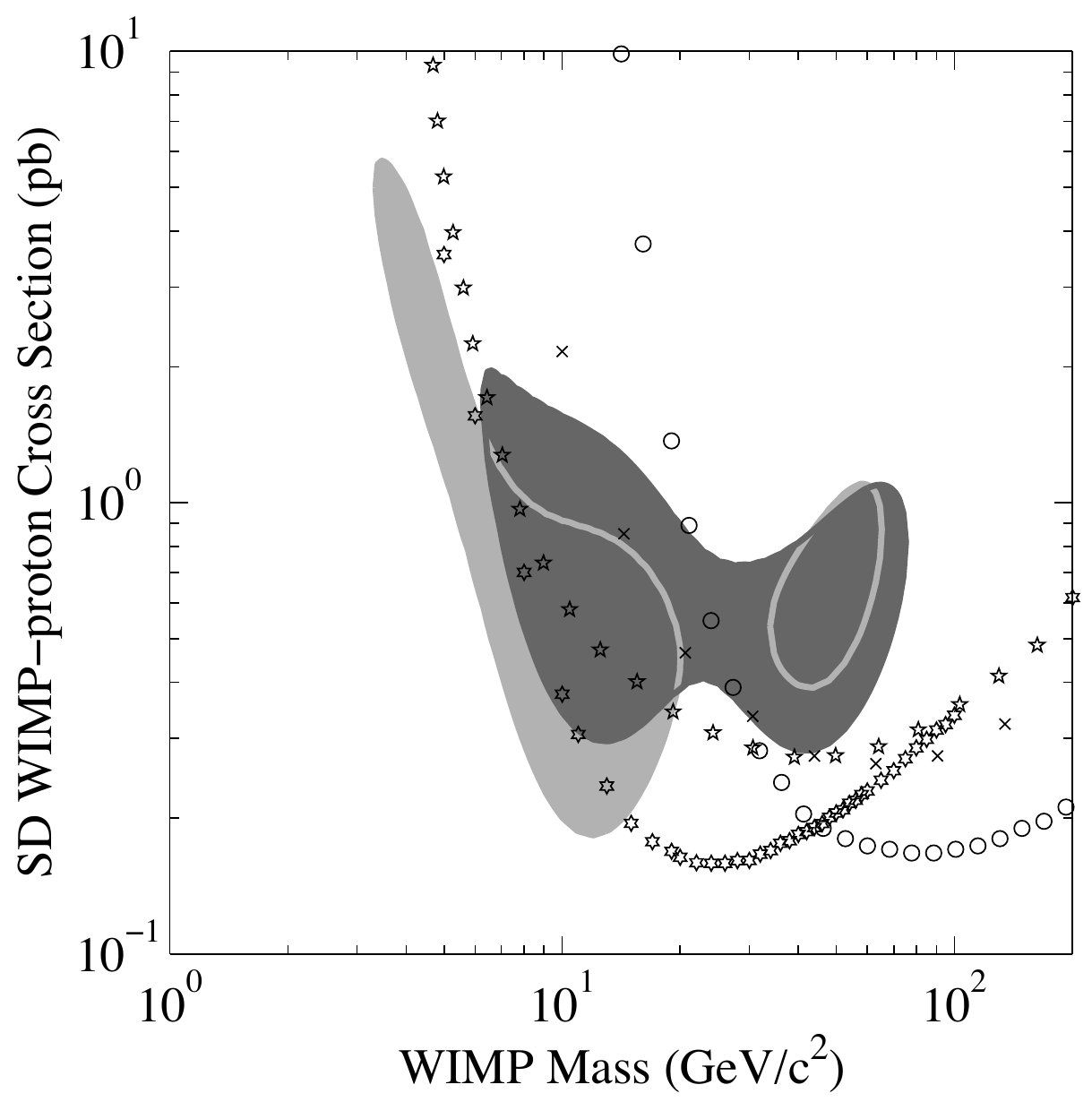,width=2.22in}
\end{center}
  \caption{Comparison of DAMA (3$\sigma$) allowed regions as interpreted in Ref.~\refcite{savageDAMA} to 
  published (90\% C.L.) upper limits  on the spin-independent (left) and proton-spin-dependent (right) WIMP-nucleon cross section 
as functions of WIMP mass under the standard assumptions~\cite{lewinsmith}
  about the WIMP distribution in the Galaxy described in Section~\ref{section:signal}.
For DAMA, interactions of WIMPs with mass $\gapprox25$\,\gev are dominated by interactions on  I, 
while those $\lapprox25$\,\gev are dominated by interactions on Na.
  Upper limits of CDMS~\cite{CDMSscience} (solid), XENON-10~\cite{xenon2007} (dashes), and KIMS~\cite{kims2007} (circles) are clearly incompatible with spin-independent interactions on I, while limits from PICASSO~\cite{picasso2009} (6-pointed stars), COUPP~\cite{coupp2008SDlimits} (5-pointed stars),  and KIMS (circles) are clearly incompatible with spin-dependent interactions on I.  
  Parts of the regions including the effects of ion channeling as presented in 
  Ref.~\refcite{DAMAchanneling} (light shade) are allowed at low masses.  
Almost the entire region assuming insignificant channeling (dark shade) for both spin-independent and spin-dependent interactions, even at low masses,
is excluded by combinations of limits of 
CDMS on Si~\cite{r119prl} (dash-dots), CoGeNT~\cite{cogent2008} (dots), and PICASSO. } 
\label{DAMAlimits}
\end{figure}

If interpreted as a standard WIMP interaction on iodine, the results are clearly inconsistent with limits from other experiments, as shown in Fig.~\ref{DAMAlimits} for spin-independent and proton-spin-dependent interactions (neutron-spin-dependent interactions are excluded even more strongly, since Na and I isotopes are odd-p). 
The results are in conflict with other experiments if they  are due to standard low-mass WIMP scattering on sodium instead of iodine, whether the dominant interaction is spin-independent or spin-dependent, if the standard halo model is correct.  Since alternate halo models may produce a larger modulation for the same WIMP rate~\cite{copikrauss},
a non-standard halo may improve consistency between DAMA and other experiments.  

Figure~\ref{DAMAlimits} also shows DAMA allowed regions if ion channeling is significant in NaI, as presented in   
Ref.~\refcite{DAMAchanneling}.
Ion channeling is an effect observed experimentally for nuclei sent into lattice from outside.
If the nucleus travels down the channel between lattice sites, it transfers more of its energy to electrons rather than to other nuclei, producing much more light than usual.
If WIMP-induced recoiling nuclei are channeled, the 
sensitivity of DAMA would be increased especially for low-mass WIMPS since the channeling would allow a low-energy recoil to  produce enough light to be detectable. 
However, channeling has not been observed, nor is it expected, for nuclei that start on a lattice site, as would be the case for nuclei recoiling from a WIMP interaction.
In a perfect lattice, no nucleus would be channeled by the rule of reversibility.
Accounting for thermal vibrations, Bozorgnia, Gelmini, and Gondolo find channeling's effect on the DAMA energy spectrum is $<1$\%~\cite{DAMAnochanneling}.

Many non-standard interactions have been proposed (\eg~\cite{DAMAiDM2009,DAMAwimpless,DAMAmirror,DAMAscalarAndreas2008}) 
that may explain the DAMA signal.  Since these are described  in  Neil Weiner's contribution to these proceedings~\cite{weinerTASI2009}, they will not be discussed here.
Instead, it is worthwhile considering in detail whether the annual modulation signal could be caused by something other than dark matter, in order to help evaluate how seriously alternate WIMP models should be considered.  Could some background be causing the annual modulation?  Such a background would have to fulfill the annual modulation characteristics of a WIMP.  Its rate would have to vary cosinusoidally over the course of the year, with period $T=1$\,year and phase $t_0\approx152.5$\,days.  The modulation should appear only in the lowest-energy, single-detector hits, and should produce a consistent amplitude, $A\leq7$\%,  between the NaI/LIBRA experiments and between different detectors in the experiments.

Although these requirements are extensive, they are not as strenuous at they might at first seem. 
The date of expected maximum signal, June~2, corresponds roughly (but not exactly) to summer, and there are of course many systematic differences between summer and winter.  
The lowest-energy events are the very ones most likely to be affected by a systematic effect, and the requirement that the multiple-scatter events not show modulation is not a very strong test, since there are not many multiple-scatter events. 
The requirement that the amplitude $A\leq7$\% is barely restrictive at all.

\begin{figure}[t]
\begin{center}
\psfig{file=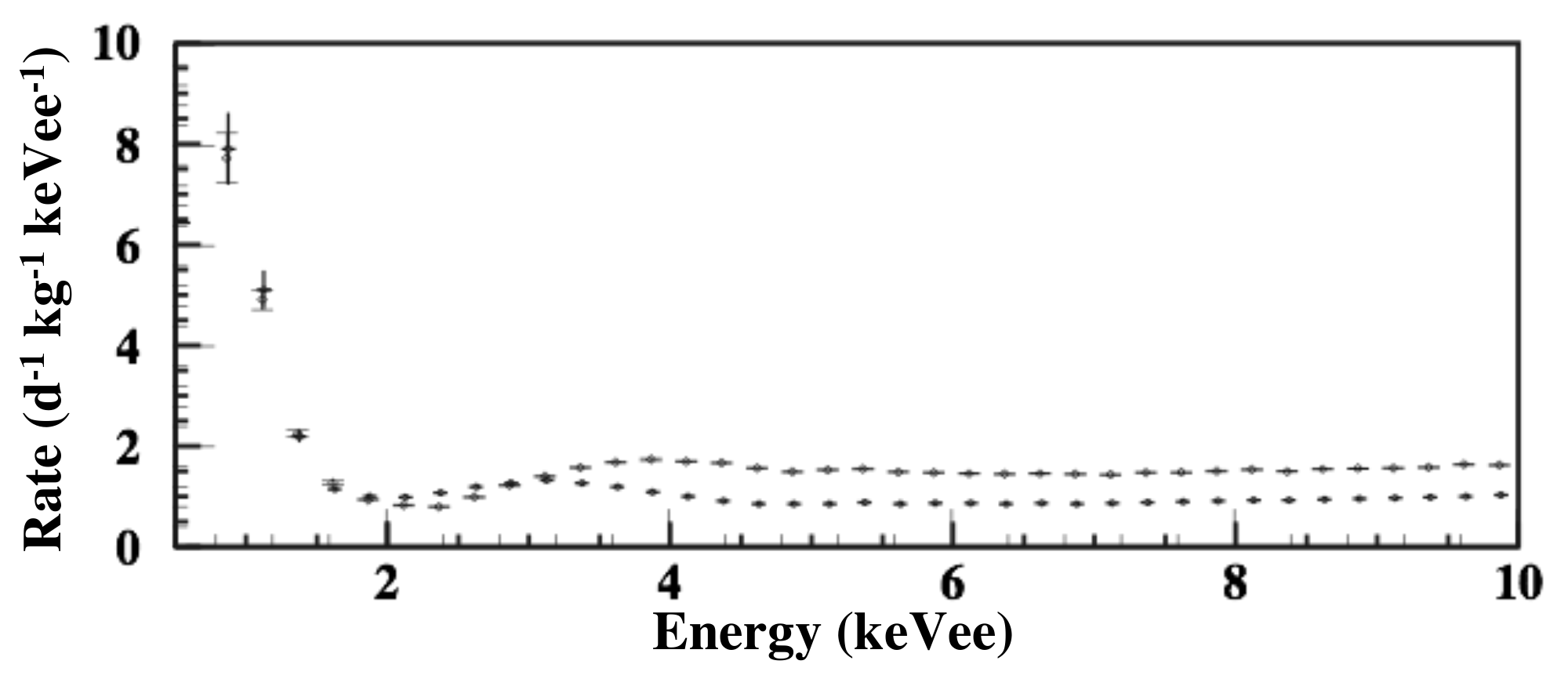,width=4.5in}
\end{center}
 \caption{
Overall single-hit rate in the DAMA/NaI (higher curve at high energies) and DAMA/LIBRA (lower curve at high energies) experiments as a function of the electron-equivalent energy.
Between 2--3\,keVee the spectra cross.
The small peak (0.7\,perkgd spread across 1.5\,keVee) evident in the DAMA/LIBRA spectrum near 3\,keVee is presumably due to $^{40}$K. 
Figure based on those in 
Refs.~\refcite{dama2004,damalibra2008}.
 }
 \label{fig:DAMAdc}
\end{figure}

Because the overall, unmodulated background was designed to be lower in DAMA/LIBRA than in DAMA/NaI, the consistency check between the two setups is interesting.  As shown in Fig.~\ref{fig:DAMAdc}, the continuum background $\gapprox$4\,keVee is $\sim$2$\times$ lower in the LIBRA setup.  Below about 3.5\,keVee, the backgrounds are surprisingly similar, with a slightly higher (statistically significant) rate in DAMA/LIBRA than in DAMA/NaI  in the range 2--2.5\,keVee.  
Any explanation of the modulation in terms of backgrounds would need to be consistent with this behavior of the overall rate.  

As can be seen by eye from Fig.~\ref{fig:DAMAmodulation}, the amplitude of the annual modulation is somewhat smaller for the DAMA/LIBRA setup than for the DAMA/NaI setup (although the larger uncertainties on the DAMA/NaI data make the difference appear larger than it is).  Table~\ref{tab:libra2}, from 
Ref.~\refcite{damalibra}, compares the fits to  the annual modulation with the function $A \cos \omega(t-t_0)$ for the two individual set-ups and together.
The worst consistency is for the full 2--6\,keVee region.
DAMA/NaI shows a modulation amplitude nearly twice as large as that for DAMA/LIBRA, $A = 0.0200 \pm 0.0032$ in comparison to $A= 0.0107 \pm 0.0019$.  This difference is $0.0093\pm0.0037$, about $2.5\sigma$, which happens by chance 1.2\% of the time.  While such a probability is certainly not so low as to indicate that the two experiments are inconsistent, it does not provide especially strong constraints on the possibility that some background may be causing the modulation.  In particular, the fact that the NaI modulation is larger primarily in the same region of the energy spectrum ($>5$\,keVee, and to a lesser extent $>4$\,keVee) for which its background is larger makes it possible that the reduction in background between DAMA/NaI and DAMA/LIBRA is what has caused the reduction in modulation amplitude.  

\begin{table}[!ht]
\tbl{Results of fits to $A \cos \omega(t-t_0)$ with the residual rates of the single-hit scintillation events,
collected by DAMA/NaI, by DAMA/LIBRA and by the two experiments together in the 2-- 4, 2--5, 
and 2--6\,keVee energy intervals~\cite{damalibra2008}.
The last column shows the C.L.\ obtained from the fitted modulation amplitudes.
}
{\begin{tabular}{lcclc}
\toprule
                 & $A$ (\perkeekgd)    & $T=\frac{2\pi}{\omega}$ (yr) & $t_0$ (day) & C.L. \\ 
                 \colrule
 DAMA/NaI        &                     &                   &              &             \\
 2--4 keVee      & $0.0252 \pm 0.0050$ & $1.01 \pm 0.02$   & $125 \pm 30$ & $5.0\sigma$ \\ 
 2--5 keVee      & $0.0215 \pm 0.0039$ & $1.01 \pm 0.02$   & $140 \pm 30$ & $5.5\sigma$ \\
 2--6 keVee      & $0.0200 \pm 0.0032$ & $1.00 \pm 0.01$   & $140 \pm 22$ & $6.3\sigma$ \\ 
 \colrule
 DAMA/LIBRA      &                     &                   &              &             \\
 2--4 keVee      & $0.0213 \pm 0.0032$ & $0.997 \pm 0.002$ & $139 \pm 10$ & $6.7\sigma$ \\ 
 2--5 keVee      & $0.0165 \pm 0.0024$ & $0.998 \pm 0.002$ & $143 \pm  9$ & $6.9\sigma$ \\
 2--6 keVee      & $0.0107 \pm 0.0019$ & $0.998 \pm 0.003$ & $144 \pm 11$ & $5.6\sigma$ \\ 
 \colrule
 DAMA Combined      &           &                   &              &             \\
 2--4 keVee      & $0.0223 \pm 0.0027$ & $0.996 \pm 0.002$ & $138 \pm  7$ & $8.3\sigma$ \\ 
 2--5 keVee      & $0.0178 \pm 0.0020$ & $0.998 \pm 0.002$ & $145 \pm  7$ & $8.9\sigma$ \\
 2--6 keVee      & $0.0131 \pm 0.0016$ & $0.998 \pm 0.003$ & $144 \pm  8$ & $8.2\sigma$ \\ 
\botrule
\end{tabular}
}
\label{tab:libra2}
\end{table}

The DAMA collaboration have taken strong steps to assure and to check that no systematic effect could be causing the modulation. 
No suggested background process appears likely to yield modulation as large as the observed signal~\cite{dama2000sys,dama2003long}.  
Although no possibility appears plausible, the most likely ones (not discussed in 
Ref.~\refcite{dama2000sys}) 
are from a possible modulation in the rate of 3.2\,keV x-rays from potassium~\cite{sunilprivate}, in the rate of some muon-induced x-ray~\cite{DAMAmuonidea}, or in the effectiveness of the phototube noise cut. Ref.~\refcite{gaitskellreview2004} describes additional possible systematic effects that also do not appear likely to be the cause of the modulation.

Potassium-40 is a common cause of background events in dark matter experiments due to its high abundance and $10^9$-year halflife.  About 10\% of the time, $^{40}$K decays by electron capture to an excited state of $^{40}$Ar, which subsequently decays by emission of a 1461\,keV gamma ray.  Since the daughter Ar atom is still missing a K-shell electron, 3.2\,keV emerges  
when an L-shell electron drops down to fill the vacancy. 
Since low-energy X-rays and electrons have short interaction lengths, this low-energy deposition essentially always appears in the parent detector where the  $^{40}$K decays, but the 1461\,keV gamma ray has a longer interaction length and so may deposit some or all of its energy in the parent detector, other detectors, or passive material such as copper.
The DAMA collaboration have counted events with the 1461\,keV gamma ray in one detector and 3.2\,keV in another, and compared the rate to simulations of this process in their detectors to infer the natural K contamination in their detectors is 13\,ppb~\cite{bernabeiprivate}.  
This rate is about right~\cite{DAMAvitaly} to be responsible for the small peak  in the overall DAMA/LIBRA event spectrum (0.45\,\perkeekgd over 1.5 keVee) at about 3\,keVee (see Fig.~\ref{fig:DAMAdc}).
A 4\% modulation in the rate of the accepted single-scatter 3.2\,keVee events could cause DAMA's observed modulation.
This size modulation appears too large to be able to be caused by a modulation of the trigger efficiency and analysis cuts~\cite{dama2000sys}, but given the similarity of the $^{40}$K background to the modulation signal it would be prudent for DAMA to check the stability of triggering, analysis cuts, and overall rate of the 3.2\,keVee events that are coincident with 1461\,keV gamma rays.  It is also barely imaginable that changing detector response or actual movement of the $K$ (due to changing humidity, for example) could result in a modulation of the single-scatter rate if such a change alters the probability that the 1461\,keV gamma rays deposit any energy in the parent detector. Therefore, it would also be useful for the DAMA collaboration to check whether reasonable changes in their simulation could result in a different rate of 3.2\,keV single-hit events.  

The known physical side reaction whose modulation best matches that of DAMA's signal is the annual modulation of the muon rate due to seasonal changes in the atmospheric temperature.
An increase in temperature of the stratosphere 
decreases its density, reducing the chance of meson interactions, resulting in a larger fraction of mesons decaying
into muons. From 1991--1994, 
MACRO measured the annual modulation of the muon rate at Gran Sasso~\cite{macroMuonSeasonal}, finding the phase nearly matches that of the DAMA signal.  The maximum occurs in mid-June, a few weeks later than the best-fit phase of DAMA's signal, perhaps consistent within uncertainties.  The muon modulation also appears less regular than that of DAMA's signal.
DAMA claims that the modulation is too small for muon-induced neutron interactions to be a significant source of background modulation~\cite{dama2000sys}.  
An alternate possibility that has not yet been tested is that 
muon spallation in NaI detectors may create a metastable isotope that decays with emission of 3\,keV~\cite{DAMAmuonidea}.
The lifetime of the metastable state would have to be $> 500$\,\micros\ to avoid DAMA's trigger holdoff time.  A beam test of NaI could see if such a line exists.

Finally, it is imaginable that an instrumental effect could be modulating annually.
For example, DAMA analysis includes a cut on the sharpness of a pulse that is used to remove ÒPMT noiseÓ events, which otherwise would pollute the lowest energy bins.
It is not possible to know what the tail of the PMT noise distribution looks like, since measurement always has particle-induced backgrounds, so it is not clear that the cut removes all PMT noise events.   It is possible that a small number of noise events passes this cut, and the number that passes this cut modulates.
Although this possibility is poorly motivated, dark current in PMTs depends exponentially on temperature, and it is possible that DAMA's temperature stability is not good enough to prevent a modulation of this potential effect.

The DAMA collaboration have made some checks of effects related to a possible modulation of noise events.  They have determined that the total number of noise events does not modulate (but they cannot check whether the number passing the cut modulates), and they have checked (using a high-rate calibration source) that the 
efficiency of the cut for particle interactions does not modulate.
Studies of the effect of the cut position on modulation amplitude could help rule out a possible modulation of noise events passing the cut.
If different cut positions all showed no modulation in the particle efficiency (using the calibration source) and affected the WIMP-search modulation amplitude in a way consistent with this efficiency for particles, it would be strong evidence that the modulation is due to particle origin.

Other experiments may be able to test DAMA's results more directly soon, using similar detectors.  The KIMS experiment, situated in Yangyang Lab, Korea, has already yielded limits~\cite{kims2007} from running four 8.7\,kg CsI crystals for a total exposure of 3409\,kg\,d.  Twelve detectors, for a total mass of 104.4\,kg, are now installed.  The ANAIS collaboration performed tests on a NaI crystal to determine its background and test its energy threshold~\cite{anais2010taup}.  While a 2\,keVee energy threshold was achieved, the $^{40}$K background level in the NaI crystals was found to be too high to allow a test of DAMA's results.  Work on producing cleaner crystals is in progress.

\subsection{Directional Detectors}
\label{section:directional}

\begin{figure}[t]
\begin{center}
\psfig{file=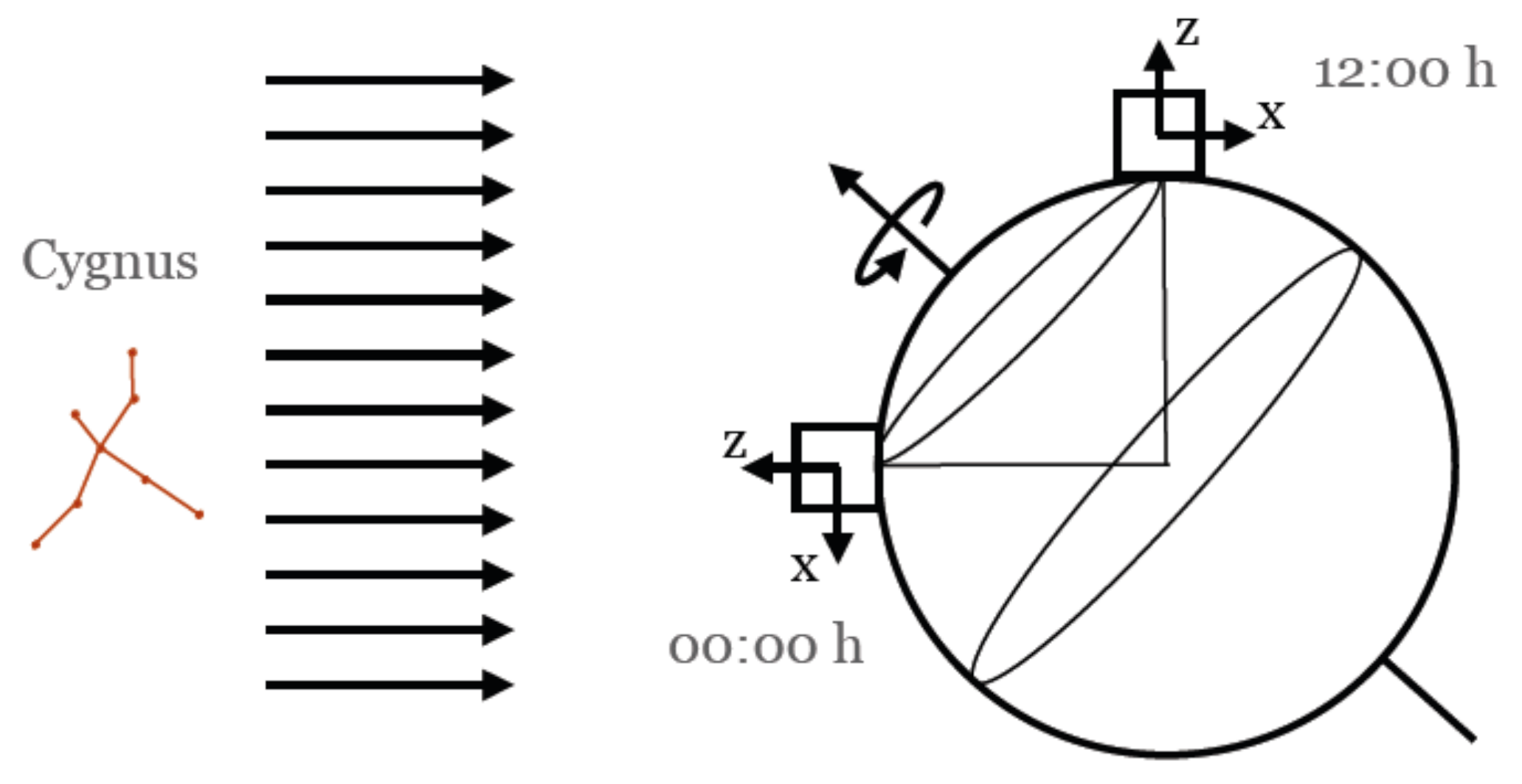,width=4.5in}
\end{center}
 \caption{
Approximate spin axis of the Earth relative to the WIMP ``wind'' due to the motion of the Earth through the Galaxy.  A detector at latitude $+60^{\circ}$ would see most scattering WIMPs come predominantly from the lab's ceiling at one time during the day and from the northern side wall twelve hours later, with the pattern repeating each sidereal day.  Figure based on that in
Ref.~\refcite{spoonerPDM}.
 }
 \label{fig:directional}
\end{figure}

Although inherently difficult, detection of the direction of scattering WIMPs would provide significant additional information.  First of all, such a detection would provide a clear signature of the non-terrestrial nature of the signal, since the motion of the Earth through the Galaxy results in a predominant WIMP ``wind'' in the opposite direction, from the constellation Cygnus. 
As shown in Fig.~\ref{fig:directional}, a detector would see 
this wind vector oscillate over the course of a day.  Over the course of a year, this directionality would repeat each sidereal day, hence rapidly going out of phase with the solar day. 
With a detector able to distinguish the full direction of recoils (\ie\ both the axis and which ends of the track are the head and tail), $\sim10$ events would be sufficient to determine that the WIMP signal is not isotropic~\cite{copi1999direction,copi2000direction,MorganDirectional1}.
As mentioned above in Section~\ref{detection}, directional detection would provide
excellent information on the WIMP velocity distribution~\cite{MorganDirectional1,MorganDirectional2}.

Unfortunately, recoil track lengths are too short in solid or liquid detectors ($\sim0.1$\,\micron\ in a condensed matter target) for direct measurement.  While the possibility of using a detector with non-uniform directional response has been considered (\eg\ Ref.~\refcite{spoonerPDM} and references therein), gaseous detectors appear the most promising, despite the difficulty in making them massive enough to detect a WIMP signal.  In order to make the recoil track length long enough to be imaged ($\sim1$\,mm), traditional gas time-projection chambers may be used if operated at low pressure $(<100$\,Torr).  As with the liquid noble time-projection chambers, position in the $x-y$ plane is determined by the position of charges collected (by crossed planes of wires~\cite{drift2nim2005} , micropixels~\cite{newage2009}, micromegas~\cite{mimac2009}, or via detection of electroluminescence~\cite{dmtpc2009}), while position along the drift direction is determined by drift time.  However, since no prompt signal is detected, the interaction time is unknown, and position information along the drift direction is only relative, not absolute.

Directional detectors potentially provide the best observables of any dark matter experiments.  The total charge indicates the energy of the recoil.  The tracklength of electron recoils is much larger than that of nuclear recoils, so comparison of the observed tracklength to the energy allows excellent  rejection of electron-recoil backgrounds.   The track itself easily indicates the axis of the recoil.  Measurement of the amount of charge released along the track can allow inference of the sense of the direction of the track (\ie\ can allow the head of the track to be distinguished from its tail).  For nuclear recoils, energies of interest are below the Bragg peak,
so the energy deposition is larger at first, then decreases as the recoiling nucleus loses energy.
Unfortunately, increased scattering near the end of the track, as well as straggling in the ion drift, can make correct identification of the head and tail of a track difficult~\cite{majewski2009headtail}.

The four directional detectors listed in Table~\ref{TableExperiments} are all in research and development phase, as their low masses do not yet permit competitive sensitivities.  Chosen gases typically maximize sensitivity to spin-dependent interactions. 
The NEWAGE collaboration has run a small, 10-g test cell with micropixel readout underground, achieving the best limits so far of any directional detector~\cite{newage2010}, albeit still 5 orders of magnitude less sensitive than the world's best proton-spin-dependent limits.
The MIMAC collaboration has run test chambers with micromegas~\cite{mimac2009}; much effort has been concentrated on low-energy measurements of the quenching factors for gases of interest. 
Two of the experiments, DRIFT~\cite{drift2nim2005} and DMTPC~\cite{dmtpc2009}, have demonstrated head-tail discrimination of the track.  The DMTCP measurement was at somewhat higher energies and tracklengths than needed for a dark matter search ($>200$\,keV)~\cite{dmtpc2008headtail500,dmtpc2008headtail200}, while the DRIFT measurement shows some discrimination down to 50\,keV~\cite{drift2009headtail}. 

It remains to be seen
if the WIMP-nucleon cross section is large enough that one can build 
gas detectors with enough mass to detect the signal.  
A 1-ton gas detector, with spin-independent sensitivity reach $\sim10^{-8}$\,pb, would be about the size of the largest accelerator-beam detectors.  To obtain sensitivity to $\sim10^{-10}$\,pb would require a detector $\sim5\times$ the size of SuperKamiokande, smaller than proposed next-generation neutrino and proton-decay experiments.
Identifying a technology to achieve adequate spatial resolution over very large areas at reasonable cost will be a challenge.  The benefits of a directional detection certainly warrant continued development to attempt to reach this lofty goal.

\section{Conclusions and Prospects}

Currently, CDMS~II and XENON-10 are the most sensitive WIMP-search experiments, with limits on the spin-independent WIMP-nucleon cross sections below $10^{-7}$\,pb,  probing the MSSM region.
Threshold detectors and KIMS, a room-temperature scintillator, have the best direct sensitivity to spin-dependent interactions on protons, with indirect detection experiments SuperKamiokande and IceCube more sensitive for some models.
Truly, a variety of technologies has been successful and continue to show promise for the future.
Significant improvements by the end of 2010 are likely from XENON-100, EDELWEISS, and COUPP, with many experiments (LUX, WArP, XMASS, SuperCDMS, and MiniCLEAN, in addition to XENON-100 and EDELWEISS) likely to surpass $10^{-8}$\,pb sensitivity for spin-independent interactions by 2012. 
The next five years should see spin-independent sensitivity improve to $10^{-9}$\,pb, or possibly even $10^{-10}$\,pb
if promising technologies can achieve the necessary background rejection. 
In particular, it is unclear what is the best technology for the multiton scale needed to achieve  $10^{-11}$\,pb sensitivity. 

\begin{figure}[t]
\begin{center}
\psfig{file=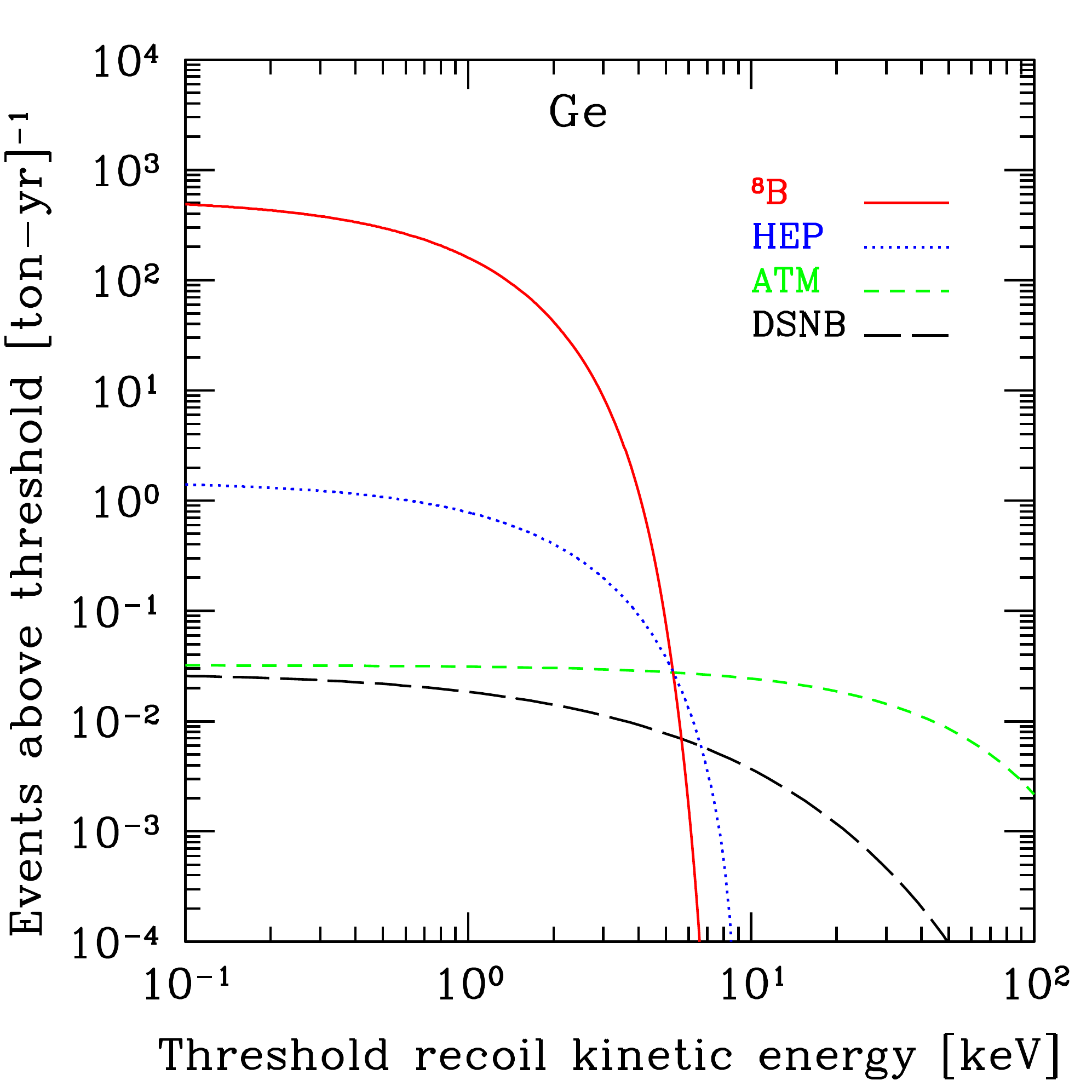,width=2.22in}
\psfig{file=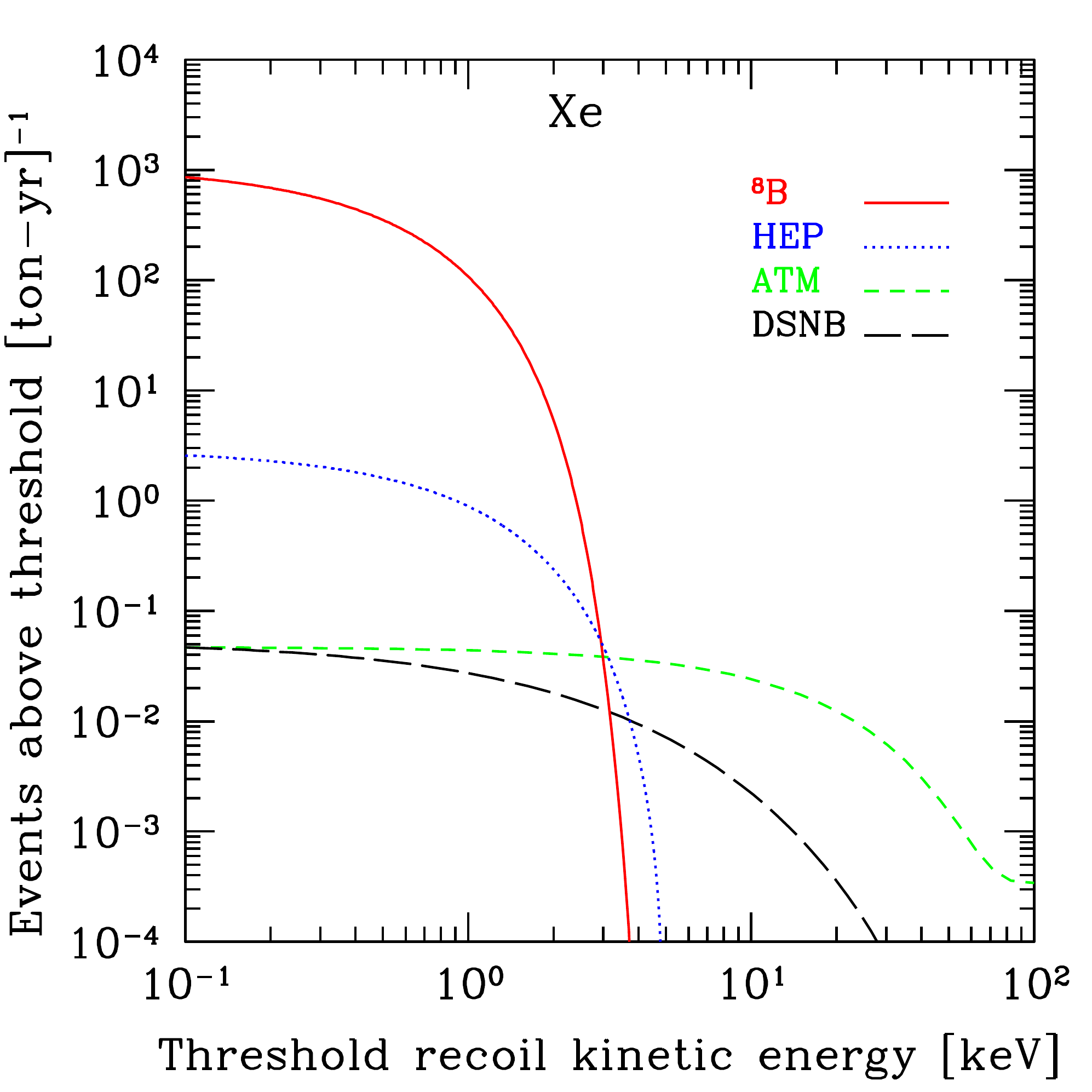,width=2.22in}
\end{center}
 \caption{
Rate of coherent neutrino interactions above a given threshold  energy for Ge (left) and Xe (right). For both the
diffuse supernova (DSNB, large black dashes) and atmospheric (ATM, small green dashes) event rates, the sum of all contributing neutrino
flavors is shown.  The significant background from $^8$B neutrinos (red solid curve) will prevent experiments from using threshold energies $\lapprox3$\,keV for Xe, $\lapprox5$\,keV for Ge, 
$\lapprox10$\,keV for Ar, or $\lapprox20$\,keV for F.
  Figures from 
Ref.~\refcite{StrigariNu}.
 }
 \label{fig:nu}
\end{figure}

The ultimate unrejectable, unshieldable  background for future WIMP-search experiments will be coherent neutrino-nucleus interactions (see Fig.~\ref{fig:nu})~\cite{StrigariNu}.  A large rate of interactions of solar $^8$B neutrinos will prevent ton-scale experiments from using threshold energies $\lapprox5$\,keV for Ge, $\lapprox3$\,keV for Xe, $\lapprox10$\,keV for Ar or Si, and $\lapprox15$\,keV for Ne or F (although subtraction of the expected energy spectrum would be possible, and rejection would be possible for directional detectors~\cite{HendersonMaximumPatch}.)  
For yet larger detectors, atmospheric neutrinos will provide background events over the entire energy range of the WIMP search, likely limiting the sensitivity reach of ultimate dark matter experiments to no better than $\sim10^{-12}$\,pb.

A direct detection of WIMPs would warrant follow-up in order to learn as much as possible about them.
Measurement of the WIMP recoil spectrum with good statistics would constrain the WIMP's mass,
potentially demonstrating that a particle produced at accelerators indeed comprises the dark matter in the galaxy. Such large statistics (provided experimental operation is sufficiently stable)
could also take advantage of the expected annual modulation of the WIMP signal to 
confirm the extraterrestrial origin of the WIMPs and to learn more about their distribution in the galaxy.
In addition, detection using different target nuclei would potentially allow determination of both the spin-independent and spin-dependent cross sections.
A measure of the direction of the recoiling nucleus would provide additional information on
the distribution of WIMPs in the galaxy, allowing WIMP astronomy.

Ultimately, the combination of WIMP direct and indirect detection with studies at colliders and
of the cosmic microwave background could answer fundamental questions beyond whether WIMPs are the dark matter and neutralinos exist.  
These combinations can determine whether the WIMPs are stable and whether there is non-baryonic dark matter other than WIMPs.
As thermal relics, the WIMPs could provide a window to the early universe, or we may learn that the WIMPs must have been generated out of thermal equilibrium.
WIMP astronomy could teach us about galaxy formation.
Furthermore, the combination of information from  these several methods may constrain 
the particle properties significantly better than colliders alone~\cite{Baltz2006LCC,Bertone2010}.
With the LHC taking data and more sensitive indirect and detect detection experiments in operation, the next five years may see an answer to the fundamental mystery of the nature of dark matter, as well as to other fundamental questions about the universe.

\section*{Acknowledgments}

I would like to thank Scott Dodelson and Csaba Csaki for organizing TASI 2009.
This work was supported in part by the National Science Foundation grant PHY-0855525.

\bibliographystyle{utphys}
\bibliography{schnee}

\end{document}